\documentclass[prd,nofootinbib,11pt,tightenlines,floatfix,eqsecnum,superscriptaddress,showkeys]{revtex4-1}

\usepackage[utf8]{inputenc}          
\usepackage{graphicx,xcolor}         
\usepackage{array,dcolumn,longtable} 
\usepackage{amsmath,amssymb,slashed} 
\usepackage{bm,xfrac}

\providecommand\lfstyle{}                   
\providecommand\romanup[1]{\text{#1}}       
\providecommand\greekup[1]{#1}              
\renewcommand\textsc{\MakeUppercase}

\usepackage{siunitx}                                  
\newcommand\rhic{\textsc{rhic}}

\newcommand\lhc{\textsc{lhc}}

\newcommand\STAR{\textsc{star}}
\newcommand\phenix{\textsc{phenix}}

\newcommand\cms{\textsc{cms}}

\newcommand\dabmod{\textsc{dab-mod}}

\newcommand\mckln{\textsc{mckln}}

\newcommand\qgp{\textsc{qgp}}

\newcommand\adscft{\textsc{a}d\textsc{s}/\textsc{cft}}

\declareslashed{}{\divslash}{0.08}{0}{\mathcal{D}}
\renewcommand\vec[1]{\ensuremath{\bm{#1}}}
\newcommand\dimensions[1]{#1\romanup{D}}

\newcommand\econst{\romanup{e}}
\newcommand\imaginary{\romanup{i}}

\newcommand\dd{\mathop{}\!\romanup{d}}

\DeclareSIUnit{\fm}{\femto\metre}

\newcommand\qcharm{{\romanup{c}}}

\newcommand\qbottom{{\romanup{b}}}
\newcommand\lelectron{{\romanup{e}}}
\newcommand\lmuon{{\greekup{\mu}}}

\newcommand\electronpositron{{\lelectron^\pm}}

\newcommand\muonpm{{\lmuon^\pm}}
\newcommand\Bmeson{{\romanup{B}}}
\newcommand\Dmeson{{\romanup{D}}}

\newcommand\Bzero{{\Bmeson^0}}
\newcommand\Dzero{{\Dmeson^0}}
\newcommand\Bplus{{\Bmeson^+}}
\newcommand\Dplus{{\Dmeson^+}}

\newcommand\Dstar{{\Dmeson^{*}}}

\newcommand\PbPb{{\romanup{PbPb}}}
\newcommand\AuAu{{\romanup{AuAu}}}

\newcommand\snn[1][]{\sqrt{s_\text{NN}}\ifx\\#1\\\else=\SI{#1}{\TeV}\fi}
\newcommand\snnGeV[1][]{\sqrt{s_\text{NN}}\ifx\\#1\\\else=\SI{#1}{\GeV}\fi}

\newcommand\pt{p_\text{T}}

\newcommand\raa{R_\text{AA}}
\newcommand\vn[1]{v_{#1}}
\newcommand\vnn{\vn{n}}
\newcommand\psin[1]{\psi_{#1}}
\newcommand\psinn{\psin{n}}

\newcommand\Vn[1]{V_{#1}}
\newcommand\Vnn{\Vn{n}}
\newcommand\Td{T_\text{d}}

\renewcommand\Re{\ensuremath{\operatorname{Re}}}
\newcommand\llangle{\ensuremath{\langle\!\langle}}

\newcommand\gammaflow{\varGamma_\text{flow}}
\newcolumntype{C}[1]{>{\vspace{1.5pt}\centering\let\newline\\\arraybackslash\hspace{0pt}}m{#1}}

\begin{document}
\title{\dabmod\ sensitivity study of heavy flavor \textmd{$\raa$} and azimuthal anisotropies based on beam energy, initial conditions, hadronization, and suppression mechanisms.}
\date{\today}

\author{Roland Katz}
\affiliation{SUBATECH, Universit\'e de Nantes, EMN, IN2P3/CNRS, 44307 Nantes, France}
\author{Caio A.~G.~Prado}
\affiliation{Institute of Particle Physics, Central China Normal University (CCNU), Wuhan, Hubei 430079, China}
\author{Jacquelyn Noronha-Hostler}
\affiliation{Department of Physics and Astronomy, Rutgers University, Piscataway, NJ 08854, USA}
\author{Jorge Noronha}
\affiliation{Instituto de F\'{i}sica, Universidade de S\~{a}o Paulo, C.P. 66318, 05315-970 S\~{a}o Paulo, SP, Brazil}
\author{Alexandre A.~P.~Suaide}
\affiliation{Instituto de F\'{i}sica, Universidade de S\~{a}o Paulo, C.P. 66318, 05315-970 S\~{a}o Paulo, SP, Brazil}

\begin{abstract}
    Heavy flavor probes provide important information about the in-medium properties of the quark-gluon plasma produced in heavy-ion collisions. In this work, we investigate the effects of $\dimensions{2}+1$ event-by-event fluctuating hydrodynamic backgrounds on the nuclear suppression factor and momentum anisotropies of heavy flavor mesons and non-photonic electrons. Using the state-of-the-art $\Dmeson$ and $\Bmeson$ mesons modular simulation code (called ``\dabmod''), we perform a systematic comparison of different transport equations in the same background, including a few energy loss models --- with and without energy loss fluctuations --- and a relativistic Langevin model with different drag parametrizations. We present the resulting $\Dmeson$ and $\Bmeson$ mesons $\raa$, $\vn2$, $\vn3$, and $\vn4$ as well as multi-particle cumulants, in $\AuAu$ collisions at $\snnGeV[200]$ and $\PbPb$ collisions at $\snn[2.76]$ and $\snn[5.02]$, and compare them to the available experimental data. The $\vn2\{4\}/\vn2\{2\}$ ratio, which is known to be a powerful probe of the initial conditions and flow fluctuations in the soft sector, is also studied in the context of heavy flavor. We also investigate the correlations between the transverse anisotropies of heavy mesons and all charged particles to better understand how heavy quarks couple to the hydrodynamically expanding quark-gluon plasma. We study the influence that different initial conditions and the implementation of heavy-light quark coalescence has on our results.
\end{abstract}
\maketitle

\section{Introduction}

The quark-gluon plasma (QGP) --- a deconfined state of mater composed of strongly interacting quarks and gluons --- has been successfully reproduced in the laboratory both at the Large Hadron Collider (\lhc) and at the Relativistic Heavy Ion Collider (\rhic). While it remains unknown how the strongly fluctuating color fields present in the initial stages of heavy-ion collisions~\cite{Schenke:2012wb} may later evolve to a state that displays hydrodynamic behavior (for a review, see~\cite{Romatschke:2017ejr}), it is well established by now that the soft, low transverse momentum $\pt$ sector of the QGP determined by the particle spectra and flow harmonics of light hadrons can be generally well-described by event-by-event relativistic viscous hydrodynamic simulations~\cite{Niemi:2015voa,Noronha-Hostler:2015uye,Eskola:2017bup,Giacalone:2017dud,Gale:2012rq,Bernhard:2016tnd}. Some of the properties of the early stages of heavy-ion collisions can be investigated using hard probes, such as light flavor jets and heavy flavor hadrons, which are created at very early times via hard scattering processes and then propagate through the evolving medium. Because of the large mass of the heavy quarks, many times larger than the cross-over temperature from the QGP to the hadron resonance gas~\cite{Borsanyi:2010bp}, they are not likely to be created within the hydrodynamic evolution of the QGP and, thus, heavy flavor content is preserved from the initial stages well into hadronization~\cite{Moore:2004tg}. Therefore, heavy flavor observables can provide key information about the early time dynamics of heavy-ion collisions.

Furthermore, heavy mesons in the intermediate/high $\pt$ region lose energy mostly from radiative processes whereas the low $\pt$ regime is dominated by collisional processes that may be described via Langevin-like equations~\cite{Andronic:2015wma, Moore:2004tg}. The two main experimental measurements involving open heavy flavor D and B mesons that we are interested in are the nuclear modification factor, $\raa$, and the azimuthal anisotropies, $v_n(\pt)$. $\raa$ mainly encodes how much energy is lost in the medium compared to elementary pp collisions and, since the farther the heavy quark travels through the QGP medium the more energy it loses, one expects that heavy quarks lose more energy in media formed by heavy-ion collisions at high energies. Additionally, because each event has on average certain geometrical features (an almond-like shape especially for mid-central collisions), one expects that heavy quarks lose more energy along the long axis of the almond shape (in-plane) vs. the short axis (out-of-plane), which leads to a final elliptical azimuthal anisotropy $v_2$ defined by the second  Fourier coefficient of the particle spectra ~\cite{Wang:2000fq,Gyulassy:2000gk}.

However, quantum mechanical fluctuations involving the position of the incoming nucleons, and likely also the fluctuations of quark and gluon fields, lead to other initial geometrical patterns such as, e.g., a triangular geometry, producing $v_3$ and other even higher order harmonics~\cite{Nahrgang:2014vza,Prado:2016szr}, which have been measured in the heavy flavor sector at \rhic~\cite{Lomnitz:2016rpz} and the \lhc~\cite{Aaboud:2018bdg}. Furthermore, we note that the overall flow of the medium has been shown to influence energy loss mechanisms~\cite{Brewer:2017fqy} and hard probes can provide important information about the properties of the initial state~\cite{Cao:2014fna,Andres:2019eus}. More specifically, it was recently shown in~\cite{Andres:2019eus}  that the time $\tau_0$ after which hard probes begin to lose energy appears to be larger than was initially expected. Further proof of the influence of event-by-event fluctuations of the initial conditions on the heavy flavor sector can be seen in event-shape engineering calculations~\cite{Prado:2016szr} and measurements~\cite{Acharya:2018bxo} of heavy flavor flow harmonics.

Theoretical comparisons to the measured nuclear modification factor demonstrate that energy loss models alone significantly underpredict $\raa$ in the low $\pt$ sector (azimuthal anisotropies experience the same issue) though they can describe the high $\pt$ regime quite well~\cite{Prado:2016szr}. On the other hand, Langevin or Boltzmann based models perform quite well in the low $\pt$ sector but the addition of coalescence is needed to reproduce experimental data~\cite{Greco:2003xt,Greco:2003mm,Fries:2003vb,Greco:2003vf,Oh:2009zj,Nahrgang:2014vza,Cao:2016gvr}. In this manner, the $\pt$ dependence of $\raa$ can be a useful tool to understand different regimes of heavy flavor energy loss. The heavy flavor $v_n$'s depend on the assumptions regarding the heavy flavor model as well but they are also sensitive to the choice of initial conditions, which are strongly related to final result for the flow harmonics (proven by the strong correlation quantified by the Pearson coefficient between the initial eccentricity vector and the final $V_n$ vector). There is likely also a sensitivity to the choice of shear viscosity to entropy density ratio, $\eta/s$, see~\cite{Betz:2016ayq} and~\cite{Esha:2016svw}. Thus, special care must be taken to first fully constrain the initial conditions and medium viscosity using the soft sector before calculating heavy flavor flow harmonic observables. This has been done in recent years in~\cite{Prado:2016szr} and also by DUKE/SUBATECH~\cite{Cao:2015hia,Nahrgang:2014vza}, which has led to a better theoretical understanding of heavy flavor momentum anisotropies.

It is important to note that event-by-event initial state fluctuations are not the only source of fluctuations that can affect heavy flavor $\raa$ and $v_n$. In fact, energy loss fluctuations can also occur in a given event and that has already been shown to affect light flavor high $\pt$ flow harmonics~\cite{Zapp:2013zya,Betz:2014cza,Betz:2016ayq}. Thus, a systematic study of the effect of energy loss fluctuations should also be performed in the heavy flavor sector. Furthermore, the advent of multi-particle flow cumulants involving hard probes (where the hard probe particle of interest is correlated with other reference soft particles~\cite{Sirunyan:2017pan,Betz:2016ayq}) provides a unique opportunity to study heavy flavor from an entirely new angle.

In order to confront this complicated emerging picture of the heavy flavor sector where experimental observables are influenced by multiple competing factors, in this paper we systematically study
 the effects of $\dimensions{2}+1$ (i.e., longitudinally boost invariant) event-by-event fluctuating hydrodynamic backgrounds on the nuclear suppression factor and momentum anisotropies of heavy flavor mesons and non-photonic electrons. The bulk dynamics of the medium is simulated using the event-by-event relativistic viscous hydrodynamic model, v-USPhydro~\cite{Noronha-Hostler:2013gga,Noronha-Hostler:2014dqa}, coupled to either \mckln\ initial conditions~\cite{Drescher:2006pi,Drescher:2007ax,Drescher:2006ca} or Trento initial conditions (tuned to IP-Glasma~\cite{Gale:2012rq})~\cite{Moreland:2014oya}, with hydrodynamic parameters constrained to describe the relevant properties of the soft sector. The heavy flavor sector is described using the state-of-the-art $\Dmeson$ and $\Bmeson$ mesons modular simulation code (called ``\dabmod'')\footnote{A preliminary version of \dabmod, which did not yet include coalescence and Langevin dynamics, was used in~\cite{Prado:2016szr}.}, with which we perform a systematic study of different transport equations, including a few energy loss models (with and without energy loss fluctuations) and a relativistic Langevin model with different drag parametrizations, and investigate their effect on $\raa$ and $v_n$.
We present calculations of $\Dmeson$ and $\Bmeson$ meson $\raa$, $\vn2$, $\vn3$, and $\vn4$ as well as multi-particle flow cumulants in $\AuAu$ collisions at $\snnGeV[200]$ and $\PbPb$ collisions at $\snn[2.76]$ and $\snn[5.02]$, and compare them to the available experimental data. Heavy flavor multi-particle flow cumulants are investigated via the $\vn2\{4\}/\vn2\{2\}$ ratio, which plays an important role in determining  the initial conditions and flow fluctuations in the soft sector. We also study the correlations between the momentum anisotropies of heavy mesons and the flow of all charged particles to better understand how heavy quarks couple to the expanding medium.

This paper is organized as follows. In the next section we explain the different assumptions and details regarding the modeling of heavy flavor dynamics included in \dabmod. Sec.\ \ref{SecIII} investigates how the transport model assumptions and variations in the decoupling temperature affect the observables considering the case of \mckln\ initial conditions without coalescence. We study how the results change when one replaces \mckln\ by Trento initial conditions in Sec.\ \ref{SecIV}. Sec.\ \ref{Section: coalescence} describes how coalescence is now implemented in \dabmod\ and how it affects our results. Our final remarks are presented in Sec.\ \ref{conclusions}.

\section{Details of the \dabmod\ simulation}\label{Section:DABMODingredients}

To study the propagation of charm and bottom quarks inside the medium created in high energy heavy-ion collisions, we developed a modular Monte Carlo simulation, \dabmod, that allows for a variety of backgrounds and transport models to be implemented.

\subsection{The background hydrodynamically expanding medium}\label{sec:back}

Heavy quark transport models require background medium profiles that provide the temperature and flow velocity of the fluid cells at each time step along the heavy quark trajectories. In this work, we use either the event-by-event Monte Carlo Kharzeev-Levin-Nardi (\mckln\ ) initial conditions --- an implementation of a Color Glass Condensate $k_\text{T}$-factorization model~\cite{Drescher:2006pi,Drescher:2007ax,Drescher:2006ca} --- or the event-by-event Trento initial conditions tuned to IP-Glasma~\cite{Moreland:2014oya}.

\begin{figure}[h!]
  \centering
  \includegraphics[width=0.5\textwidth]{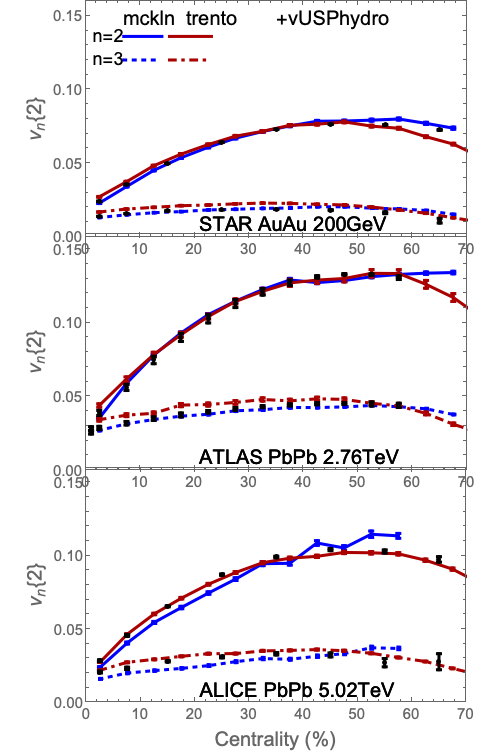}

  \caption{Centrality dependence of $v_n\{2\}$ ($n=2,3$) for $\AuAu$ 200~GeV, $\PbPb$ 2.76~TeV and $\PbPb$ 5.02~TeV comparing \mckln\ and Trento initial conditions coupled to v-USPhydro. Experimental data from STAR~\cite{Adams:2004bi}, ATLAS~\cite{Aad:2013xma,Aad:2014vba}, and ALICE~\cite{Acharya:2018lmh,Acharya:2018zuq,Adam:2016izf}.}
  \label{fig:v22}
\end{figure}

These initial profiles are evolved from an initial time $\tau_0 = \SI{0.6}{\fm}$ using the v-USPhydro code~\cite{Noronha-Hostler:2013gga,Noronha-Hostler:2014dqa,Noronha-Hostler:2015coa}, a $\dimensions{2}+1$ relativistic viscous hydrodynamical model (with a Cooper-Frye freeze-out prescription) that has passed the standard accuracy tests of the field~\cite{Marrochio:2013wla}. The viscous hydrodynamic evolution is encoded in a shear viscosity to entropy density ratio that we set for \mckln\ (Trento) initial conditions to be $\eta/s= 0.08$ ($\eta/s=0.05$) for $\AuAu$ collisions at $\snnGeV[200]$ and to $\eta/s=0.11$ ($\eta/s=0.05$) and $\eta/s=0.05$ ($\eta/s=0.047$) for $\PbPb$ collisions at $\snn[2.76]$ and $\snn[5.02]$ collisions, respectively, following Refs.~\cite{Noronha-Hostler:2015coa,Noronha-Hostler:2016eow}. This model describes experimental data in the soft sector reasonably well and, thus, all the hydrodynamic parameters that involve the medium description are fixed in the present study and are not seen as free parameters of our heavy quark analysis. To obtain sufficient statistics for the heavy quark observables, we use $\sim$ 1000--2000 hydrodynamic events per 10\% centrality range. Finally, we neglect the possible effects of heavy quark energy loss on the evolution of the medium~\cite{Andrade:2014swa}.

In Fig.\ \ref{fig:v22} we show the results for the $\pt$-integrated two particle flow cumulant, $v_n\{2\}$, computed using either \mckln\ or Trento initial conditions coupled to v-USPhydro for $\PbPb$ 5.02 TeV, $\PbPb$ 2.76 TeV, and $\AuAu$ 200 GeV collisions.
We find that \mckln\ initial conditions work best at \rhic\ energies but they can work reasonably well even at \lhc\ if one is interested only in two particle correlations. Additionally, at \rhic\ \mckln\ leads to a slightly smaller $v_3\{2\}$ compared to Trento, which in turn is slightly closer to the experimental data from STAR~\cite{Adams:2004bi}. This small difference may be alleviated with a temperature dependent $\eta/s$. Considering the results for $\PbPb$ 2.76 TeV collisions compared to ATLAS data, we find a reasonably good agreement to experimental data for both \mckln\ and Trento initial conditions. However, at the top \lhc\ energy more deviations from experimental data are seen for \mckln\ initial conditions while Trento's description improves. Fig.\ \ref{fig:v22} shows for the first time a comparison between these results and \lhc\ run 2 $\PbPb$ data from ALICE at 5.02 TeV~\cite{Acharya:2018lmh,Acharya:2018zuq,Adam:2016izf}. Trento initial conditions provide the best fit to ALICE/ATLAS data, although, again, \mckln\ is relatively near the data for $v_2\{2\}$ and $v_3\{2\}$.

\begin{figure}[h!]
  \centering
  \includegraphics[width=0.5\textwidth]{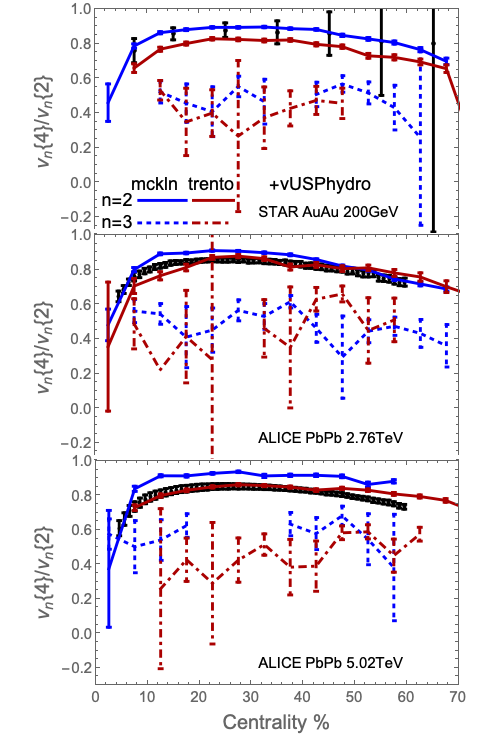}

  \caption{Centrality dependence of $v_2\{4\}/v_2\{2\}$ for $\AuAu$ 200 GeV, $\PbPb$ 2.76 TeV, and $\PbPb$ 5.02 TeV comparing \mckln\ and Trento initial conditions coupled to v-USPhydro. Experimental data from STAR~\cite{Adams:2004bi} and ALICE~\cite{Acharya:2018lmh,Acharya:2018zuq,Adam:2016izf}.}
  \label{fig:soft_fluctuations}
\end{figure}

The main caveat is  that 4-particle correlations and, more specifically, the ratio between $v_n\{4\}$ and $v_n\{2\}$ which is known to be a good constraint on initial conditions~\cite{Giacalone:2017uqx} (and is nearly medium independent), paint a different story. In Fig.\ \ref{fig:soft_fluctuations} the ratio $v_n\{4\}/v_n\{2\}$ for $n=2,3$ is plotted for different beam energies across centrality for Trento vs \mckln\ initial conditions~\cite{Alba:2017hhe}. The ratio $v_2\{4\}/v_2\{2\}$ approaches 1 when there are less $v_n$ fluctuations (i.e.~a narrower distribution) and it decreases with increasing $v_n$ fluctuations. Thus, we find that \mckln\ generally predicts less fluctuations than Trento tuned to IP-Glasma. At \rhic\ this appears to be a good predictor for $v_2\{4\}/v_2\{2\}$ and \mckln\ captures the fluctuations data from STAR well, whereas Trento predicts too much fluctuations. In contrast, at \lhc\ energies we find that \mckln\ predicts too few fluctuations whereas Trento provides a nearly perfect fit to experimental data\footnote{Note that we are avoiding issues found in ultracentral collisions since we do not expect heavy flavor observables to have sufficient statistics to be considered in that case.}. Thus, we conclude that \mckln\ works best at \rhic\ and reasonably well at \lhc\ run 1 but it misses \lhc\ run 2 data whereas Trento manages to do the best job on a global analysis level. However, we note that its predictions for the $v_2\{4\}/v_2\{2\}$ fluctuations at \rhic\ is on the low side.

Finally, we compare our results for $v_3\{4\}/v_3\{2\}$ as well. Even with $\sim 35,000+$ events, statistical error bars (computed using jackknife resampling) are an issue on our side. That being said, we find that both \mckln\ and Trento are within error bars of each other. It does appear that \mckln\ may have less $v_3$ fluctuations compared to Trento but it is difficult to say from hydrodynamics at this level. In~\cite{Giacalone:2017uqx} it was found that $\varepsilon_3\{4\}/\varepsilon_3\{2\}$ was consistently larger for \mckln\ as well.

\subsection{Heavy quark initial conditions}\label{Sec:HQInitialConditions}

Because of their large masses, heavy quarks are produced at the very beginning of the collisions in parton scatterings that can be described by perturbative QCD (pQCD). Neglecting the possible effects of shadowing at mid-rapidities, which are observed to be non negligible for $\pt<10$ GeV~\cite{Jena:2014xga,Song:2015ykw,Kusina:2017gkz}, we assume that the amount of heavy quarks produced per binary nucleon-nucleon collisions and their initial momentum distributions to be the same as in the reference proton-proton collisions. Using Monte Carlo, we then sample the heavy quark initial transverse momenta by using the distributions coming from pQCD FONLL calculations~\cite{Cacciari:1998it,Cacciari:2001td} in proton-proton collisions, choosing their central predictions in the renormalization scale range. The initial azimuthal directions of the heavy quark momenta are chosen randomly. We also sample the spatial distribution of the heavy quarks following the medium initial energy density of each hydrodynamic event, which reflects the initial parton scatterings of the considered collision. Even though in reality the number of heavy quarks at mid-rapidities per heavy-ion collision is on the order of ten, we largely oversample the number of heavy quarks to $\sim 10^7$ for each hydrodynamic event. This large oversampling is necessary for the statistics of some of the observables under study (e.g. the higher order particle cumulants) and it can be justified by the limited number of hydrodynamic events used in the simulation as compared to the large number of collisions in the experiments. Each of our hydrodynamic events can then be seen as a typical event with geometrical properties that correspond to a large number of collisions in the experiment.

\subsection{Evolution of the heavy quarks in the medium}

During the first stage of the collision ($\tau \lesssim 1$ fm/c) we neglect the effects of heavy quark energy loss in the cold nuclear matter of the colliding ions~\cite{Arleo:2014oha}. During the deconfined stage of the collision ($ 1 \lesssim  \tau \lesssim 10 $ fm/c), because the heavy quark typical energy scale (such as its mass $m>1$ GeV) is much larger than the medium scale (temperature $T \sim 100$--400 MeV), we assume each heavy quark to propagate and interact inside the expanding medium either through an energy loss process along a straight line or via Brownian motion described by relativistic Langevin dynamics.

\subsubsection{Energy loss models}\label{Subsection:EnergyLoss}

To easily study the dependence of the observables on some common variables, we use a simple parametric model for the heavy quark energy loss per unit length previously introduced in Refs.~\cite{Betz:2014cza,Horowitz:2011gd}, which is given by
\begin{equation}\label{eqn:Eloss}
\frac{\dd E}{\dd x}(T,v_\text{Q}) = -f(T,v_\text{Q}) \,\zeta\, \gammaflow,
\end{equation}
where $T$ is the local medium temperature experienced by the heavy quark, $v_\text{Q}$ is the heavy quark velocity in the global laboratory frame, $f(T,v_\text{Q})$ is a function encoding the energy loss parametrization, $\zeta$ is a random variable related to the energy loss fluctuations, and $\gammaflow$ takes into account the boost from the rest frame of the moving medium cell to the global laboratory frame~\cite{Baier:2006pt}.
The $\gammaflow$ factor is given by
\begin{equation}\label{eqn:GammaFlow}
\gammaflow =\gamma\big [1-v_\text{flow}\cos(\varphi_\text{Q}-\varphi_\text{flow})\big ],
\end{equation}
where $\gamma = 1\Big/\sqrt{1-v_\text{flow}^2}$, $v_\text{flow}$ and $\varphi_\text{flow}$ are respectively the local medium cell velocity and azimuthal angle, and $\varphi_\text{Q}$ is the azimuthal angle defined by the propagating heavy quark in the transverse plane. Note that the relation (\ref{eqn:GammaFlow}) is a light quark jet ($p\gg m$) approximation to the general formula,
\begin{equation}\label{eqn:GammaFlowExact}
\gammaflow^{\rm \, exact}=\gamma \sqrt{1 - 2\frac{v_\text{flow}}{v_\text{Q}}\cos(\varphi_\text{Q}-\varphi_\text{flow}) + \frac{v_\text{flow}^2}{v_\text{Q}^2} - v_\text{flow}^2\sin^2(\varphi_\text{Q}-\varphi_\text{flow})},
\end{equation}
which can be derived following a procedure similar to the one used in~\cite{Baier:2006pt} but without assuming $v_\text{Q} \to 1$. Discrepancies between the two expressions appear when $p_\text{Q}\lesssim 2m_\text{Q}$, i.e.~when $p_\text{c} < 3$ GeV for charm and $p_\text{b} < 10$ GeV for bottom quarks. The consequences of this generalized formula on the observables will be investigated in a future work.

\begin{figure}[h!]
  \centering
  \includegraphics[width=0.7\textwidth]{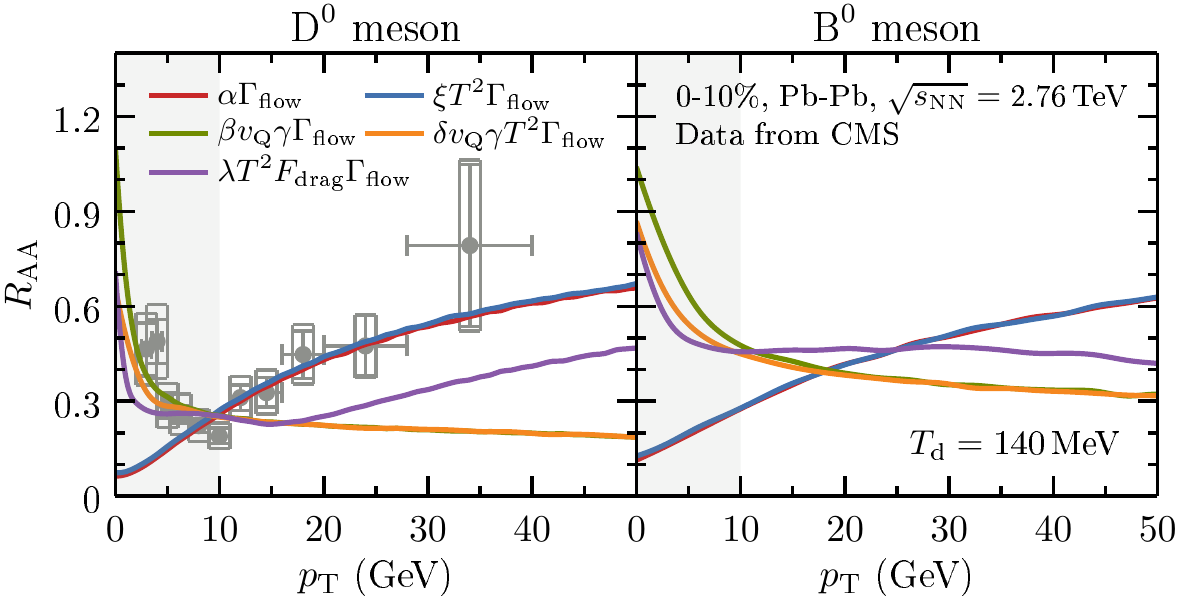}
  \caption{$\Dzero$ meson (left) and $\Bzero$ meson (right) nuclear modification factor in the 0--10\% centrality class of $\snn[2.76]$ $\PbPb$ collisions obtained with different energy loss models, compared to CMS data~\cite{CMS:2015hca}. The shaded area corresponds to the region of $\pt$ where other effects may be important.}
  \label{fig:RAA_few_parameterizations}
\end{figure}

As shown in Fig.\ \ref{fig:RAA_few_parameterizations}, we have tested a few parameterizations of the $f(T,v_\text{Q})$ function in order to select the ones that can roughly reproduce $\raa$ data in $\snn[2.76]$ $\PbPb$ collisions in the  0--10\% centrality range. The parametrizations $f = \xi T^2$ and $f = \delta \gamma_\text{Q} v_\text{Q} T^2$ (with $\gamma_\text{Q} = 1/\sqrt{1-v_\text{Q}^2}$) are inspired by conformal \adscft\ calculations~\cite{Gubser:2006bz}, whereas $f = \alpha$ and $f = \beta \gamma_\text{Q} v_\text{Q}$ are inspired by Ref.~\cite{Das:2015ana}, which showed that a non-decreasing drag coefficient near the cross-over transition is favored for a simultaneous description of heavy flavor $\raa(\pt)$ and $\vn2(\pt)$ (this is also supported by $T$-matrix calculations~\cite{vanHees:2007me,Riek:2010fk}). Finally, we also consider the temperature dependent non-conformal drag force dependent model $f = \lambda T^2 F_{\rm drag}$, where $F_{\rm drag}$ has been evaluated using holographic models that describe  lattice QCD thermodynamics in~\cite{Rougemont:2015wca}. The $\alpha$, $\beta$, $\delta$, $\lambda$ and $\xi$ factors are proportionality coefficients. Fig.\ \ref{fig:RAA_few_parameterizations} shows that the two energy loss models which are independent of the heavy quark velocity lead to very similar $\raa$, which increases with increasing $\pt$. Their temperature dependence does not seem to play any important role for $\raa$. The two energy loss models which are velocity dependent give also similar results but lead to $\raa$ continuously decreasing with $\pt$. The non-conformal drag force model has a strong dependence on the parton masses, leading to different trends for D and B mesons. Thus, the non-conformal drag force and velocity dependent models are favored by the low-$\pt$ $\raa$ data. Nevertheless, as the high $\pt$ data favors energy loss models which lead to increasing $\raa$ with increasing $\pt$, we limit our study in this paper to the two velocity independent energy loss models defined above. As further explained in Section \ref{SubSection:FreeParameters}, their two factors $\alpha$ and $\xi$ are tuned in this work via a single experimental observable.

To test the effects of energy loss fluctuations on the observables, we adopt a description put forward in~\cite{Betz:2014cza} where energy loss fluctuations in Eq.\ (\ref{eqn:GammaFlow}) are implemented via a random variable $\zeta$ for each heavy quark propagating in the medium. Three different probability
distributions for $\zeta$ have been implemented for the constant energy loss model ($f = \alpha$) in this work: a Gaussian distribution $f_\zeta^{\rm gauss}(\zeta)=1/(\sqrt{2\pi}\sigma) \,{\rm exp}[- (\zeta-1)^2/2\sigma^2]$ with the $\sigma$ parameter taken to be $0.3$, a uniform distribution $f_\zeta^{\rm uniform}(\zeta)=1/2$ with $0\leq \zeta \leq 2$, and finally a linear distribution $f_\zeta^{\rm linear}(\zeta)=2/3-2/9\,\zeta$ with $0\leq \zeta \leq 3$.

\begin{figure}[h!]
  \centering
  \includegraphics[width=0.5\textwidth]{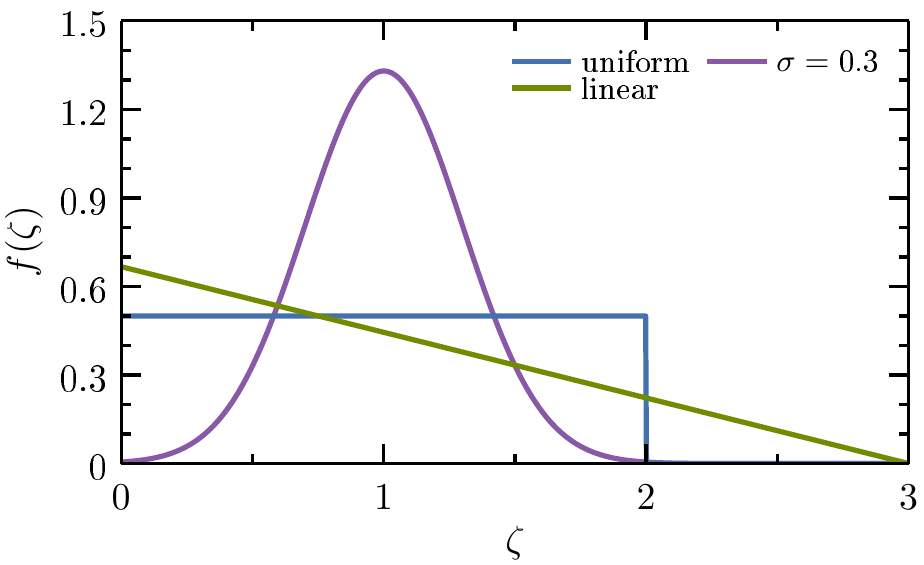}
  \caption{Energy loss fluctuation distributions used in this work.}
  \label{fig:fluctuations_distr}
\end{figure}

\subsubsection{Relativistic Langevin dynamics}

Due to the large separation of scales, $T\ll m_\text{Q}$, one can also describe the propagation of heavy quarks in the deconfined medium as a stochastic Brownian motion. The Brownian approximation is particularly valid for bottom quarks but only partially valid for charm quarks when the temperature exceeds 250 MeV~\cite{Das:2013kea}. Brownian motion is commonly described via the Fokker-Planck equation and simulated in practice for each individual heavy quark through a relativistic Langevin equation, which naturally encodes both some energy gain through a fluctuating force and some energy loss through a drag term~\cite{Moore:2004tg, vanHees:2005wb, He:2011qa, He:2013zua, Akamatsu:2008ge, Horowitz:2015dta, Cao:2015hia, Cao:2011et, Nahrgang:2018rlf, Xu:2017hgt, Nahrgang:2015zoa}. Assuming the diffusion coefficients to be isotropic and the momentum space diffusion coefficient $\kappa$ to be independent of the heavy quark momentum $\vec{p}$, one can write the relativistic Langevin equation as
\begin{eqnarray}
&&\dd x_i = \frac{p_i}{E}\dd t,\\
&&\dd p_i = -\Gamma(\vec{p})p_i \dd t + \sqrt{\dd t} \sqrt{\kappa}\rho_i ,
\end{eqnarray}
where the index $i=x,y$ corresponds to transverse plane coordinates, $\Gamma$ is the drag coefficient, and $\rho_i$ is the fluctuating force described classically by white noise in a Markovian process. For the heavy quarks to be able to reach the correct thermal equilibrium~\cite{He:2013zua}, the relativistic Einstein fluctuation-dissipation relation between the diffusion coefficients sets
\begin{eqnarray}
\kappa = 2 E\Gamma T= 2T^2/D,
\end{eqnarray}
where $D$ is the spatial diffusion coefficient. As the deconfined medium is rapidly expanding, one needs to perform the necessary Lorentz boosts between the local rest frame of the moving medium (in which the heavy quark interacts) and the global laboratory frame (where it propagates). To do so, we first boost the heavy quark 4-momentum from the global frame $p=(E,\vec{p})$ to the local rest frame of the medium cell $p'=(E',\vec{p'})$, where we calculate the momentum update via the Langevin equation with a pre-point Ito implementation of the stochastic scheme,
\begin{eqnarray}
p'_i = p'_i-\Gamma(\vec{p'})p'_i \Delta t' + \sqrt{\Delta t'} \sqrt{\kappa}\rho_i,
\end{eqnarray}
for a time step $\Delta t'$ of the implementation in the local rest frame of the medium, given by
\begin{eqnarray}
\Delta t'=\Delta t*\displaystyle\frac{p^{\,\mu}v^{\rm \,flow}_{\,\mu}}{E}=\Delta t*\gamma\left[1-\displaystyle\frac{\|\vec{p}\|\|\vec{v}_{\rm flow}\|}{E}\cos{(\varphi_{\rm Q}-\varphi_{\rm flow})}\right],
\end{eqnarray}
where $\Delta t$ is the related time step in the global frame, $v^{\rm \,flow}=(\gamma,\gamma\vec{v}_{\rm flow})$ is the local 4-velocity of the medium cell, and $\gamma=1/\sqrt{1-\vec{v}_{\rm flow}^2}$ is the corresponding Lorentz factor. We then boost the heavy quark 4-momentum back to the global frame, where we compute its propagation in the transverse plane via
\begin{eqnarray}
x_i(t+\Delta t) = x_i(t)+\frac{p_i}{E}\Delta t.
\end{eqnarray}
Finally, we repeat these operations until reaching the hadronization temperature.

In this work, two parametrizations of the diffusion coefficients have been chosen. The first one, denoted by ``M\&T'' in the following, is inspired by Moore and Teaney's leading order QCD description of the scattering process and a Debye mass correction of the gluon propagator~\cite{Moore:2004tg}. In this model, the spatial diffusion coefficient reduces to
\begin{eqnarray}
D_{\rm M\&T}=k_{\rm M\&T}/(2\pi T),
\end{eqnarray}
where $k_{\rm M\&T}$ is a factor --- estimated to be around 6 in their work --- that we use in this paper as a tunable parameter. The second parametrization adopted in this work, denoted by ``G\&A'' in the following, comes from Gossiaux and Aichelin's collisional model based on a running coupling constant and an optimized hard thermal loop correction of the gluon propagator~\cite{Gossiaux:2009qf}. To obtain a tunable parameter as in the other transport models explored in this work, we multiply the drag $A_{\rm G\&A}(T,p)$ [c/fm] directly obtained from this model by a factor $k_{\rm G\&A}$, such that we obtain
\begin{eqnarray}
\Gamma_{\rm G\&A}=k_{\rm G\&A}*A_{\rm G\&A}.
\end{eqnarray}

\begin{figure}[h!]
  \centering
  \includegraphics[width=0.45\textwidth]{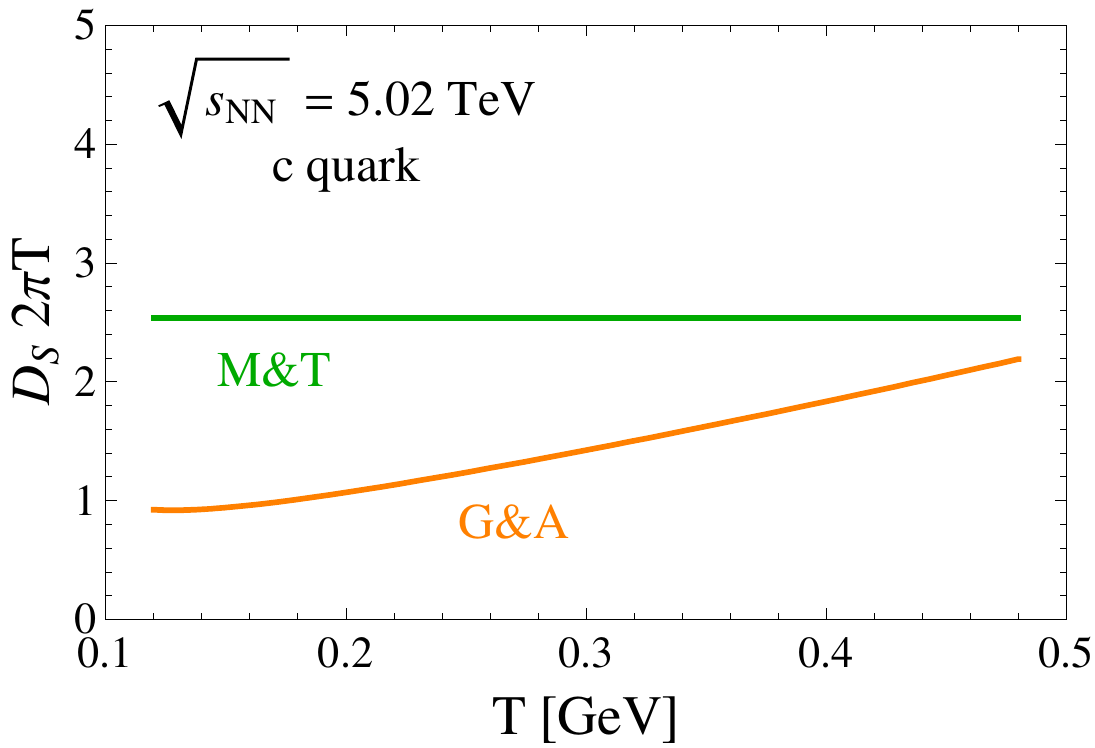}
\vspace{5mm}
  \includegraphics[width=0.45\textwidth]{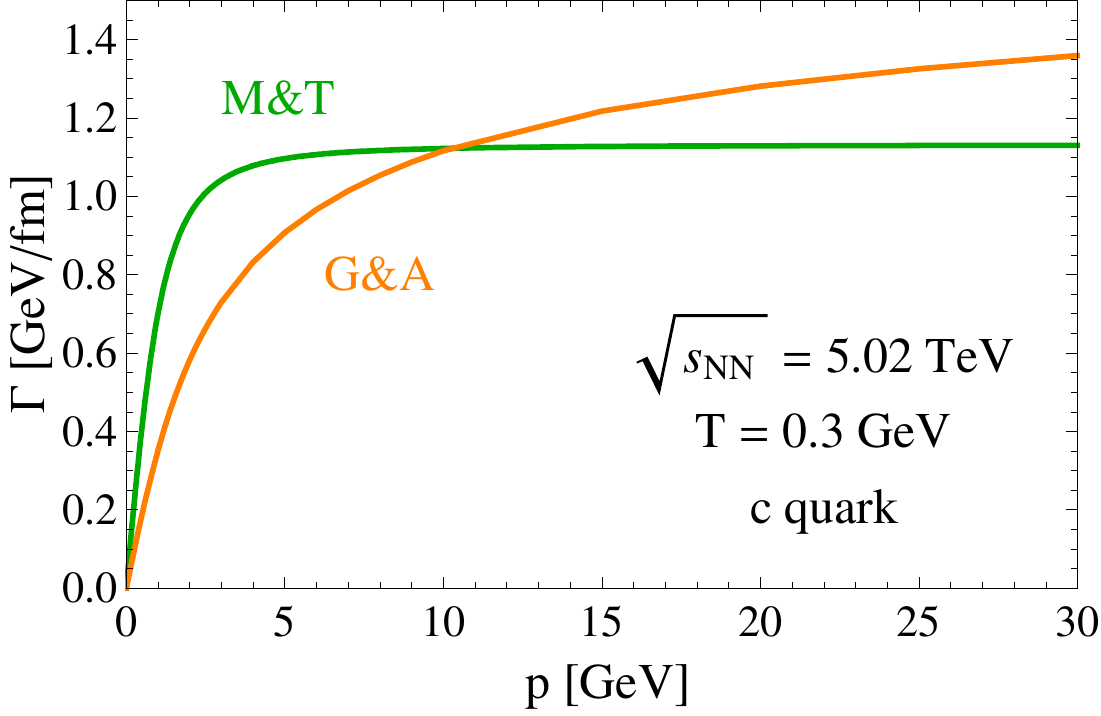}
  \caption{Comparison of the spatial diffusion coefficients $D_s=D(p=0)$ as a function of temperature (left) and of the drags as a function of momentum (right) obtained with the two chosen parameterizations of the diffusion coefficients. Here we use $k_{\rm M\&T}=0.5$ and $k_{\rm G\&A}=0.62$ which correspond to $\snn[5.02]$ $\PbPb$ collisions with decoupling temperature $\Td=120$ MeV (see definition below and Tab.\ \ref{TableFactors}).}
  \label{fig:comparison_drags}
\end{figure}

A comparison between the two corresponding drag coefficients can be found in Fig.\ \ref{fig:comparison_drags} (right). At fixed temperature $T=0.3$ GeV, the M\&T parametrization gives a significantly larger drag at low momentum whereas it saturates at larger momentum where the G\&A becomes larger and increases linearly. The Fig.\ \ref{fig:comparison_drags} (left) illustrates the temperature dependence of the dimensionless quantity $ D (2\pi T)$ computed at zero momentum in the two models. While this quantity is constant by definition in the M\&T parametrization, for G\&A it acquires a temperature dependence approaching the results found using M\&T from below. As compared to other models~\cite{Andronic:2015wma,Xu:2017obm,Cao:2018ews}, the present drag and diffusion coefficients are rather small.

\subsection{Fragmentation and decay}

We assume each heavy quark to propagate until it reaches a cell where the temperature of the medium is lower or equal to a decoupling temperature $\Td$. Inspired by other works~\cite{Betz:2016ayq,Nahrgang:2016lst}, and by lattice QCD results on the cross-over transition~\cite{Aoki:2006br,Aoki:2009sc,Borsanyi:2010bp,Mukherjee:2015mxc}, we choose a range of $\Td$ from 120 MeV to 160 MeV in order to assess part of the uncertainties related to the complicated process of hadronization. In this work, fragmentation is assumed to be the same as in pp collisions (for medium modified fragmentation functions see~\cite{Arleo:2008dn,Cacciari:2012mu}). To perform the fragmentation, we use the Peterson fragmentation function
\begin{eqnarray}
f(z)\propto\frac{1}{z(1-1/z-\epsilon_\text{Q}/(1-z))^2}
\end{eqnarray}
as a probability distribution to obtain the fraction $z$ of the fragmenting heavy quark $E_\text{Q}+p_\text{Q}$ taken by the daughter hadron $E_\text{H}+p_\text{H}=z(E_\text{Q}+p_\text{Q})$. The Peterson function parameters $\epsilon_\qcharm$ and $\epsilon_\qbottom$ are chosen such as to reproduce $\Dzero$ and $\Bzero$ meson FONLL spectra in pp collisions~\cite{Cacciari:1998it,Cacciari:2001td}. Following the fragmentation, we do not consider the possible interactions between the produced heavy mesons and the hadronic gas~\cite{Torres-Rincon:2015ibt}. Finally, we perform the decays of the $\Dzero$ and $\Bzero$ mesons using Pythia 8~\cite{Sjostrand:2007gs}, focusing only on their semileptonic channels.

\subsection{Determination of the tunable coupling factor of each transport model} \label{SubSection:FreeParameters}

In each of the transport models described above there is a free parameter, i.e.~the coupling factors $\alpha$ and $\xi$ for the chosen energy loss models and the factors $k_{\rm M\&T}$ and $k_{\rm G\&A}$ for the Langevin models, which must be fixed. This is done by finding the value that gives the best description of differential $\raa$ for $\pt\gtrsim 10$ GeV at \lhc\ energies or $\pt\gtrsim 5$ GeV at \rhic\ obtained from the simulation compared to available experimental data for each model, heavy quark type, limiting values of the decoupling temperature considered ($\Td=120$ and $160$ MeV), and collision energy. To set the factors for the charm quark simulations, we use $\Dzero$ meson differential $\raa$ data in one centrality range (0--10\% unless specified otherwise). The obtained values of these constants are summarized in table \ref{TableFactors}. As almost no data for $\Bzero$ mesons is available, to obtain the constants for bottom quarks we use heavy flavor electron differential $\raa$ data in most central collisions. The idea is to use the already fixed coupling factor for charm quarks and vary the bottom factor to find its value such that the total contribution to the electron $\raa$ matches the data. Note that here we use more recent experimental data in $\PbPb$ collisions at $\snn[5.02]$ than in our previous paper~\cite{Prado:2016szr}. The values of the constants for bottom quarks can be found in table \ref{TableFactorsBottom}. The values for the overall coupling factors for charm and bottom quarks are of the same order of magnitude and their ratios are not related to the quark mass ratio. For the energy loss models, bottom quark constants are generally smaller than charm quark factors, i.e. bottom quarks require a smaller coupling to the medium to fit the most central $\raa$ data than charm quarks.

\begin{table}[h!]
\begin{center}
    \begin{tabular}{|C{5cm}||C{3cm}|C{3cm}|C{3cm}|}
    \hline
Coupling factors for charm quarks at $\Td=120$ $\backslash$ 160 MeV & \rhic\ $\AuAu$ $\snnGeV[200]$ & \lhc\ $\PbPb$ $\snn[2.76]$ & \lhc\ $\PbPb$ $\snn[5.02]$\\
    \hline
    \hline
$\alpha$ without fluctuations & 0.393 $\backslash$ 0.623 & 1.0 $\backslash$ 1.624 & 0.708 $\backslash$ 1.011 \\
    \hline
$\alpha$ with uniform fluct. & 0.649 $\backslash$ none & 1.7 $\backslash$ none & 0.993 $\backslash$ none  \\
    \hline
$\alpha$ with linear fluct. & 0.77 $\backslash$ none & 2.024 $\backslash$ none &  1.130 $\backslash$ none \\
    \hline
$\alpha$ with gaussian fluct. & 0.43 $\backslash$ none & 1.1 $\backslash$ none &  0.751 $\backslash$ none \\
    \hline
$\xi$ & 11.57 $\backslash$ 15.16 & 30.28 $\backslash$ 40.05 & 14.76 $\backslash$ 17.16 \\
    \hline
$k_{\rm M\&T}$ & 0.48 $\backslash$ 0.34 & 0.227 $\backslash$ 0.169 & 0.5 $\backslash$ 0.41 \\
    \hline
$k_{\rm G\&A}$ & 0.639 $\backslash$ 0.921 & 1.039 $\backslash$ 1.577 & 0.622 $\backslash$ 0.828 \\
    \hline
    \end{tabular}
\caption {\label{TableFactors}
\small Values of the coupling factors for charm quarks determined for each transport model, collision energy, and decoupling temperature. These values are obtained using \mckln\ initial conditions.}
\end{center}
\end {table}

\begin{table}[h!]
\begin{center}
    \begin{tabular}{|C{5cm}||C{3cm}|C{3cm}|C{3cm}|}
    \hline
Coupling factors for bottom quarks at $\Td=120$ $\backslash$ 160 MeV & \rhic\ $\AuAu$ $\snnGeV[200]$ & \lhc\ $\PbPb$ $\snn[2.76]$ & \lhc\ $\PbPb$ $\snn[5.02]$\\
    \hline
    \hline
$\alpha$ without fluctuations & 0.264 $\backslash$ 0.4 & 0.72 $\backslash$ 1.12 & 0.667 $\backslash$ 0.823 \\
    \hline
$\alpha$ with uniform fluct. & 0.316 $\backslash$ none & 0.857 $\backslash$ none & 0.824 $\backslash$ none  \\
    \hline
$\alpha$ with linear fluct. & 0.339 $\backslash$ none & 0.921 $\backslash$ none &  0.913 $\backslash$ none \\
    \hline
$\alpha$ with gaussian fluct. & 0.265 $\backslash$ none & 0.76 $\backslash$ none &  0.624 $\backslash$ none \\
    \hline
$\xi$ & 7.6 $\backslash$ 10 & 21.52 $\backslash$ 27.06 & none $\backslash$ none \\
    \hline
$k_{\rm M\&T}$ & 0.648 $\backslash$ 0.486 & 0.32 $\backslash$ 0.226 & 0.516 $\backslash$ 0.411 \\
    \hline
$k_{\rm G\&A}$ & 0.606 $\backslash$ 0.808 & 3.21 $\backslash$ 2.26 & 0.681 $\backslash$ 0.884 \\
    \hline
    \end{tabular}
\caption {\label{TableFactorsBottom}
\small Values of the coupling factors for bottom quarks determined for each transport model, collision energy, and decoupling temperature. These values are obtained using \mckln\ initial conditions.}
\end{center}
\end {table}

\subsection{Examples of heavy quark spatial evolution}

\begin{figure}[h!]
  \centering
  \includegraphics[width=0.75\textwidth]{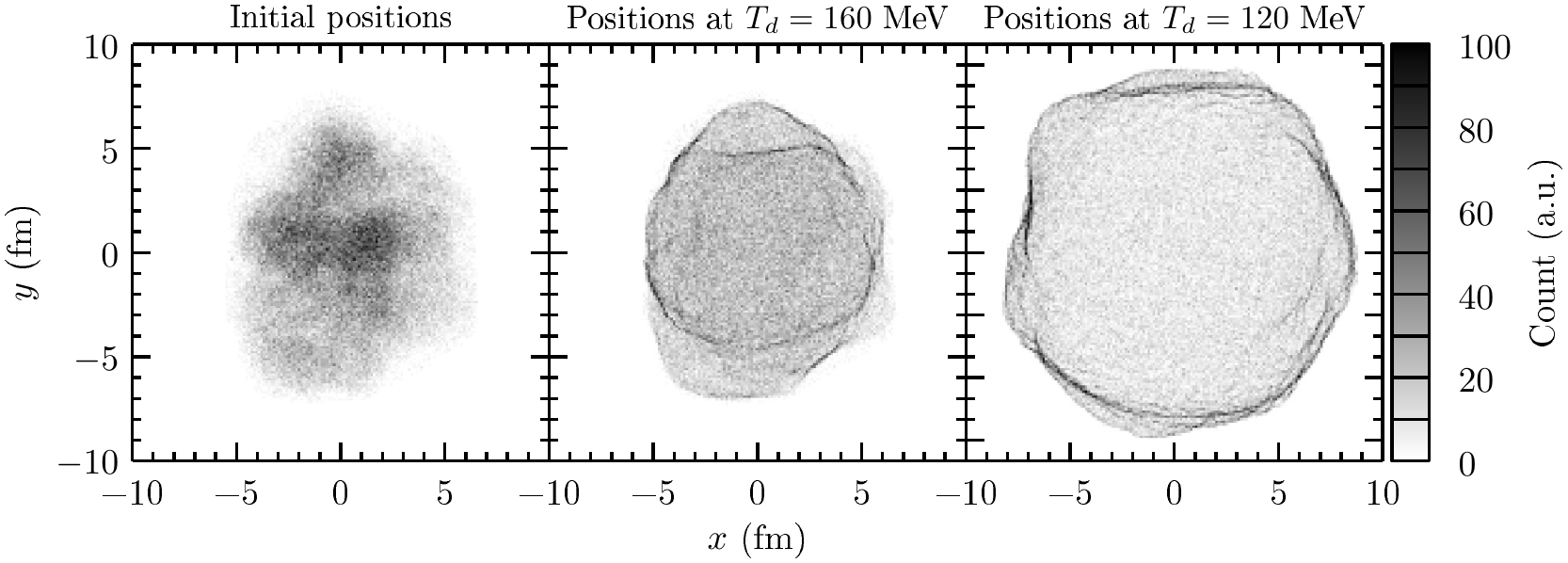}
  \includegraphics[width=0.75\textwidth]{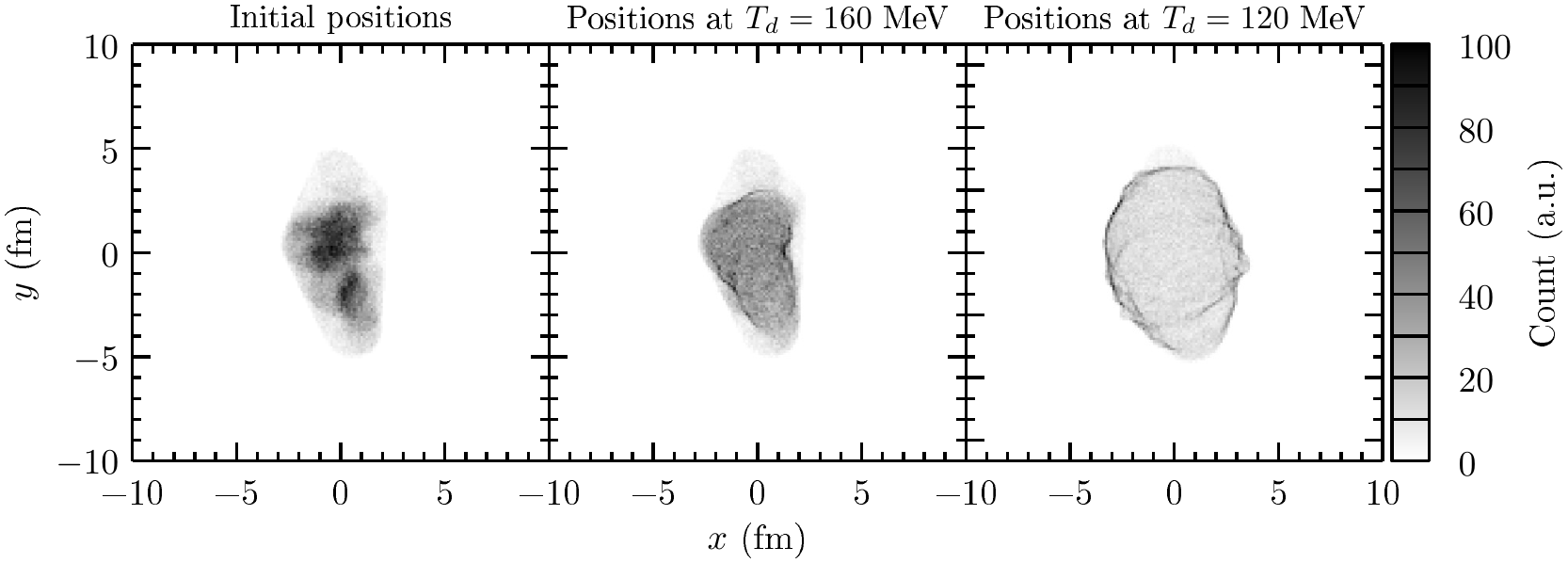}
  \caption{Examples for a chosen central 0--10\% event (top) and a more peripheral 40--50\% event (bottom) of the heavy quarks initial spatial distribution (left) and final distribution at $\Td=160$ MeV (center) and $\Td=120$ MeV (right). The evolution of the heavy quarks is here obtained via the Langevin equation with the M\&T parametrization.}
  \label{fig:HQdistributions}
\end{figure}

As explained above, in \dabmod\ we oversample each initial condition with a large number of heavy quarks. An example of the heavy quarks initial coordinates in the transverse plane for a central collision is shown at the top-left plot of Fig.\ \ref{fig:HQdistributions}. The distribution of heavy quarks follows the spatial fluctuations in energy density of the underlying medium. Each heavy quark then evolves and interacts with the bulk following a given transport model until it reaches a cell where the temperature of the medium is lower or equal to the decoupling temperature $\Td$. The corresponding final distributions for the decoupling temperatures $\Td=160$ MeV and $\Td=120$ MeV are shown, respectively, on the center-top and right-top of Fig.\ \ref{fig:HQdistributions}. At $\Td=160$ MeV, the heavy quark spatial distribution has a size similar to the initial distribution and is quite homogeneous. One can also note the presence of a relatively thick peripheral area including some high density ``filaments'' which are mainly composed of high $\pt$ heavy quarks and which density patterns are clearly correlated to the initial density patterns. At $\Td=120$ MeV, the distribution is wider and mainly homogeneous. The peripheral area is now thinner and more ring-shaped with the filaments being closer to each other as compared to $\Td=160$ MeV. Going to smaller $\Td$ seems therefore to partially wash out the initial density fluctuations. The same observations can be made in the case of a more peripheral event as shown at the bottom of Fig.\ \ref{fig:HQdistributions} and with other transport models.

\section{Effect of transport model and decoupling temperature on observables in the case of \mckln\ initial conditions and no coalescence}\label{SecIII}

In this section we use \mckln\ initial conditions and only fragmentation as the hadronization mechanism. A systematic study is performed where we:
\begin{itemize}
 \item compare different energy loss parameterizations vs.\ different diffusion coefficients within the Langevin model
 \item vary the distributions of energy loss fluctuations
 \item determine the effect of the decoupling temperature $\Td$
 \item investigate the differences between D and B mesons, $\electronpositron$, and muons
 \item study the centrality and beam energy dependence of our calculations.
\end{itemize}

\subsection{Nuclear modification factor}

The nuclear modification factor is essentially the ratio between the particle spectrum in AA collisions, $\dd N_\text{AA}/\dd \pt$, and the spectrum in pp collisions, $\dd N_{pp}/\dd \pt$, with a normalization factor $\mathcal{N}$  defined in terms of the number of binary collisions~\cite{Miller:2007ri}. This gives:
\begin{equation}
\raa(\pt,\phi) = \frac{1}{\mathcal{N}}\frac{\dd N_\text{AA}/\dd \pt \dd \phi}{\dd N_{pp}/\dd \pt},
\end{equation}
where we leave our calculations dependent on $\phi$ (the azimuthal angle in the plane transverse to the beam direction) in order to calculate the Fourier harmonics. Here, in practice, the $N \times \dd N_{pp}/\dd \pt$ spectrum is obtained using the same ingredients as described in Sec.\ \ref{Section:DABMODingredients} but in this case one turns off the heavy quark transport equations while hadronization is still done via fragmentation. In this work, the calculations are assumed to be boost invariant so comparisons are made only to mid-rapidity experimental data. Integrating over $\phi$ then reproduces the typical $\raa(\pt)$ that can be compared to experiment. Because we oversample the number of heavy quarks for each hydrodynamic background, we are able to reconstruct the entire $\raa(\pt,\phi) $ for each event.

As explained in Sec.\ \ref{SubSection:FreeParameters}, the values of the free parameters of the transport models are determined by the best fit to most central high $\pt$ differential $\raa$ experimental data and, thus, the magnitude of $\raa(\pt)$ in most central collisions obtained with \dabmod\ is not a prediction. However, its centrality and $\pt$ dependence are legitimate predictions of the model.

\subsubsection{$\Dzero$ meson}

\begin{figure}[!htb]
  \centering
  \includegraphics[width=0.48\textwidth]{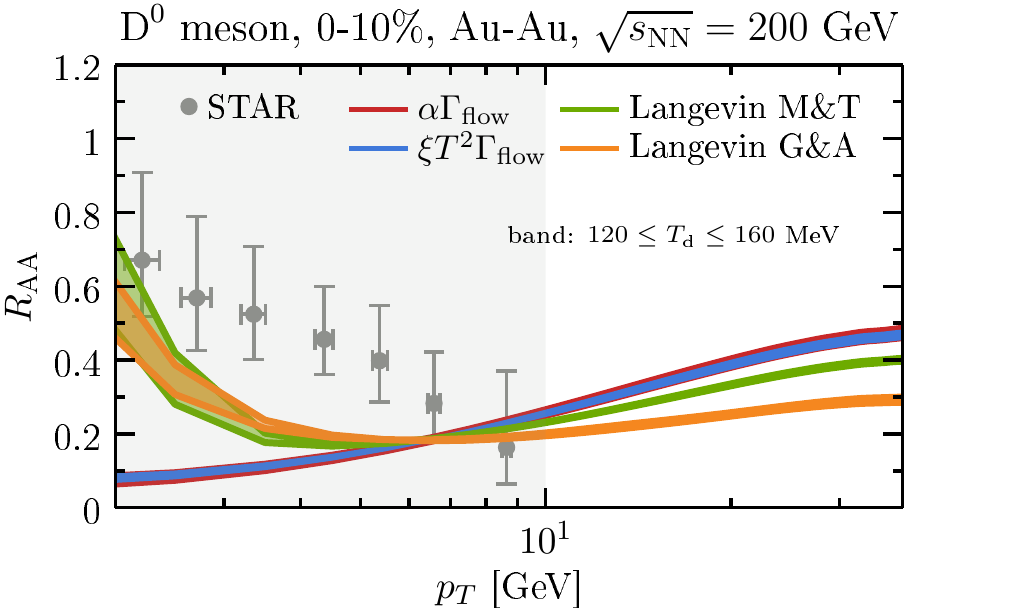}
  \includegraphics[width=0.50\textwidth]{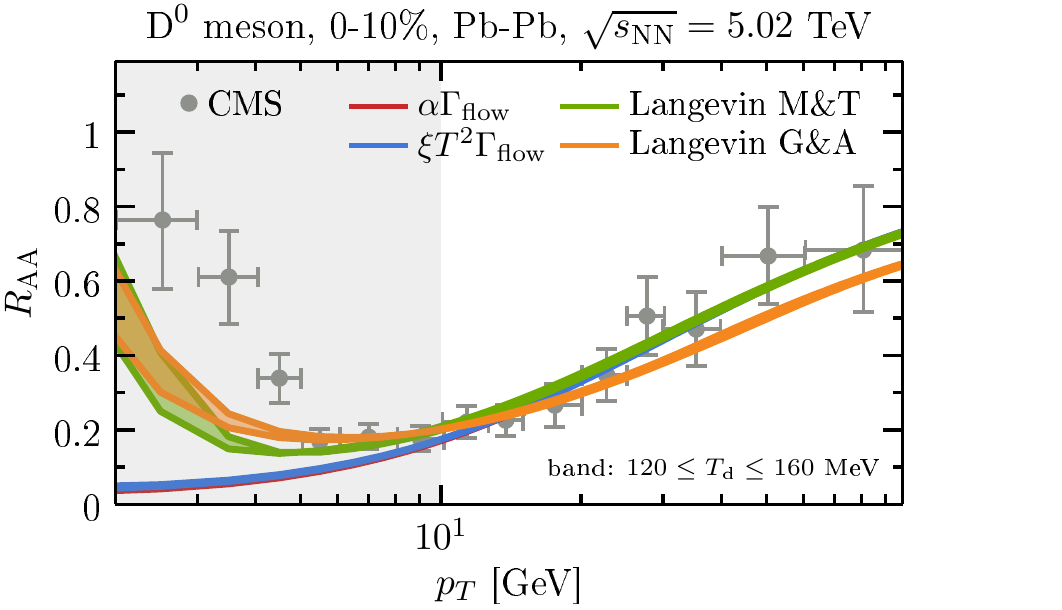}
  \caption{$\Dzero$ meson nuclear modification factor $\raa$ in the  0--10\% centrality class of $\snnGeV[200]$ $\AuAu$ (left) and $\snn[5.02]$ $\PbPb$ (right) collisions, computed using different transport models. The gray area indicates the $\pt$ region where coalescence and initial/final state effects may be important. Experimental data from the STAR ($|y|<1$)~\cite{Bruna:2019lcu} and CMS ($|y|<1$)~\cite{Sirunyan:2017xss} collaborations.}
  \label{fig:RAAmodels}
\end{figure}

In Fig.\ \ref{fig:RAAmodels} we compare the results from different transport models  for $\Dzero$ mesons in the 0--10\% centrality of $\AuAu$ 200 GeV (left) and $\PbPb$ 5.02 TeV (right) collisions. At first sight, the four chosen transport models exhibit correct trends at intermediate/high $\pt$ when compared to data. However at low $\pt$, the energy loss and Langevin models lead to very different behaviors, clearly favoring Langevin models in the comparison to data. This difference originates in the fluctuating force within the Langevin approach: the balance between this force and the energy loss brought by the drag term maintains a certain amount of heavy quarks within the typical momentum range of the plasma particles. On the other hand, the energy loss models, where there is no energy gain, lead to a continuous decrease of $\raa$ towards lower $\pt$ until a certain value (much) smaller than 1 GeV where a large amount of heavy quarks ``accumulate'' (not shown in the present figures).

Looking more carefully at Fig.\ \ref{fig:RAAmodels} one can see that in Langevin models the choice of the diffusion coefficient leads to different results at high $\pt$. M$\&$T is able to reasonably capture the high $\pt$ behavior of $\raa(\pt)$ data whereas G$\&$A is found to be below the data\footnote{If we would have compared these predictions to the ALICE collaboration $\raa(\pt)$ data~\cite{Acharya:2018hre}, the conclusion might have been different since that data is significantly lower at high $\pt$.}. At low $\pt$ they exhibit nearly identical behavior. While the results are somewhat lower than experimental data, we find (see  Sec.~\ref{Section: coalescence}) that the addition of coalescence shifts the $\raa(\pt)$ curve upwards at low $\pt$ matching experimental data. The results for the transport models for different beam energies are also shown in Fig.\ \ref{fig:RAAmodels}. As explained in Sec.\ \ref{Subsection:EnergyLoss}, for the energy loss models we consider either a constant energy loss function, $f(T,v_\text{Q})=\alpha$, or a temperature dependent one where $f(T,v_\text{Q}) =\zeta T^2$. Despite this difference, the $\Dzero$ $\raa(\pt)$ results are almost identical, which shows that this temperature dependence is not actually realized in practice in our calculations. Additionally, we do not find a strong dependence with beam energy when it comes to the qualitative differences between different types of heavy quark evolution. We consistently find that the Langevin model produces the largest $\raa(\pt)$ at low $\pt$ and that all the transport models behaves correctly at high $\pt$.

The point where a heavy quark decouples from the medium is still a source of uncertainty in heavy flavor modeling. In this work we investigate this issue by considering a range of decoupling temperatures between 120 and 160 MeV, which creates the bands in our theoretical calculations in following sections.  In Fig.\  \ref{fig:RAAmodels} $\raa(\pt)$ is plotted\footnote{We have checked that all experimental observables show either a monotonic increase or decrease with $\Td$ so it is sufficient to plot just the extrema to obtain the bands.} for $\Td=120$ MeV and $\Td=160$ MeV. $\raa(\pt)$ for both beam energies does not exhibit a strong dependence on the decoupling temperature, which can be explained within our model by the magnitude of the high $\pt$ $\raa$ being calibrated with the data for each decoupling temperature. Nevertheless, a significant difference can be seen between the decoupling temperatures at low $\pt$ within the Langevin model: $\raa$ at $\Td=120$ MeV is larger than at $\Td=160$ MeV. In other words, in this situation the low $\pt$ heavy quarks are less suppressed if they are coupled to the plasma for longer. This ordering can be explained by the fluctuating force having a longer time to act when $\Td=120$ MeV to increase the momenta of (the larger amount of) heavy quarks initially at lower $\pt$. This effect outclasses an expected countereffect: at lower temperatures the thermal distribution of the heavy quarks shifts towards lower momenta, such that the $\raa$ at $\Td=120$ MeV should be smaller than at $\Td=160$ MeV. This shows that quasi-equilibrium is far from being reached for the heavy quarks when $\Td=160$ MeV. As we will see below, the flow harmonics exhibit a different behavior and, in general, one finds that lower decoupling temperatures produce larger azimuthal anisotropies.

\begin{figure}[!htb]
  \centering
  \includegraphics[width=0.45\textwidth]{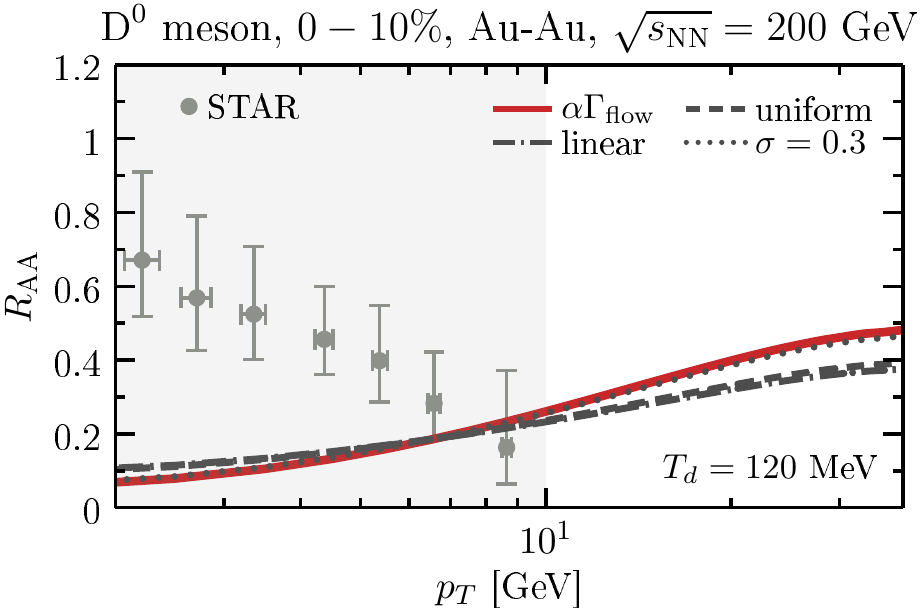}
\hspace{0.6cm}
  \includegraphics[width=0.45\textwidth]{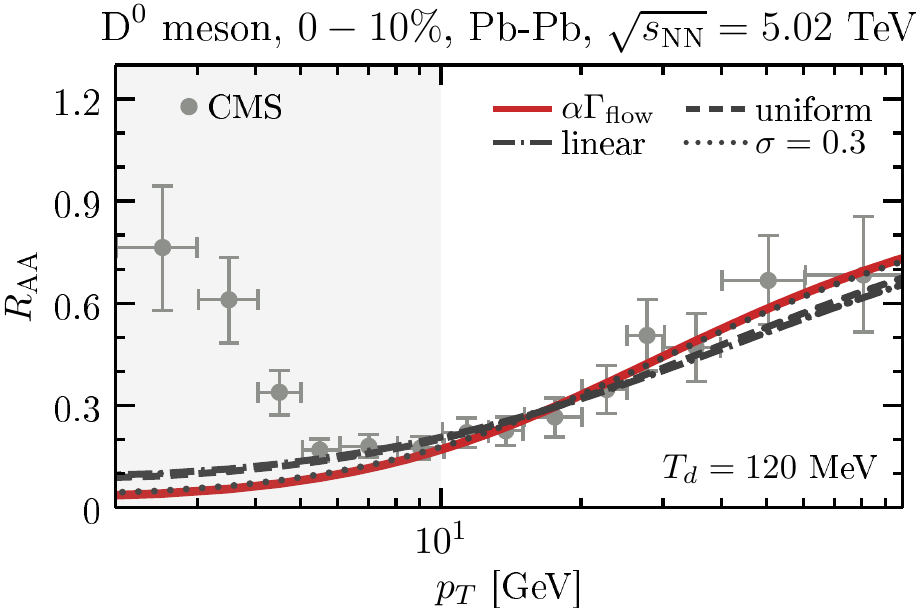}
  \caption{Comparison between the constant energy loss model with and without different types of energy loss fluctuations to experimental data (as in Fig.\ \ref{fig:RAAmodels}).}
  \label{fig:RaaElossFluctu}
\end{figure}

We also include different types of energy loss fluctuations in the case of the constant energy loss model  (see Sec.\ \ref{Subsection:EnergyLoss} for details). As shown in Fig.\ \ref{fig:RaaElossFluctu}, Gaussian fluctuations do not make a significant difference. However, linear or uniform fluctuations suppress  $\raa$ at high $\pt$, showing a disagreement with the experimental data, although they increase $\raa$ at low/intermediate $\pt$ improving the overall agreement with the data. However, all the energy loss results are within the error bars of the experimental data points so, unfortunately, no conclusion can be drawn for now regarding the specific form of the distribution of energy loss fluctuations within our model.

\begin{figure}[!htb]
  \centering
  \includegraphics[width=0.48\textwidth]{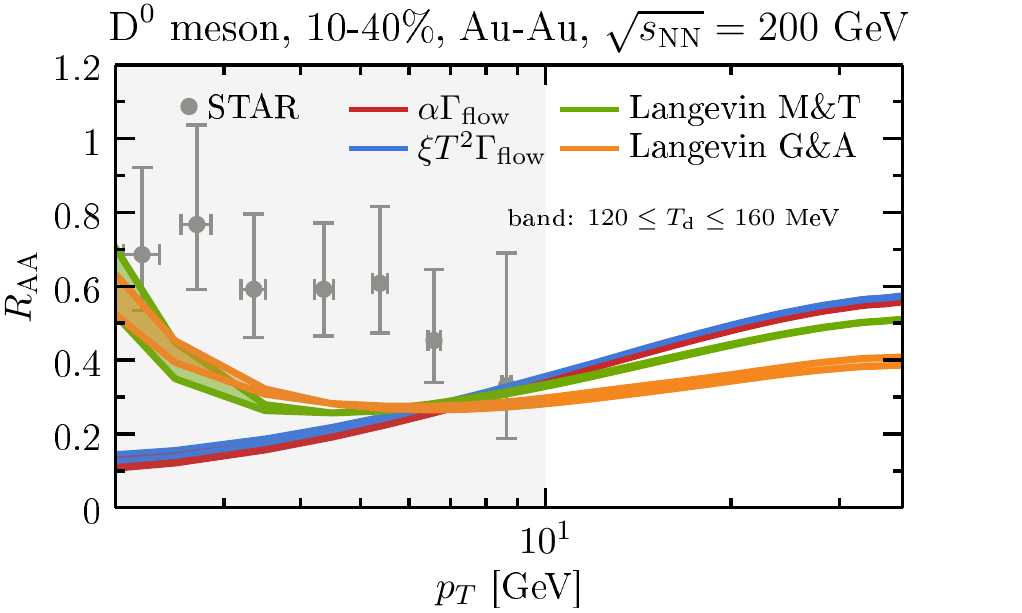}
  \includegraphics[width=0.50\textwidth]{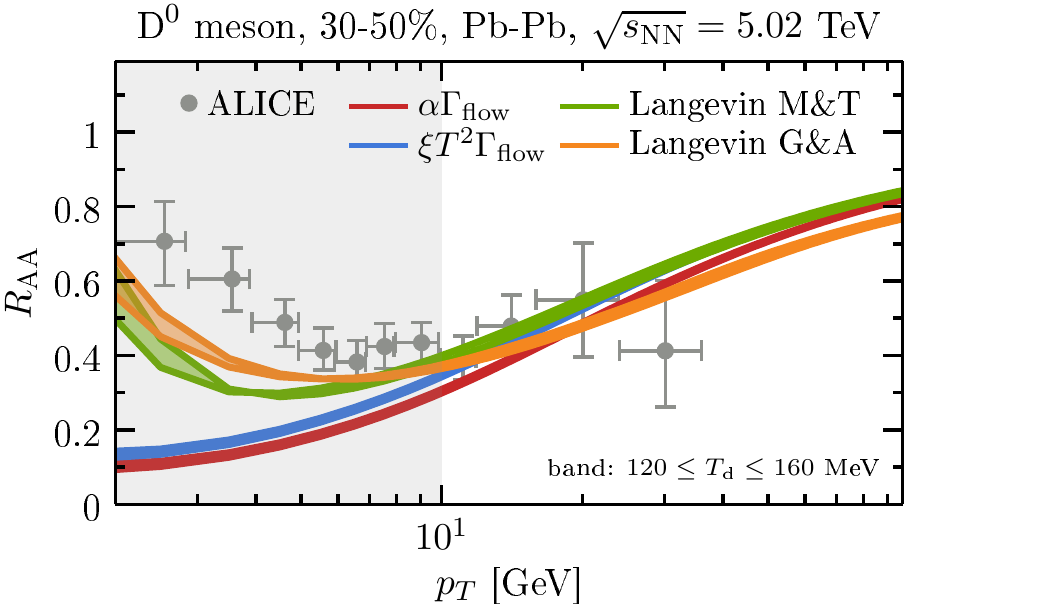}
  \caption{$\Dzero$ meson nuclear modification factor $\raa$ in the 10--40\% centrality range of $\snnGeV[200]$ $\AuAu$  collisions (left) and in the 30--50\% centrality range of $\snn[5.02]$ $\PbPb$  collisions (right), computed using different transport models. Experimental data from the STAR ($|y|<1$)~\cite{Bruna:2019lcu} and ALICE ($|y|<0.5$)~\cite{Acharya:2018hre} collaborations, respectively.}
  \label{fig:RAAmodelsPeripheral}
\end{figure}

To test if the calibration made in the most central collisions gives reasonable results for the different models in other centralities, some predictions for more peripheral collisions are shown in Fig.\ \ref{fig:RAAmodelsPeripheral}. The different models show similar features found in 0--10\% central collisions and fit correctly the less suppressed $\raa$ data at intermediate/high $\pt$.

\subsubsection{$\Bzero$ meson}

\begin{figure}[h]
  \centering
 \includegraphics[width=0.48\textwidth]{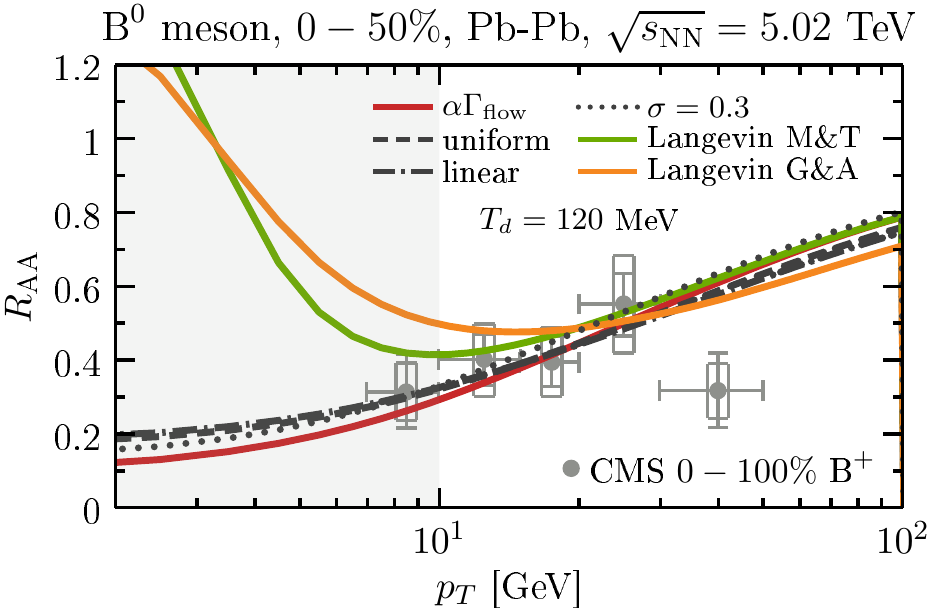}
  \caption{$\Bzero$ meson $\raa$ in the 0--50\% centrality range of $\snn[5.02]$ $\PbPb$ collisions. Experimental data for $\Bplus$ meson from the CMS collaboration ($|y|<2.4$)~\cite{Sirunyan:2017oug}.}
  \label{fig:RaaB0}
\end{figure}

Because $\Bzero$ production is less subject to initial state effects and final hadronic rescaterings compared to $\Dzero$, its observables are cleaner probes of the deconfined medium making it possible to study how changing the heavy quark mass affects its dynamics and hadronization. As explained in Sec.\ \ref{SubSection:FreeParameters}, we use heavy flavor electron $\raa$ data in the most central collisions to calibrate the bottom quark transport model coupling factors, such that all the $\Bzero$ meson results are predictions of the model. As shown in Fig.\ \ref{fig:RaaB0}, the (limited) comparison between $\Bplus$ 0--100\% experimental data and the $\Bzero$ meson $\raa(\pt)$ obtained with \dabmod\ in 0--50\% centrality range illustrates the consistency of our calibration.

\begin{figure}[h]
  \centering
 \includegraphics[width=0.48\textwidth]{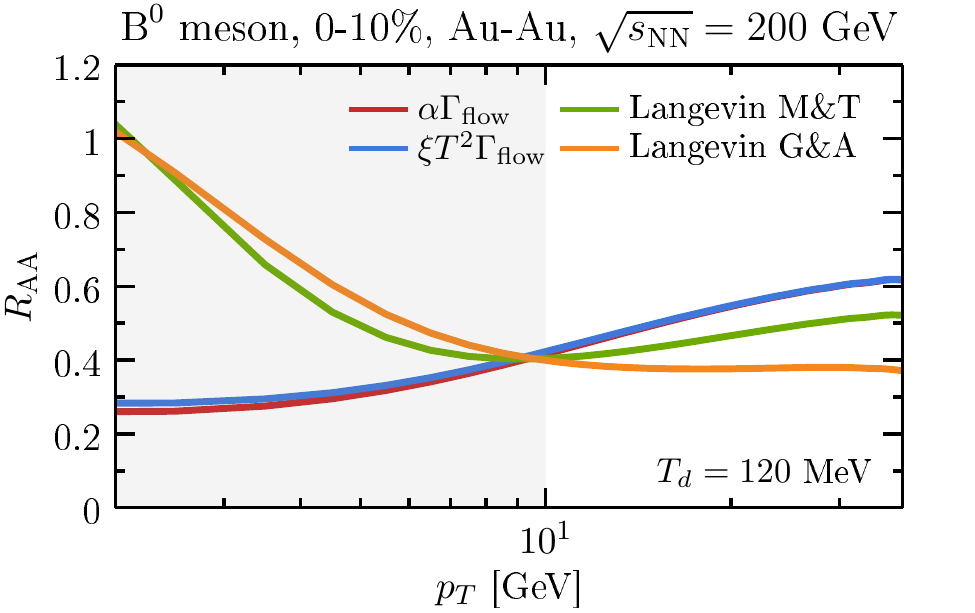}
 \includegraphics[width=0.48\textwidth]{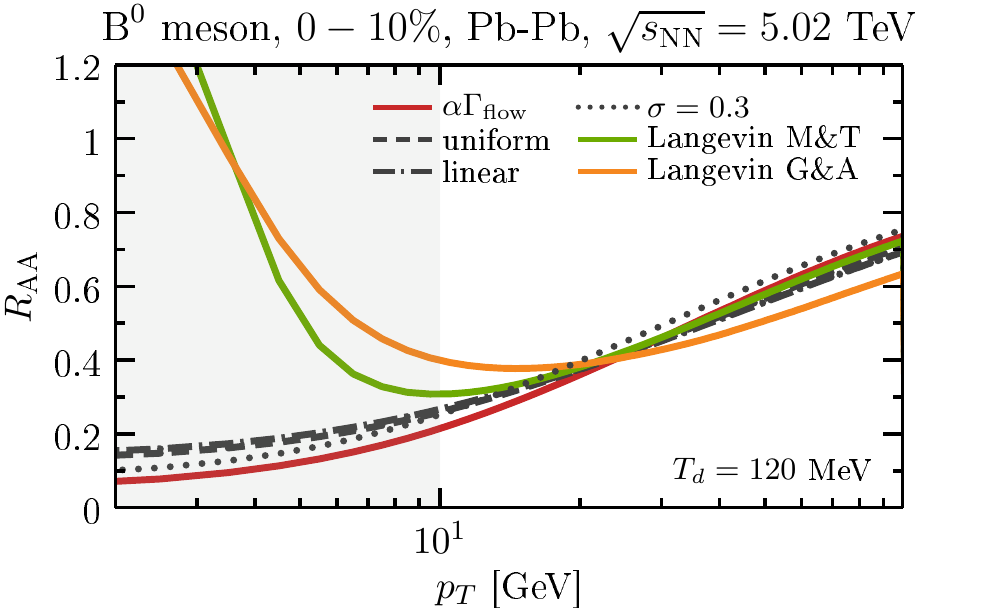}
  \caption{$\Bzero$ meson $\raa$ in the 0--10\% centrality class of $\snnGeV[200]$ $\AuAu$ (left) and $\snn[5.02]$ $\PbPb$ (right) collisions.}
  \label{fig:RaaB0cent010}
\end{figure}

In Fig.\ \ref{fig:RaaB0cent010} we compare different energy loss parametrizations to Langevin results for $\Bzero$ mesons in the 0--10\% centrality class of $\AuAu$ 200 GeV (left) and $\PbPb$ 5.02 TeV (right) collisions.  Comparing to Fig.\ \ref{fig:RAAmodels}, one can see that the $\Bzero$ and the $\Dzero$ meson $\raa(\pt)$ exhibit similar qualitative behavior. Nevertheless, at low and intermediate $\pt$, the $\Bzero$ $\raa$ is generally less suppressed by a factor $\sim 2$ while its value at high $\pt$ is larger than that found for the $\Dzero$ meson. We note that in this $\pt$ regime the discrepancy between energy loss and Langevin approaches is much larger for $\Bzero$ than for $\Dzero$ mesons (there is also a larger difference between different Langevin parametrizations). At very high $\pt$, the values for $\Bzero$ and $\Dzero$ $\raa$ are very similar since their mass difference becomes negligible when compared to the momentum. Finally, since the heavy quark mass has no impact on energy loss fluctuations, $\raa(\pt)$ of both $\Bzero$ and $\Dzero$ mesons computed within the constant energy loss model are similarly modified by energy loss fluctuations.

\subsubsection{$\electronpositron$ and $\muonpm$ from heavy flavor}

\begin{figure}[!htb]
  \centering
\includegraphics[width=0.44\textwidth]{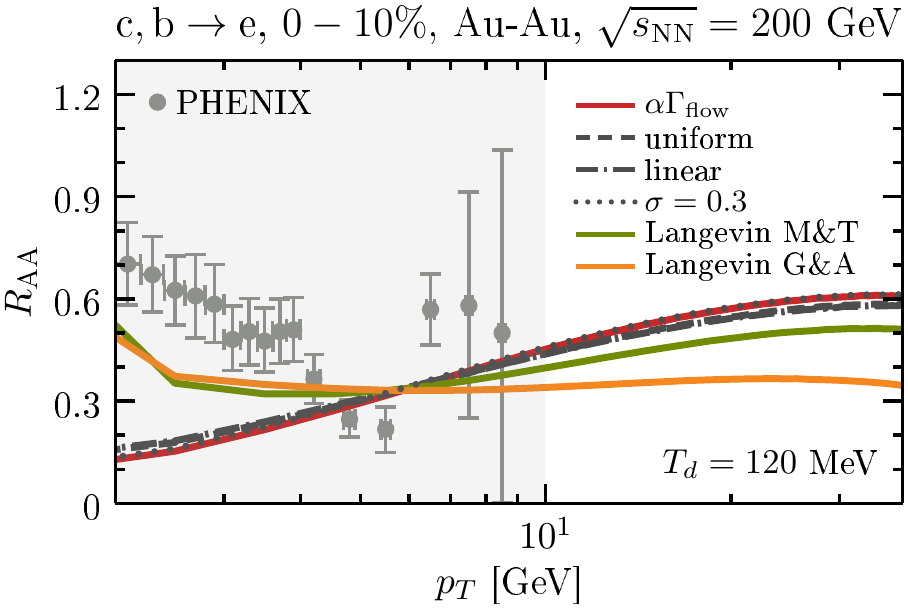}
\hspace{0.3cm}
 \includegraphics[width=0.52\textwidth]{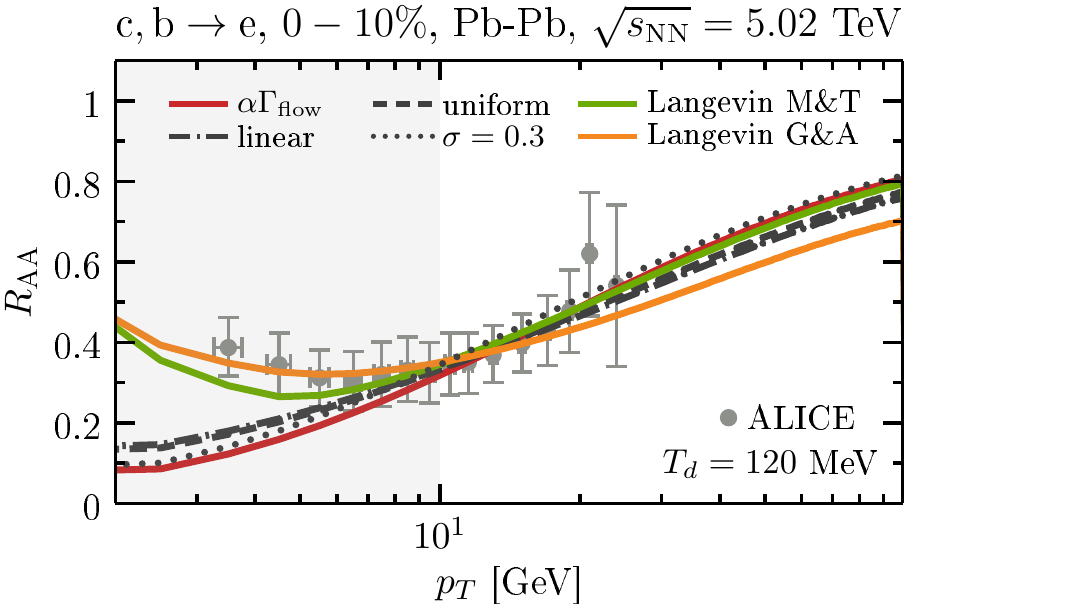}
  \caption{Heavy flavor electron $\raa$ in the 0--10\% centrality class of $\snnGeV[200]$ $\AuAu$ (left) and $\snn[5.02]$ $\PbPb$ (right) collisions. Experimental data from the PHENIX ($|y|<0.35$)~\cite{PhysRevC.84.044905} and ALICE ($|y|<0.6$)~\cite{ALICEprelHFE} collaborations, respectively.}
  \label{fig:Raa_e010}
\end{figure}

\begin{figure}[!htb]
  \centering
 \includegraphics[width=0.52\textwidth]{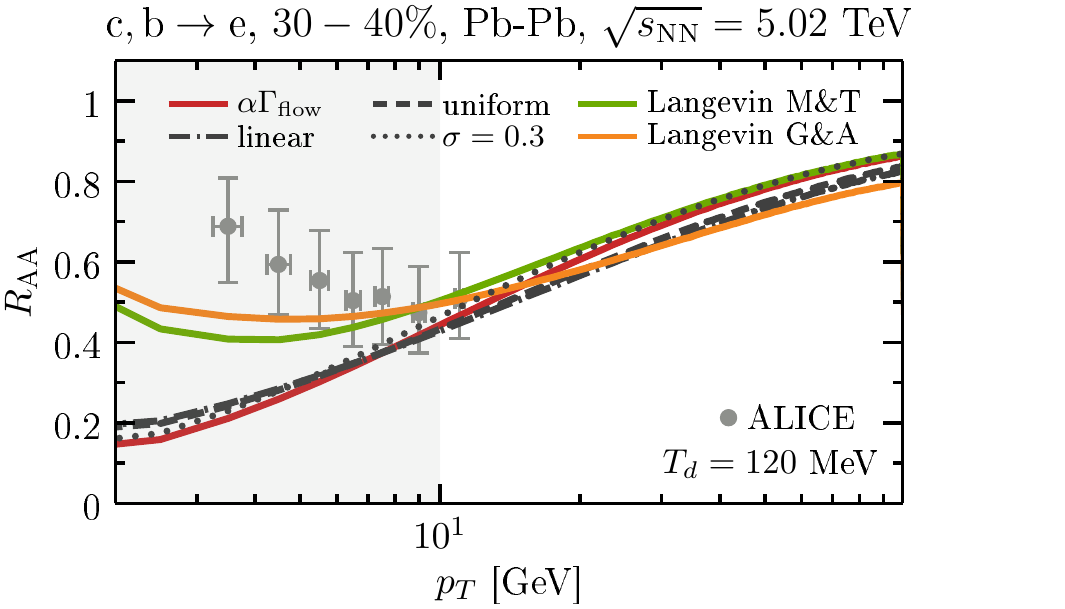}
  \caption{Heavy flavor electron $\raa$ in the 30--40\% centrality class of $\snn[5.02]$ $\PbPb$  collisions. Experimental data from the ALICE ($|y|<0.6$)~\cite{ALICEprelHFE3040} collaboration. }
  \label{fig:Raa_e3040}
\end{figure}

Since very limited $\Bmeson$ meson data is available, one can instead study electrons (or muons) from heavy flavor decays since those are more commonly measured. In Fig.\ \ref{fig:Raa_e010} the results for the heavy flavor electron $\raa$ from semileptonic decays, $\qcharm,\qbottom\rightarrow \electronpositron$, are shown for two beam energies. As with the $\Dmeson$ and $\Bmeson$ meson $\raa(\pt)$, the same hierarchy involving the heavy quark evolution is seen in the electrons and the results fit reasonably well the data at intermediate and high $\pt$ for different beam energies and centralities (see Fig.\ \ref{fig:Raa_e3040}). An exception can be observed in $\AuAu$ collisions at $\snnGeV[200]$ where the models have trouble to reproduce the sudden increase of $\raa$ in the data found around $\pt=6$ GeV. However, we note that those points are located at $\pt<10$~GeV, which is a regime that can be affected by coalescence and other initial/final hadronic effects. Additionally, the error bars in that region are quite large.

\begin{figure}[!htb]
  \centering
 \includegraphics[width=0.48\textwidth]{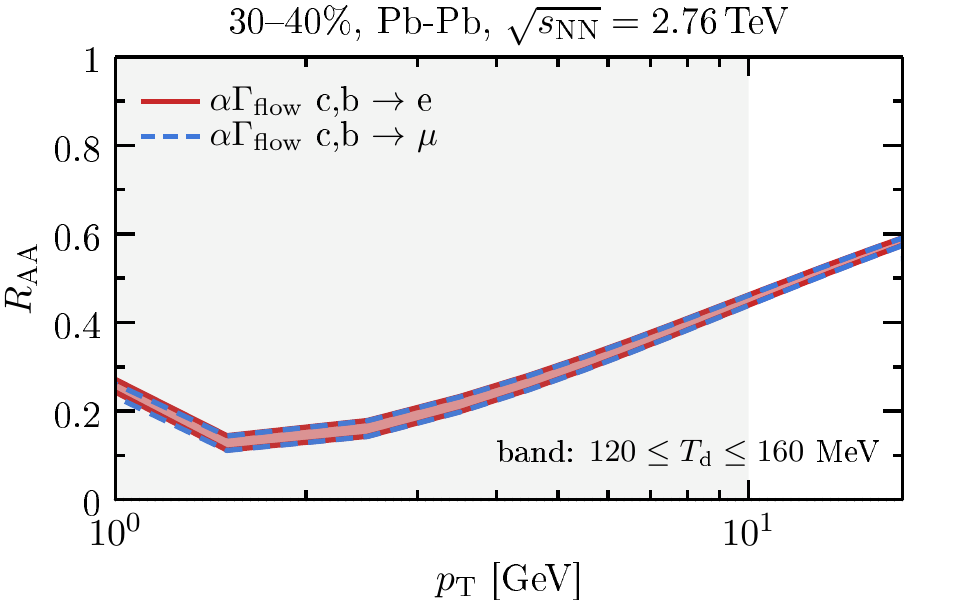}
  \caption{Heavy flavor electron and muon $\raa$ in the 30--40\% centrality class of $\snn[2.76]$ $\PbPb$  collisions.}
  \label{fig:Raa_emuon3040}
\end{figure}

The ATLAS collaboration recently compared our results to muon data in $\PbPb$ $\snn[2.76]$ collisions~\cite{Aaboud:2018bdg} and found that our constant energy loss results were able to reproduce $\raa$ for $\pt\gtrsim 10$ GeV (at that point we did not have Langevin evolution in \dabmod). Note that the results obtained in the electron and muon semileptonic decay channels are almost identical down to 1 GeV as the mass of the leptons that stem from heavy flavor decays is negligible compared to their momentum (see, for instance, Figs.\ \ref{fig:Raa_emuon3040} and \ref{fig:v4_e_LHC13040}).

\begin{figure}[!htb]
  \centering
\includegraphics[width=0.45\textwidth]{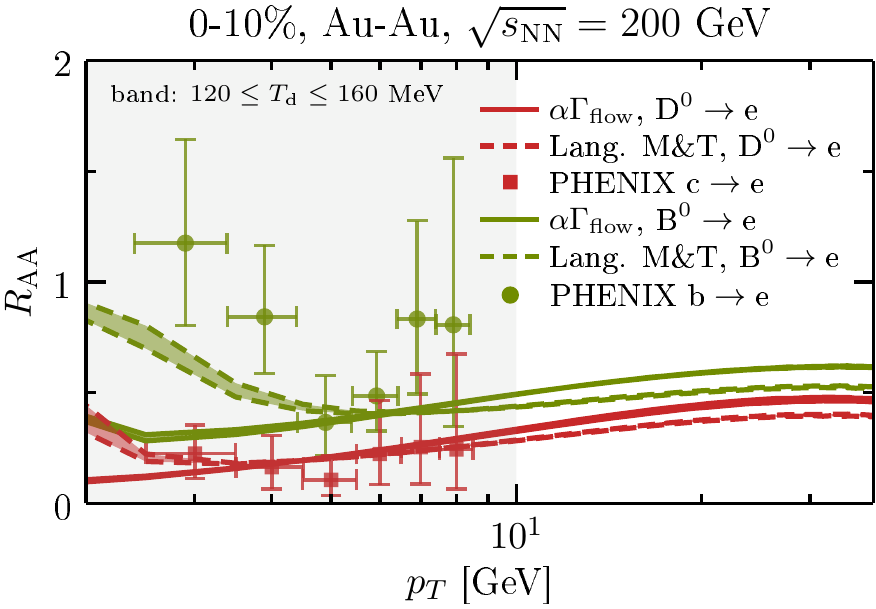}
\hspace{0.1cm}
 \includegraphics[width=0.45\textwidth]{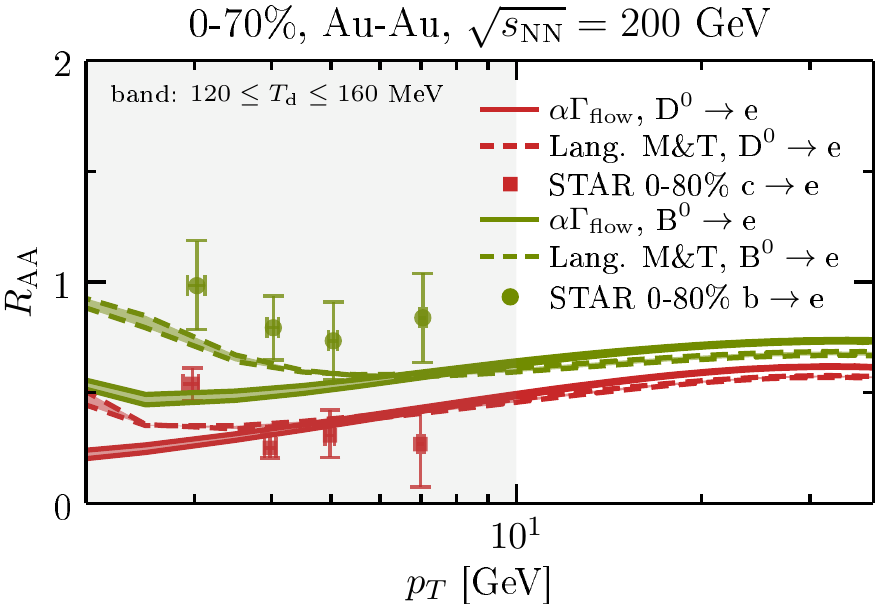}
  \caption{Electron $\raa$ from charm and bottom in $\AuAu$ collisions at $\snnGeV[200]$ for 0--10\% (left) and 0--70\% centrality classes. Experimental data from the PHENIX~\cite{Nagashima2017} (0--10\% centrality) and STAR~\cite{STARprelHFE} (0--80\% centrality) collaborations, respectively.}
  \label{fig:Raa_c_b_e_RHIC}
\end{figure}

\begin{figure}[!htb]
  \centering
\includegraphics[width=0.52\textwidth]{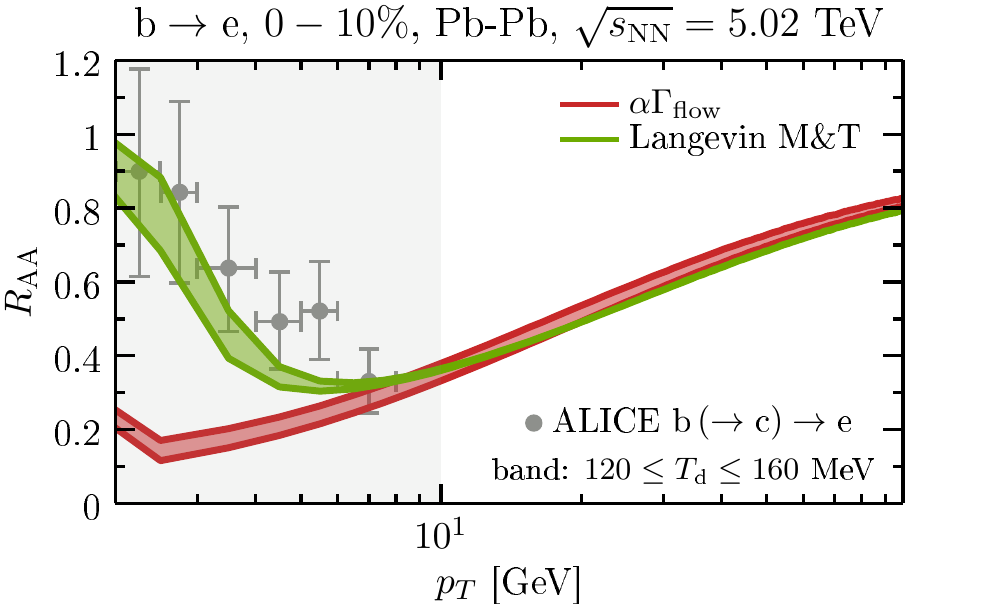}
  \caption{Electron $\raa$ from bottom in the 0--10\% centrality class of $\PbPb$ collisions at $\snn[5.02]$. Experimental data from the ALICE collaboration ($|y|<0.8$)~\cite{Dubla:2018lfa}.}
  \label{fig:Raa_b_e_LHC2}
\end{figure}

Fig.\ \ref{fig:Raa_c_b_e_RHIC} shows a comparison of the separated charm and bottom hadron decay electrons  $\qcharm\rightarrow \electronpositron$ and  $\qbottom\rightarrow \electronpositron$, respectively, for two different centrality ranges. Similarly to \rhic\ data, we observe that the electrons from bottom hadrons are less suppressed than those from charmed hadrons in the low/intermediate $\pt$ regime. The electron $\raa$ from $\Dzero$ mesons reproduces reasonably well \rhic\ data in both centrality ranges, but we underestimate the $\qbottom\rightarrow \electronpositron$ $\raa$ data. Part of this discrepancy might originate from the difficult calibration of the bottom quark transport model coefficients using the heavy flavor data at \rhic\ (because of the sudden $\raa$ increase and large error bars shown in Fig.\ \ref{fig:Raa_e010}). In Fig.\ \ref{fig:Raa_b_e_LHC2} we show a comparison to the (low $\pt$) $\qbottom\rightarrow \electronpositron$ LHC data at $\snn[5.02]$ where the heavy flavor electron data can be used with more confidence for the calibration. As for the $\Dzero$ data in Fig.\ \ref{fig:RAAmodels}, the Langevin models give a better agreement.

If one puts aside the effects of hadronization and initial/final hadronic stages on heavy meson production, a comparison between $\qcharm\rightarrow \electronpositron$ and $\qbottom\rightarrow \electronpositron$ $\raa$ illustrates the mass dependence of the energy exchanges between the heavy quarks and the QGP medium. Within the usual energy loss framework, the experimental observation $\raa^\qcharm<\raa^\qbottom$ is in agreement with the theoretical prediction of the quark mass hierarchy for radiative energy loss, $\Delta E_\qcharm > \Delta E_\qbottom$, mainly due to the dead cone effect increasing with mass~\cite{Dokshitzer:2001zm}. In \dabmod, the heavy quark mass plays a role in different parts of the simulation: 1) inherently in the formalism of the transport models: the evolution depends on mass even without any direct mass dependence in the Langevin and energy loss parametrizations, 2) in the values of the transport model free coupling factors for the charm and bottom quarks, although since they are in general of the same order (Tab.\ \ref{TableFactors} and \ref{TableFactorsBottom}) this is only a small effect, 3) in the initial spectra from FONLL and 4) embedded in the hadronization process (both in the case of fragmentation and coalescence, see Sec.\ \ref{Section: coalescence}).

\subsection{Azimuthal anisotropy and multi-particle cumulants}
\newcommand\notes[1]{\begin{quote} \tiny \color{red}#1\end{quote}}

In a previous work involving all charged high $\pt$ hadrons, multi-particle cumulants were predicted~\cite{Betz:2016ayq} and later measured by \cms~\cite{Sirunyan:2017pan}.  For a correlation of $m$~particles, because of the low statistics at high $\pt$, one hard particle is correlated with $m-1$ soft particles.  In that case, when both particles come from the same $\pt$ bin, one can write the flow vectors $\Vnn$, accounting for both  real and imaginary parts, as
\begin{equation}
    \Vnn = \vnn\, \econst^{\imaginary n \psinn}.
\end{equation}
Here we use the same notation as in~\cite{Gardim:2011xv,Gardim:2014tya,Betz:2016ayq} for the sake of consistency.  However, in the hard sector we are considering the particle of interest from separate $\pt$ bins so a $\pt$ dependence appears
\begin{equation}
    \Vnn(\pt) = \vnn(\pt)\, \econst^{\imaginary n \psinn(\pt)}.
\end{equation}
This leads to a $\pt$ dependent two particle correlation that uses the complex conjugate to produce a real-valued result as in
\begin{equation}
    \Re\{\Vnn\Vnn^* (\pt)\} = \vnn\vnn(\pt)\, \cos\big[n \big(\psinn - \psinn(\pt)\big)\big].
\end{equation}
We then consider the typical 2-particle correlations by taking one hard and one soft particle and the 4-particle correlation by taking 1 hard particle and 3 soft particles that are averaged over events within a fixed centrality window.  They are defined by
\begin{align}
    \vnn\{2\}(\pt) &= \frac{d_n\{2\}(\pt)}{(c_n\{2\})^{\sfrac12}}  \\
    \vnn\{4\}(\pt) &= \frac{d_n\{4\}(\pt)}{(-c_n\{4\})^{\sfrac34}}
\end{align}
where
\begin{align}
    d_n\{2\}(\pt) &= \llangle \Vnn\Vnn^*(\pt) \rangle_j\rangle = \big\langle\!\big\langle \vnn\vnn(\pt)\, \cos\big[n\big(\psinn - \psinn(\pt)\big)\big] \big\rangle_j\big\rangle  \\
    c_{n,j}\{2\}  &= \llangle \Vnn\Vnn^* \rangle_j\rangle      = \llangle \vnn^2 \rangle_j\rangle  \\
    d_n\{4\}(\pt) &= \big\langle 2\, \langle \Vnn\Vnn^* \rangle_j\, \langle \Vnn\Vnn^* (\pt) \rangle_j - \langle \Vnn\Vnn^* \Vnn\Vnn^* (\pt) \rangle_j \big\rangle  \nonumber\\
                  &= \big\langle2\, c_{n,j}\{2\}\, d_{n,j}\{2\}(\pt) - \langle \vnn^2 \Vnn\Vnn^* (\pt) \rangle_j \big\rangle  \\
    -c_{n,j}\{4\} &= \big\langle2\, \langle \Vnn\Vnn^* \rangle_j^2 - \langle \Vnn\Vnn^* \Vnn\Vnn^* \rangle_j \big\rangle = \big\langle2(c_{n,j}\{2\})^2 - \langle v_n^4 \rangle_j \big\rangle .
\end{align}
Here, the outer bracket $\langle\ldots\rangle$ is an artifact of centrality rebinning where in experiments finer centrality bins are taken e.g. 0.5\% centralities, which we indicate as $j$, that are then recombined using multiplicity weighing in a wider centrality bin of a width of 10\%, for example. The inner brackets indicate  averaging over the events within the $j$\textsuperscript{th} fine centrality bin.

Initially, in Ref.~\cite{Betz:2016ayq} it was expected that $\vn2\{4\} / \vn2\{2\}(\pt) \sim \text{constant}$ at low $\pt$.  This is because the ratio of $\vn2\{4\}/\vn2\{2\}(\pt)$ encapsulates  a non-trivial interplay between the event-by-event fluctuations of energy loss and the initial condition fluctuations, with the latter being typically associated only with the soft sector (though it was found in~\cite{Betz:2016ayq} that these fluctuations also influence the multi-particle cumulants at high $\pt$ as well). In fact, in Eq.\ (A2) of Ref.~\cite{Betz:2016ayq}, it was shown that the exact interplay between event-by-event initial condition fluctuations and hard physics fluctuations is given by
\begin{equation}\label{eqn:hardv24}
    \frac{\vnn\{4\}(\pt)}{\vnn\{2\}(\pt)} = \frac{\vnn\{4\}}{\vnn\{2\}} \left[
        1 + \left(\frac{\vnn\{2\}}{\vnn\{4\}}\right)^4 \left(
            \underbrace{\frac{\langle\vnn^4\rangle}{\langle\vnn^2\rangle^2}}_\text{soft fluctuations} -
            \underbrace{\frac{\langle\vnn^2\Vnn\Vnn^*(\pt)\rangle}{\langle\vnn^2\rangle \langle\Vnn\Vnn^*(\pt)\rangle}}_\text{hard fluctuations}
        \right)
    \right],
\end{equation}
where the terms $\frac{\langle\vnn^4\rangle}{\langle\vnn^2\rangle^2}$ and $\frac{\vnn\{2\}}{\vnn\{4\}}$ are determined from initial condition fluctuations, which are translated into the final flow harmonics via linear+cubic response~\cite{Noronha-Hostler:2015dbi,Sievert:2019zjr} within the soft sector, i.e., in the context of relativistic hydrodynamics. The only term that contains contributions from the hard sector is then $\frac{\langle\vnn^2\Vnn\Vnn^*(\pt)\rangle}{\langle\vnn^2\rangle \langle\Vnn\Vnn^*(\pt)\rangle}$, which can be interpreted as a measure of azimuthal anisotropy fluctuations associated with hard physics.  When there are only soft fluctuations
\begin{align}
    \text{Only soft fluctuations:}&&
        \frac{\langle\vnn^2\Vnn\Vnn^*(\pt)\rangle}{\langle\vnn^2\rangle \langle\Vnn\Vnn^*(\pt)\rangle} &\rightarrow \frac{\langle\vnn^4\rangle}{\langle\vnn^2\rangle^2}, \\
    \intertext{which implies that Eq.~(\ref{eqn:hardv24}) then becomes}
\label{eqn:obs_soft}
    \text{Only soft fluctuations:}&&
        \frac{\vnn\{4\}(\pt)}{\vnn\{2\}(\pt)} &\rightarrow \frac{\vnn\{4\}}{\vnn\{2\}}.
    \intertext{In fact, $\frac{\vnn\{4\}(\pt)}{\vnn\{2\}(\pt)}$ is only sensitive to hard physics fluctuations when the soft and hard sector have \emph{different} sources of fluctuations.  At that point one can write}
    \text{Soft + Hard fluctuations:}&&
        \frac{\langle\vnn^2\Vnn\Vnn^*(\pt)\rangle}{\langle\vnn^2\rangle \langle\Vnn\Vnn^*(\pt)\rangle} &\neq \frac{\langle\vnn^4\rangle}{\langle\vnn^2\rangle^2},
\end{align}
and the term in parentheses in Eq.~(\ref{eqn:hardv24}) is nonzero.  We should note that it is not known a priori if soft fluctuations are larger or smaller than hard fluctuations and, thus, the term $\left(\frac{\vnn\{2\}}{\vnn\{4\}}\right)^4 \left(\frac{\langle  \vnn^4\rangle}{\langle \vnn^2 \rangle^2} - \frac{\langle \vnn^2  \Vnn \Vnn^*(\pt)\rangle}{\langle \vnn^2 \rangle \langle \Vnn\Vnn^*(\pt)\rangle} \right)$ may contribute positively or negatively to the total $\frac{\vnn\{4\}(\pt)}{\vnn\{2\}(\pt)}$.  However, when a deviation from Eq.~(\ref{eqn:obs_soft}) occurs this implies that some other type of physics is occurring that does not stem from the soft sector. Therefore, this illustrates the importance of direct theory to experiment comparisons of $\frac{\vnn\{4\}(\pt)}{\vnn\{2\}(\pt)}$ in the hard sector.

Previous work had investigated 2-particle correlations coming from the same $\pt$ bin~\cite{Heinz:2013bua}, for instance, 2-particle correlations up to $\sim \SI{10}{GeV}$~\cite{ATLAS:2012at,Zhou:2014bba,Zhou:2018adv}.  However, when it comes to energy loss models that are not fully integrated into hydrodynamics, they only typically become valid above $\pt > \SI{10}{GeV}$ so those previous measurements require additional physics to interpret their results. More recently, multi-particle cumulant calculations have been extended to the heavy flavor sector~\cite{Prado:2016szr} by correlating heavy flavor particles with low $\pt$ charged particles.  However, they have not yet been systematically studied in the heavy flavor sector nor have they been measured experimentally. In the following, we systematically compare the influence of different medium transport properties and initial conditions on heavy flavor multi-particle cumulants for the first time.\footnote{We note that in~\cite{Xu:2018gux} a systematic study was done with different energy loss models for $\raa$ and $v_2$ but higher order flow harmonics and multi-particle cumulants were not yet considered.}

\subsubsection{Elliptic flow from two-particle cumulants}

\begin{figure}[h]
    \centering
    \includegraphics[width=0.6\textwidth]{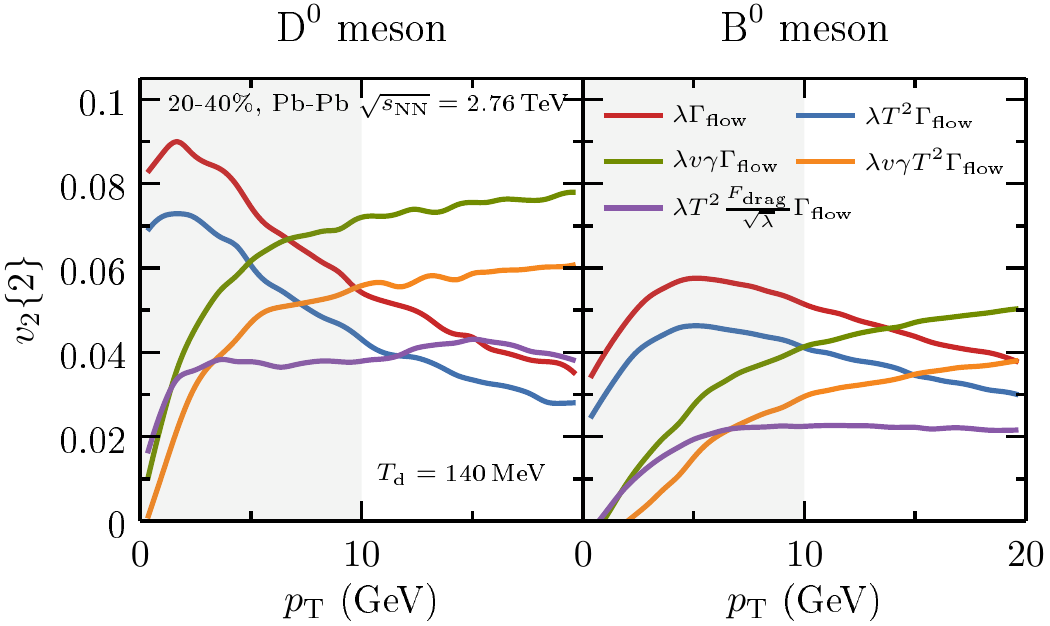}
    \caption{Elliptic flow $\vn2$ of $\Dzero$ mesons (left) and $\Bzero$ mesons (right) in the 20--40\% centrality class of $\PbPb$ collisions at $\snn[2.76]$ obtained with different energy loss models.}
    \label{fig:v2_egyloss}
\end{figure}

First we test the influence of different energy loss models. As a test bed we use $\PbPb$ collisions at $\snn[5.02]$ with \mckln\ initial conditions, which are the same hydrodynamic backgrounds used in~\cite{Betz:2016ayq}. In Fig.\ \ref{fig:v2_egyloss} results for $\Dzero$ and $\Bzero$ mesons are shown for the same energy loss models previously shown in Fig.\ \ref{fig:RAA_few_parameterizations} for $\raa$. We notice that the different models can lead to a variety of different curves for $\vn2$. At the same time, models that are indistinguishable using only the $\raa$ calculations are clearly separated when considering the $\vn2$ results, as is the case for the two energy loss parametrizations $f=\alpha$ and $f=\xi T^2$. These two particular models differ from the other three in that they do not have an explicit dependence on the heavy quark momentum and lead to a bump in the low $\pt$ region as is expected from experimental data. This observation agrees with the previous choice of energy loss parametrizations made using $\raa$ data.

\begin{figure}[h!]
    \centering
    \includegraphics[width=0.495\textwidth]{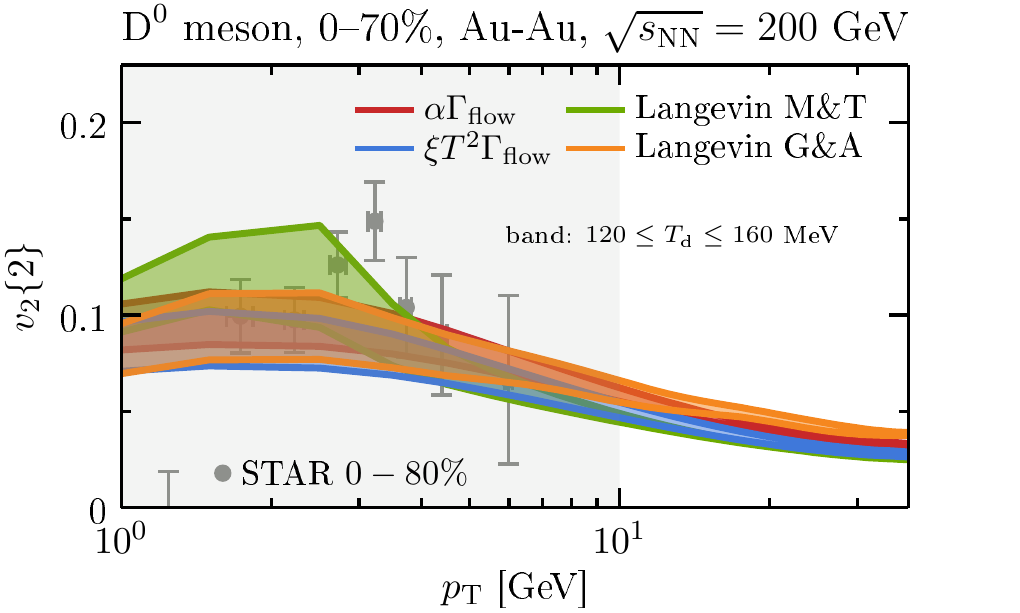}
    \includegraphics[width=0.495\textwidth]{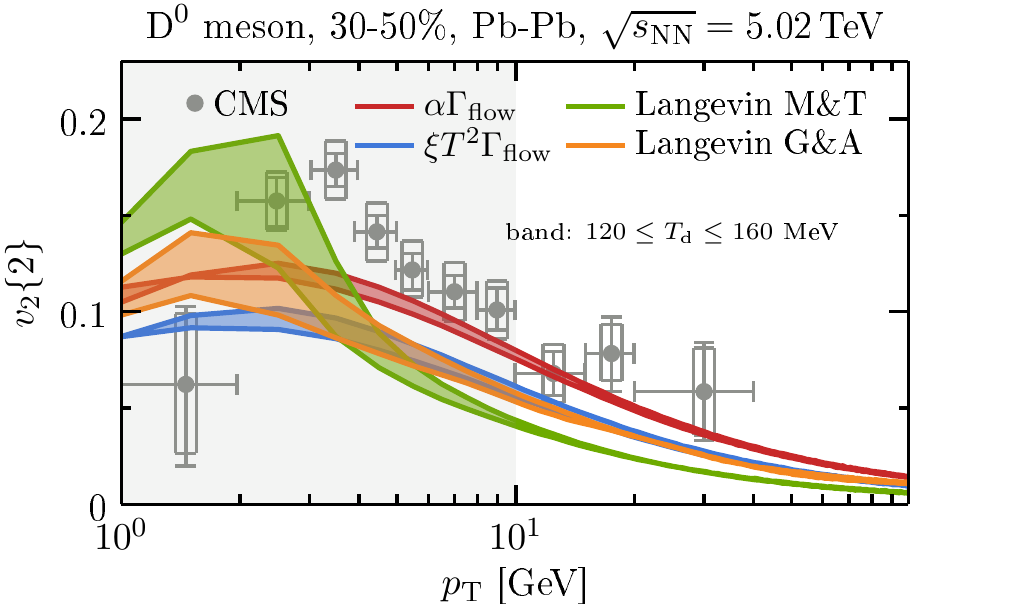}
    \caption{$\Dzero$ meson elliptic flow $\vn2$ in the 0--70\% centrality range of $\snnGeV[200]$ $\AuAu$ collisions (left) and in the 30--50\% centrality range of $\snn[5.02]$ $\PbPb$ collisions (right). The gray area indicates the $\pt$ region where coalescence may be important. Experimental data from the \STAR\ ($|y| < 1$)~\cite{Adamczyk:2017xur} and \cms\ ($|y| < 1$)~\cite{Sirunyan:2017plt} collaborations, respectively.}
    \label{fig:v2_models}
\end{figure}

The selected energy loss models are compared with the two different parametrizations within the Langevin formalism in Fig.\ \ref{fig:v2_models} for a range of decoupling temperatures. On the left side we observe that the M\&T parametrization seems to better capture the characteristic bump in the low $\pt$ regime present in the experimental data while other models seem to overlap in $\AuAu$ collisions at \rhic\ with $\snnGeV[200]$. In the case of $\PbPb$ $\snn[5.02]$ collisions at \lhc, the different models seem to be more distinguishable with the constant energy loss model leading to the largest values of $\vn2$ at large $\pt$, while the Langevin model with M\&T parametrization remains closest to experimental data at low $\pt$.  Results for $\vn2$ show better agreement with data for the lowest collision energy overall, while being slightly underestimated in the other case.

\begin{figure}[h!]
    \centering
    \includegraphics[width=0.46\textwidth]{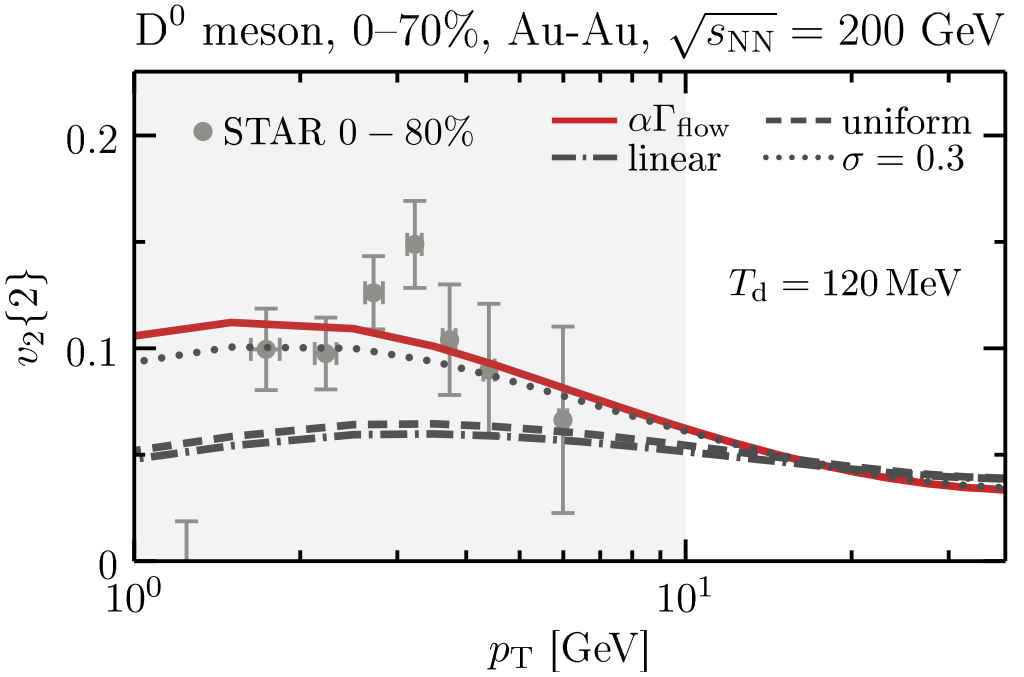}
    \hspace{0.2cm}
    \includegraphics[width=0.465\textwidth]{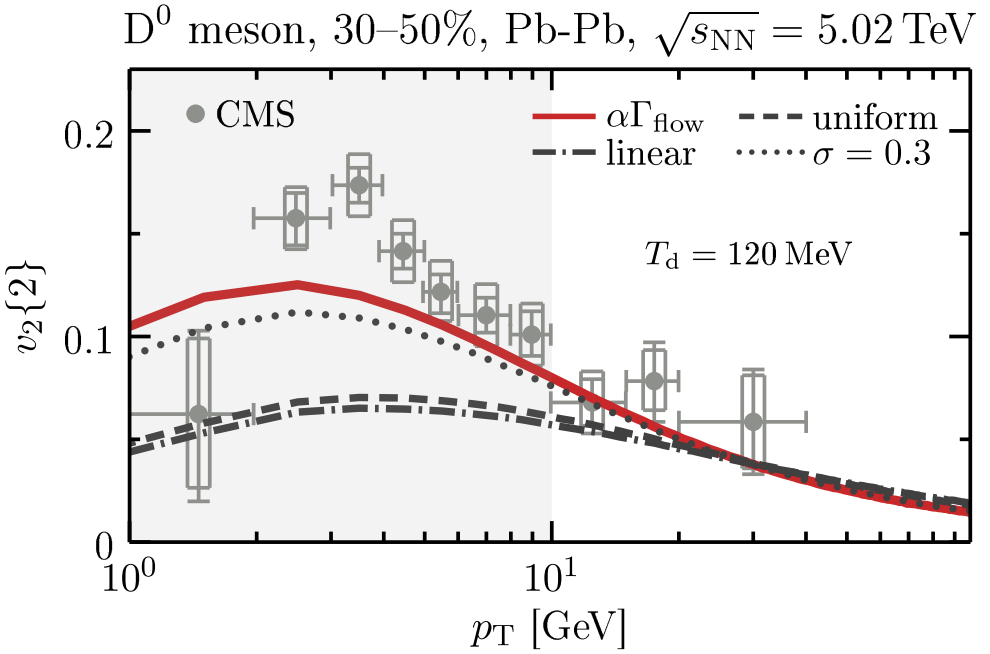}
    \caption{$\Dzero$ meson elliptic flow $\vn2$ in the 0--70\% centrality range of $\snnGeV[200]$ $\AuAu$ collisions (left) and in the 30--50\% centrality range of $\snn[5.02]$ $\PbPb$ collisions (right). Comparison of the constant energy loss model with and without different types of energy loss fluctuations. Experimental data from the \STAR\ ($|y| < 1$)~\cite{Adamczyk:2017xur} and \cms\ ($|y| < 1$)~\cite{Sirunyan:2017plt} collaborations, respectively.}
    \label{fig:v2_egylossfluc}
\end{figure}

\begin{figure}[h!]
    \centering
    \includegraphics[width=0.495\textwidth]{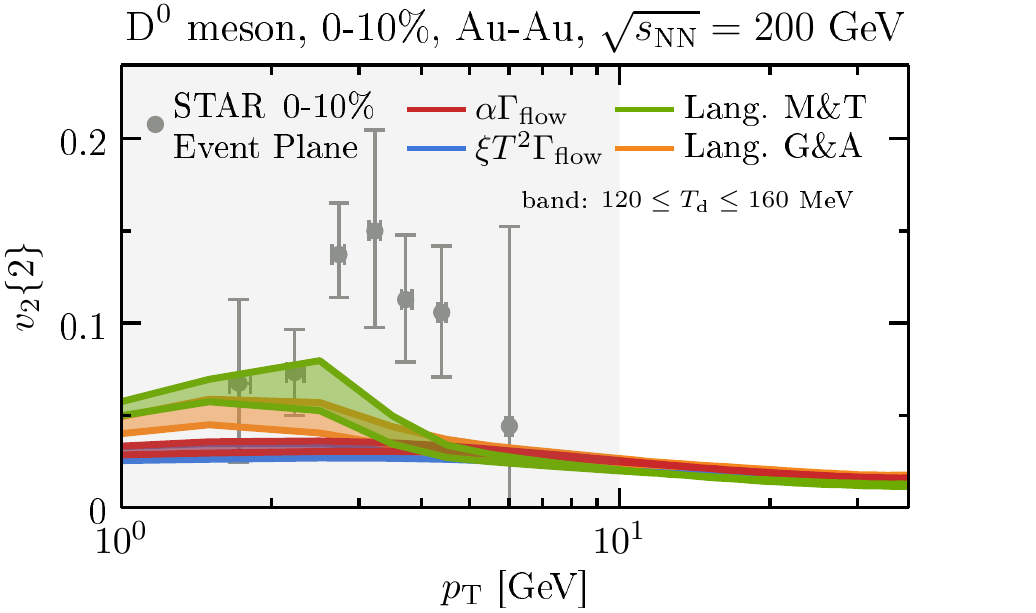}
    \includegraphics[width=0.495\textwidth]{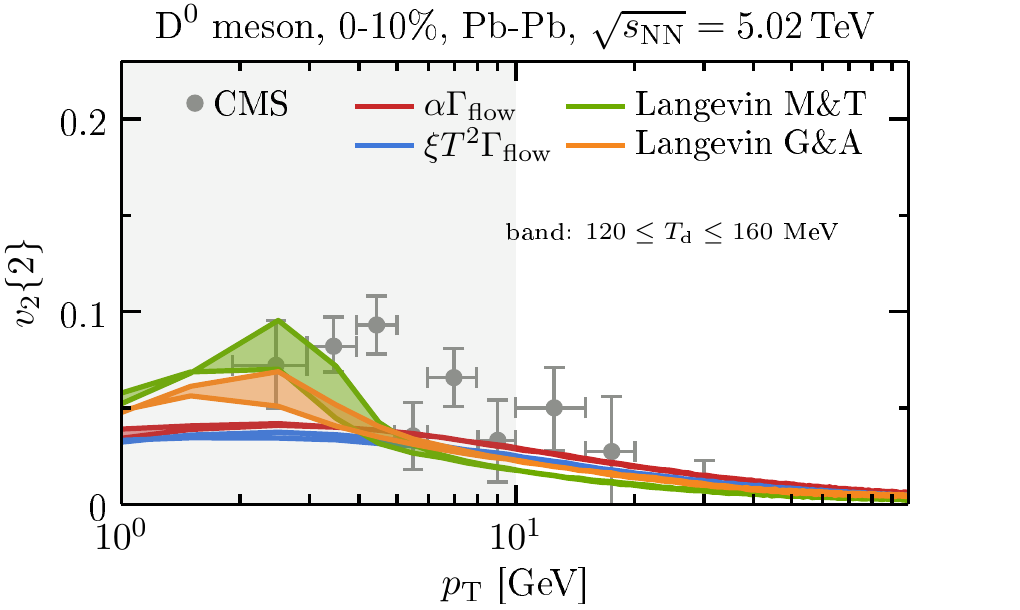}
    \caption{$\Dzero$ meson elliptic flow $\vn2$ in the 0--10\% centrality class of $\snnGeV[200]$ $\AuAu$  collisions (left) and in the 0--10\% centrality class of $\snn[5.02]$ $\PbPb$ collisions (right). Preliminary experimental data from \STAR\ ($|y|<1$)~\cite{Bruna:2019lcu} and data from the \cms\ ($|y|<1$)~\cite{Sirunyan:2017plt} collaboration are shown.}
    \label{fig:v2_central}
\end{figure}

We take a special look at how the three parametrizations of energy loss fluctuations affect the results for the anisotropic flow coefficient. These results are shown in Fig.\ \ref{fig:v2_egylossfluc}.  The Gaussian fluctuation Ansatz leads to a more subtle effect by slightly lowering $\vn2$ at low $\pt$.  The other two models lead to stronger effects and, even though they have very different functional forms, the $\vn2$ results do not show much difference among them. These results suggest that energy loss fluctuations are more relevant to the calculation of the low $\pt$ regime of the elliptic flow coefficient.

It is also possible to compare the simulation results with experimental data for central collisions, as it is shown in Fig.\ \ref{fig:v2_central}. The same behavior observed in Fig.\ \ref{fig:v2_models} is also found in this case when comparing the different models. However, the experimental data comparison shows a better agreement with \lhc\ collisions rather than \rhic's, in which case the results from the simulation underestimate the data except for very low $\pt$ regime.  Since these comparisons very much rely on the low $\pt$ region, we expect that the introduction of coalescence in the simulations to change these results.

\begin{figure}[h]
    \centering
    \includegraphics[width=0.50\textwidth]{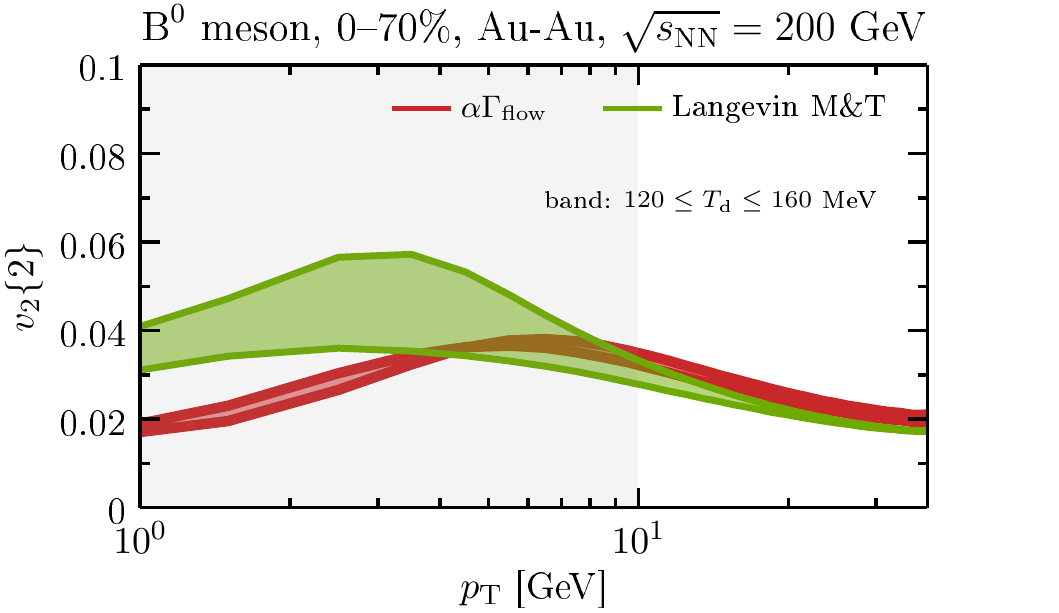}
    \includegraphics[width=0.49\textwidth]{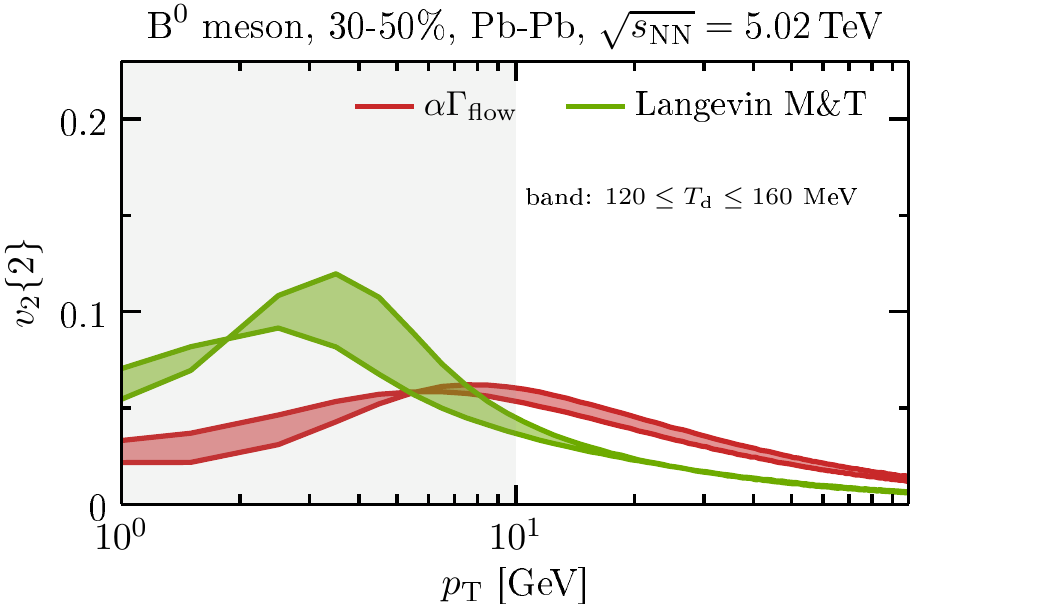}
    \caption{$\Bzero$ meson elliptic flow $\vn2$ in the 0--70\% centrality range of  $\snnGeV[200]$ $\AuAu$  collisions (left) and in the 30--50\% centrality range of $\snn[5.02]$ $\PbPb$ collisions (right).}
    \label{fig:v2_Bmeson}
\end{figure}

In Fig.\ \ref{fig:v2_Bmeson} we show predictions for $\vn2$ of $\Bzero$ mesons obtained from our simulations, using only the constant energy loss parametrization and the M\&T parametrization within the Langevin formalism, for the same collision setup that was used for the $\Dmeson$ meson results. We notice a similar behavior between both models with Langevin's leading to a larger $\vn2$ at low $\pt$ though it becomes lower at high $\pt$.

\begin{figure}[h]
    \centering
    \includegraphics[width=0.48\textwidth]{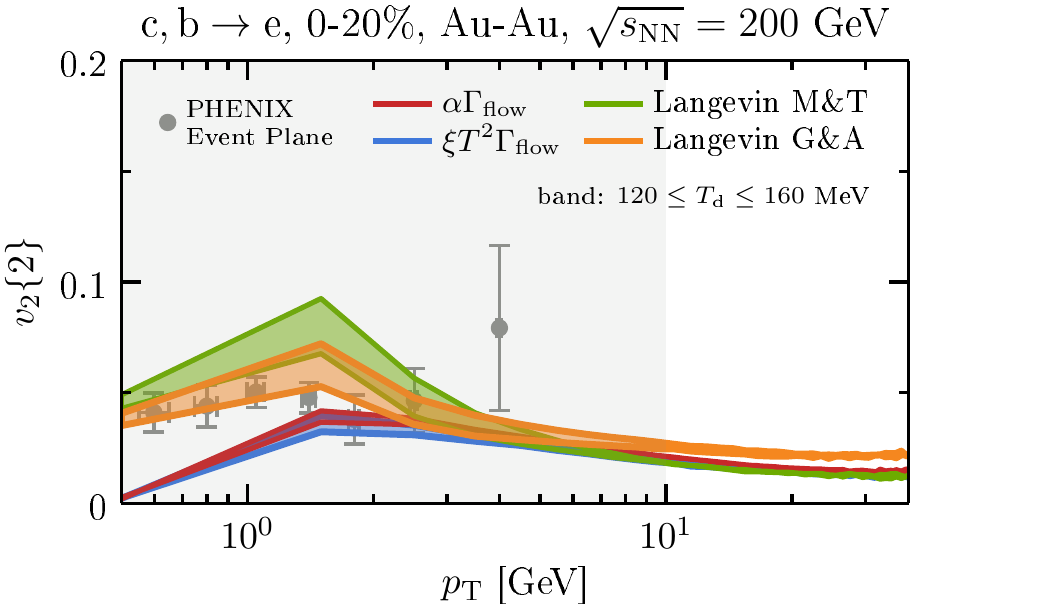}
    \includegraphics[width=0.48\textwidth]{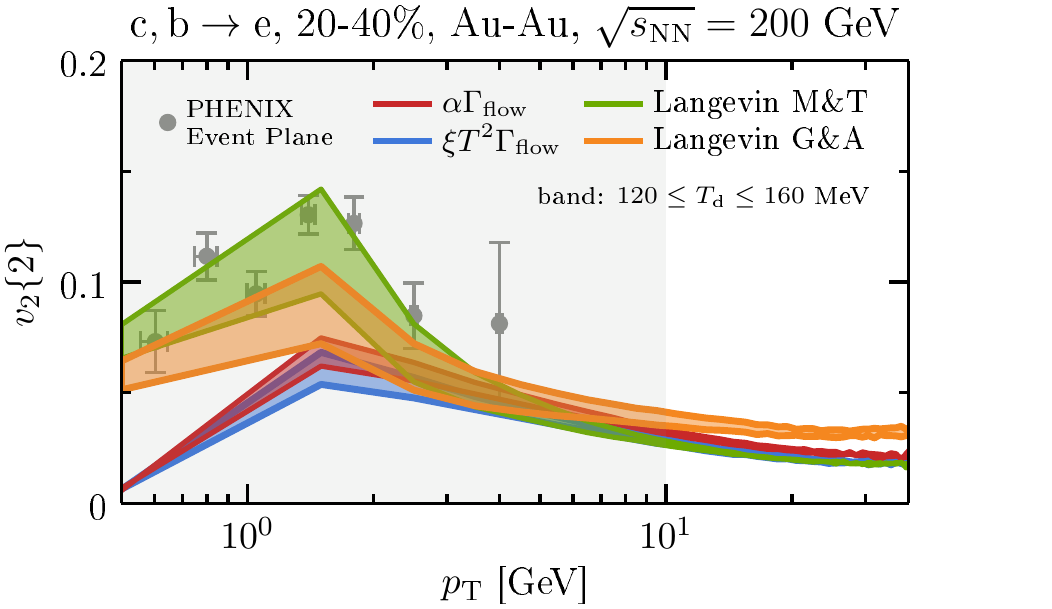}
    \caption{Heavy flavor electron elliptic flow coefficient $\vn2$ in the 0--20\% centrality class of $\snnGeV[200]$ $\AuAu$ collisions  (left) and also in the 20--40\% centrality class (right). The gray area indicates the $\pt$ region where coalescence may be important.  Experimental data from the \phenix\ ($|y| < 0.35$) collaboration~\cite{Adare:2010de}.}
    \label{fig:v2_electron}
\end{figure}

\begin{figure}[h]
    \centering
    \includegraphics[width=0.50\textwidth]{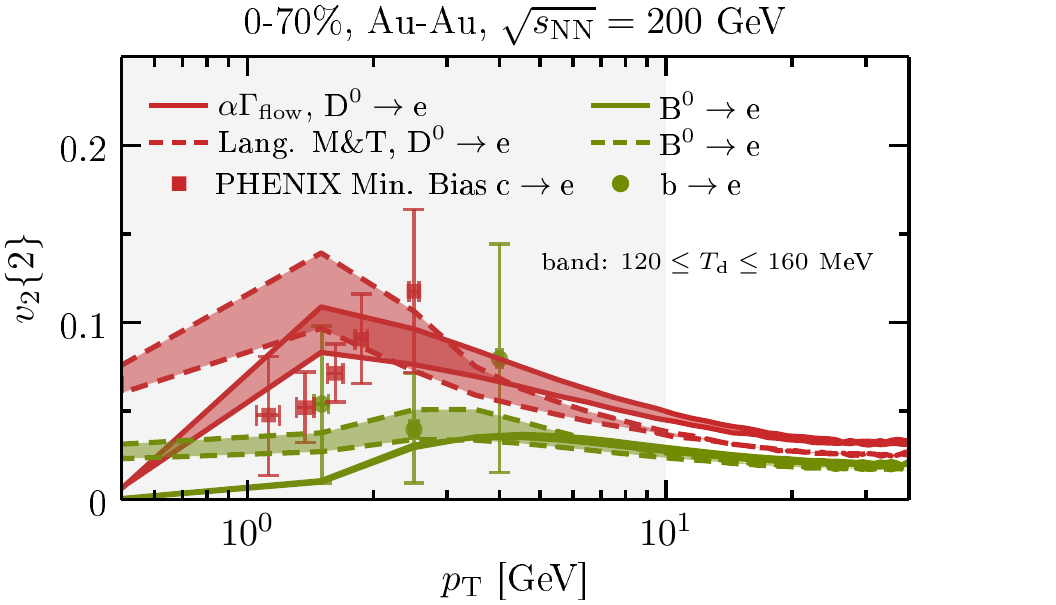}
    \caption{Electron elliptic flow coefficient $\vn2$ from charm and bottom in the 0--70\% centrality range of  $\snnGeV[200]$ $\AuAu$ collisions.  Experimental data from the \phenix\ collaboration~\cite{Bruna:2019lcu}.}
    \label{fig:v2_bcelectron}
\end{figure}

Having both results for $\Bzero$ and $\Dzero$ mesons, we can now obtain the elliptic flow coefficient for heavy flavor electrons from decays. The comparison between the different transport models for this case is shown in Fig.\ \ref{fig:v2_electron}. We observe that the Langevin parametrizations tend to better reflect the features of the experimental data, especially for non-central collisions, shown in the right plot of the figure. In the case of central collisions, although the results for the $\Dzero$ meson underestimated the data, a good agreement with data is observed for the heavy flavor electrons for both Langevin parametrizations.  Also, the difference between the models in the mid-$\pt$ range up to $\approx \SI{30}{GeV}$ is not as pronounced for electrons as it is for the heavy mesons.

We further explore the role of each heavy quark, bottom or charm, in building up the elliptic flow of heavy flavor electrons in Fig.\ \ref{fig:v2_bcelectron} where the results for $\AuAu$ collisions at $\snnGeV[200]$ are shown compared to experimental data from the \phenix\ collaboration. A good agreement is obtained considering the large uncertainties in the measured values. Unfortunately, the comparison is limited to the lower $\pt$ range where the physics is considerably more complex and other effects might be important.

\subsubsection{Elliptic flow from multi-particle cumulants}\label{sec:multi}

\begin{figure}[!b]
    \centering
    \includegraphics[width=0.9\textwidth]{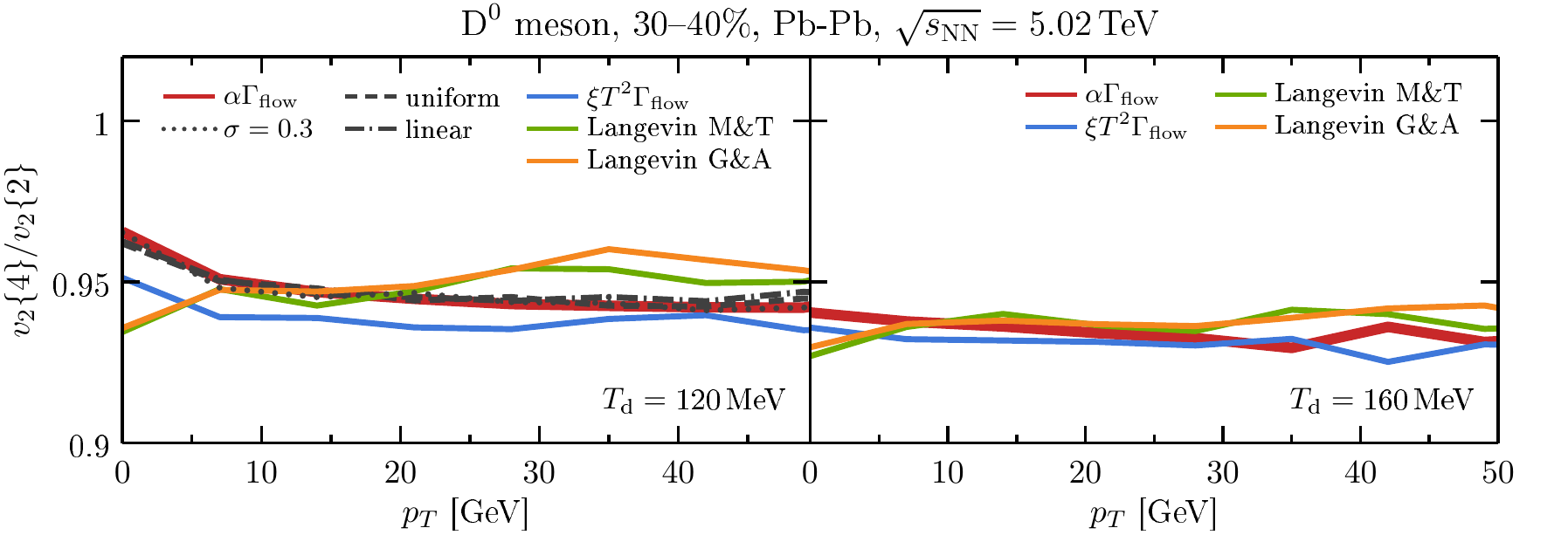}
    \caption{$\vn2\{4\}/\vn2\{2\}(\pt)$ of $\Dzero$ mesons as a function of $\pt$ for the two decoupling temperatures $\Td = \SI{120}{MeV}$ (left) and $\Td = \SI{160}{MeV}$ (right).}
    \label{fig:v24v22_ELoss}
\end{figure}

Up to this point we only explored the correlation between the particles using 2-particle cumulants. We now test the influence of different energy loss and Langevin diffusion coefficients on the calculation of the 4-particle cumulants. For the 30--40\% centrality class in the soft sector \mckln\ initial conditions produce $v_2\left\{4\right\}/v_2\left\{2\right\}\approx 0.91$. In Fig.\ \ref{fig:v24v22_ELoss} the corresponding heavy flavor predictions are shown. Similar to the high $p_T$ particles result in~\cite{Betz:2016ayq}, the heavy flavor predictions are  $v_2\left\{4\right\}/v_2\left\{2\right\}\approx 0.95$. There is a slight downward shift if a higher decoupling temperature is considered, which may provide some insight into the optimal values of the decoupling temperature parameter. Additionally, energy loss fluctuations have an opposite trend in the $\pt$ dependence compared to Langevin results. In the energy loss approach a peak in $v_2\{4\}/v_2\{2\}(\pt)$ is seen at low $\pt$ whereas in the Langevin scenario $v_2\{4\}/v_2\{2\}(\pt)$ increases at high $\pt$.

\begin{figure}[!htb]
  \centering
\includegraphics[width=0.48\textwidth]{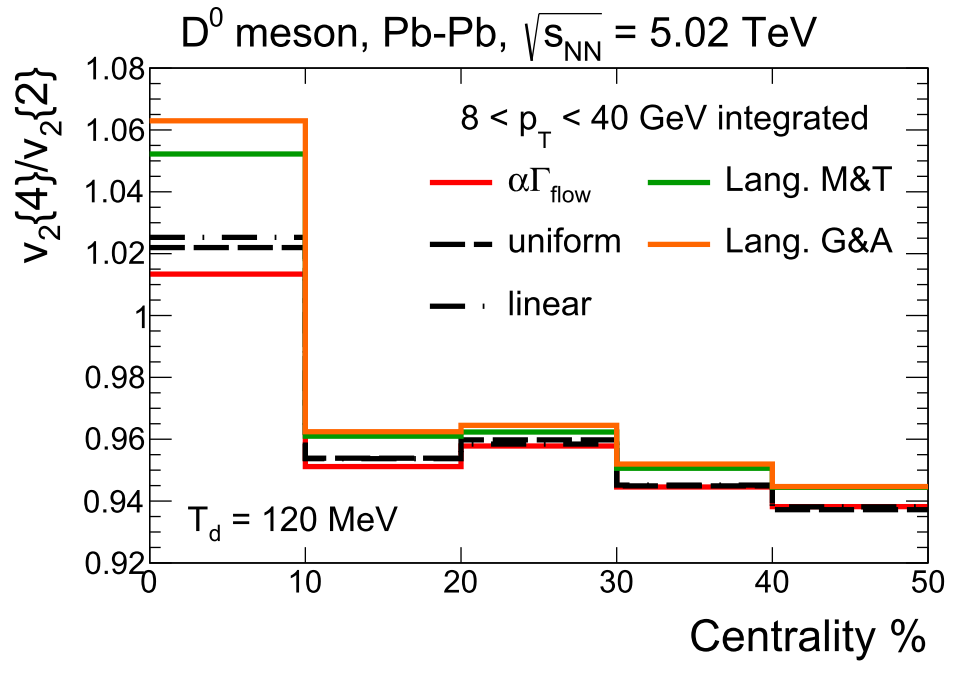}
 \includegraphics[width=0.48\textwidth]{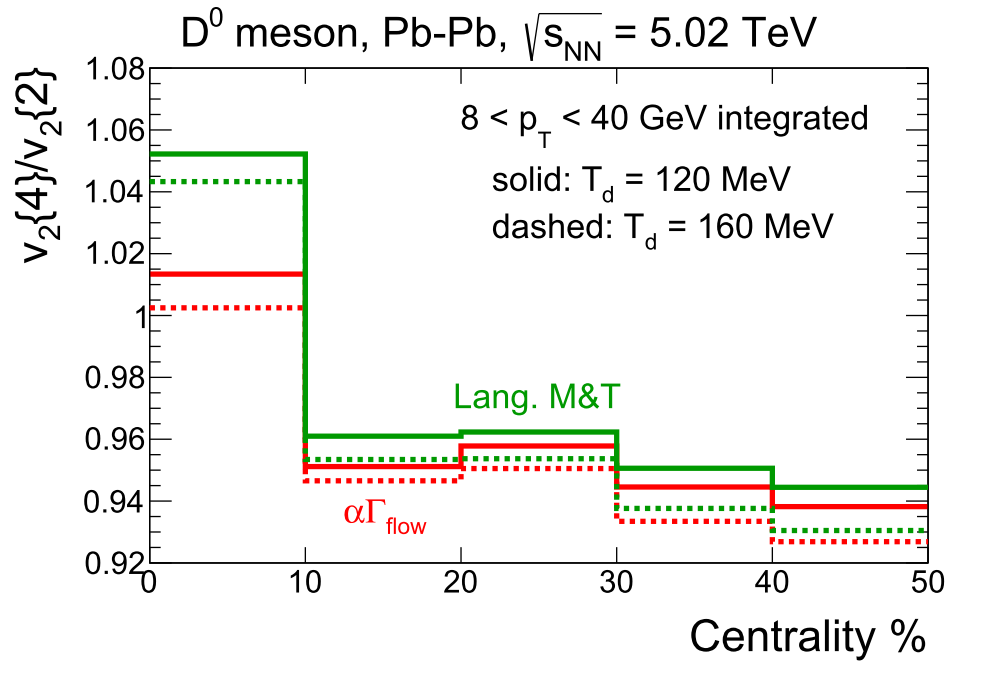}
  \caption{Centrality dependence of the $\Dzero$ meson $v_2\{4\}/v_2\{2\}$  integrated over $8 < \pt < 40$ GeV for $\snn[5.02]$ $\PbPb$ collisions. Left: Dependence with the transport model assumptions. Right: Dependence with the decoupling temperature parameter $\Td$. }
  \label{fig:fluc_cen}
\end{figure}
Furthermore, it is also instructive to investigate the centrality dependence of this observable. Thus, we also consider $v_2\{4\}/v_2\{2\}$ integrated in the range $8 < \pt < 40$ GeV. In Fig.\ \ref{fig:fluc_cen} different types of energy loss fluctuations and Langevin diffusion transport coefficients are shown (left) for a fixed decoupling temperature. On the right, the decoupling temperatures are compared for our two best fit setups.
One of the biggest takeaways from Fig.\ \ref{fig:fluc_cen} is that the central collision region is the most sensitive regime in the description of heavy flavor dynamics. More peripheral collisions are predicted to have nearly identical results regardless of the underlying assumptions made in the heavy flavor modeling. However, peripheral collisions are more sensitive to the decoupling temperature (and less sensitive to the heavy flavor description) so by investigating 0--10\% and 40--50\% centrality classes one may be able to constrain both simultaneously. The caveat relies on the ability of experiments to obtain these results with reasonable error bars since the differences are small.

\begin{figure}[!htb]
  \centering
\includegraphics[width=0.48\textwidth]{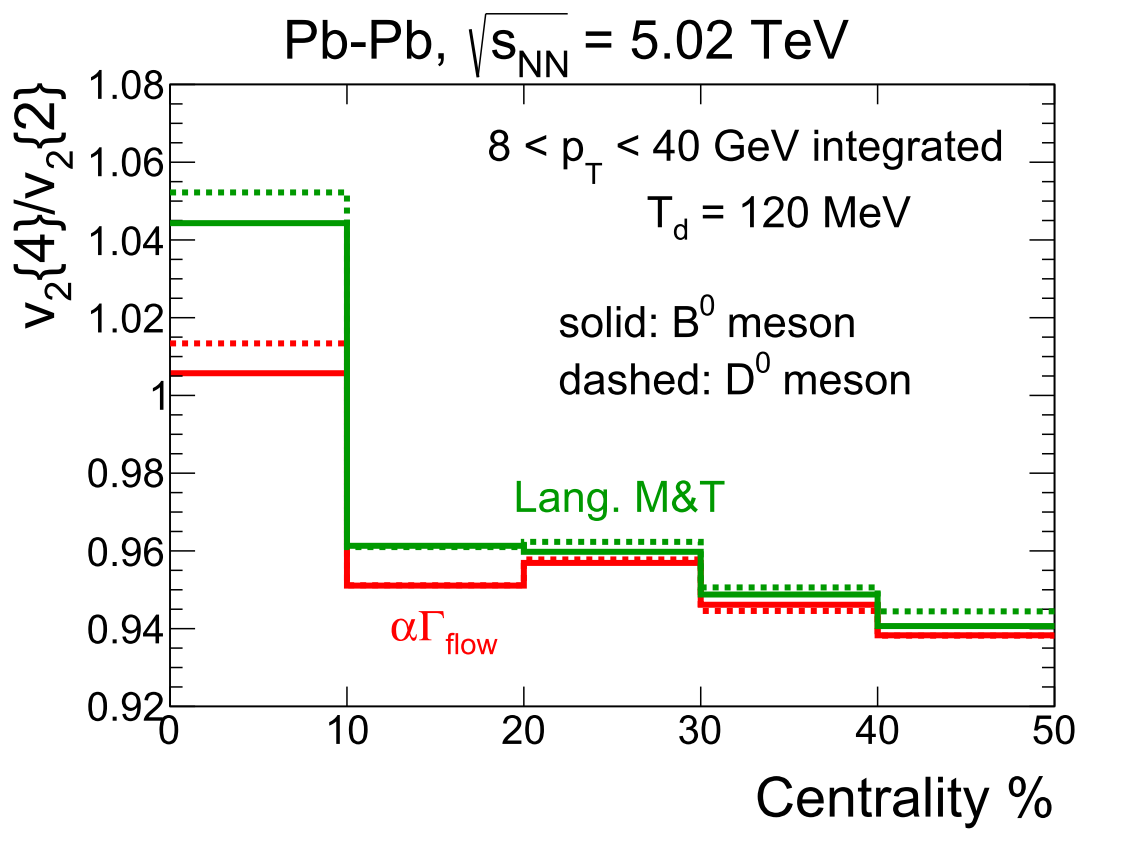}
  \caption{ Comparison between the $\Dzero$ and $\Bzero$ meson $v_2\{4\}/v_2\{2\}$ ratio integrated over $8 < \pt < 40$ GeV in $\snn[5.02]$ $\PbPb$ collisions. }
  \label{fig:DvB}
\end{figure}

\begin{figure}[!htb]
  \centering
 \includegraphics[width=0.48\textwidth]{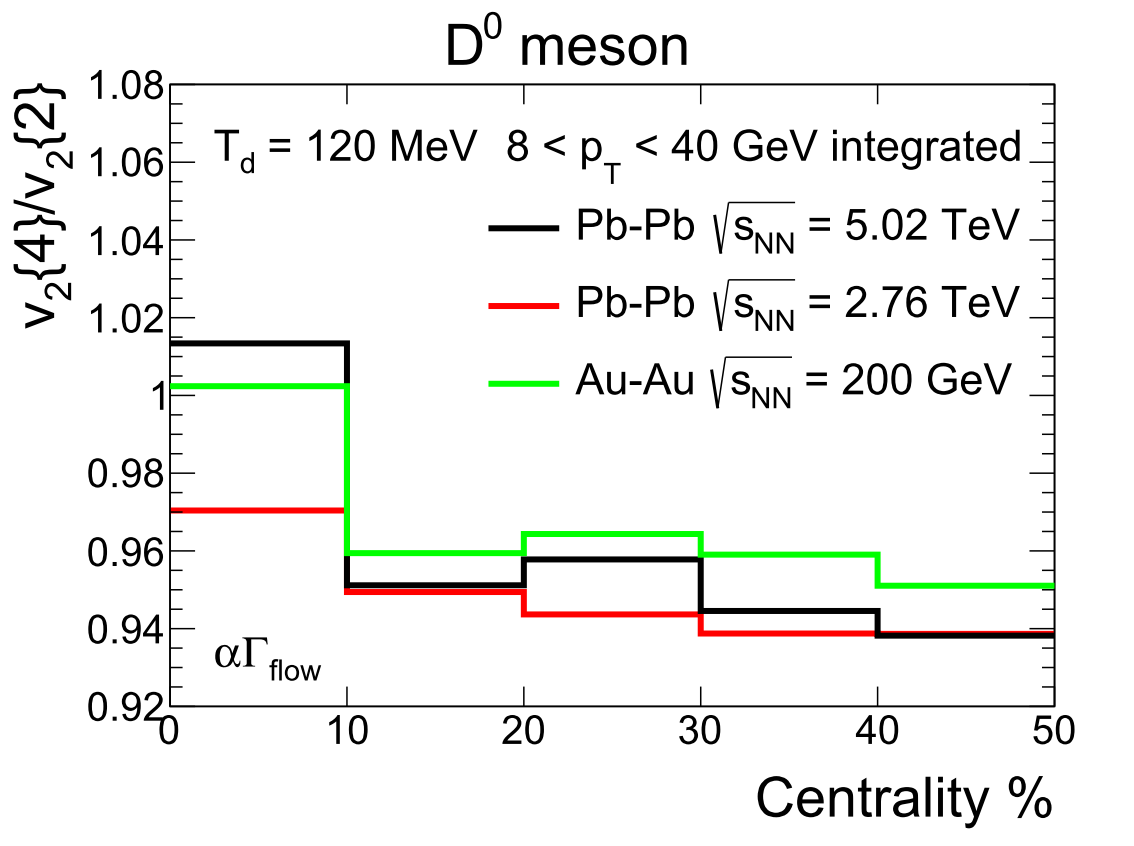}
  \caption{ Comparison between the integrated $v_2\{4\}/v_2\{2\}$ ratio in $\snnGeV[200]$ $\AuAu$ collisions and in $\PbPb$ collisions at $\snn[2.76]$ and $\snn[5.02]$.}
  \label{fig:Dsnn}
\end{figure}

It is also interesting to consider possible mass effects on the $v_2$ fluctuations of heavy flavor mesons. Up to this point, our previous results only considered $\Dzero$ mesons but here we compare $\Dzero$ and $\Bzero$ mesons in Fig.\ \ref{fig:DvB}. We find that $\Bzero$ mesons have a slightly larger  $v_2\{4\}/v_2\{2\}$ in central collisions but the effect is very small. Finally, we show how the $\Dzero$ meson $v_2\{4\}/v_2\{2\}$ changes with collision energy and system size in Fig.\ \ref{fig:Dsnn}. Again, the largest difference occurs in central collisions while, for peripheral collisions, results at \lhc\ energies become similar though still distinct from the \rhic\ result.

\subsubsection{Triangular flow with two-particle cumulants}

\begin{figure}[h]
    \centering
    \includegraphics[width=0.47\textwidth]{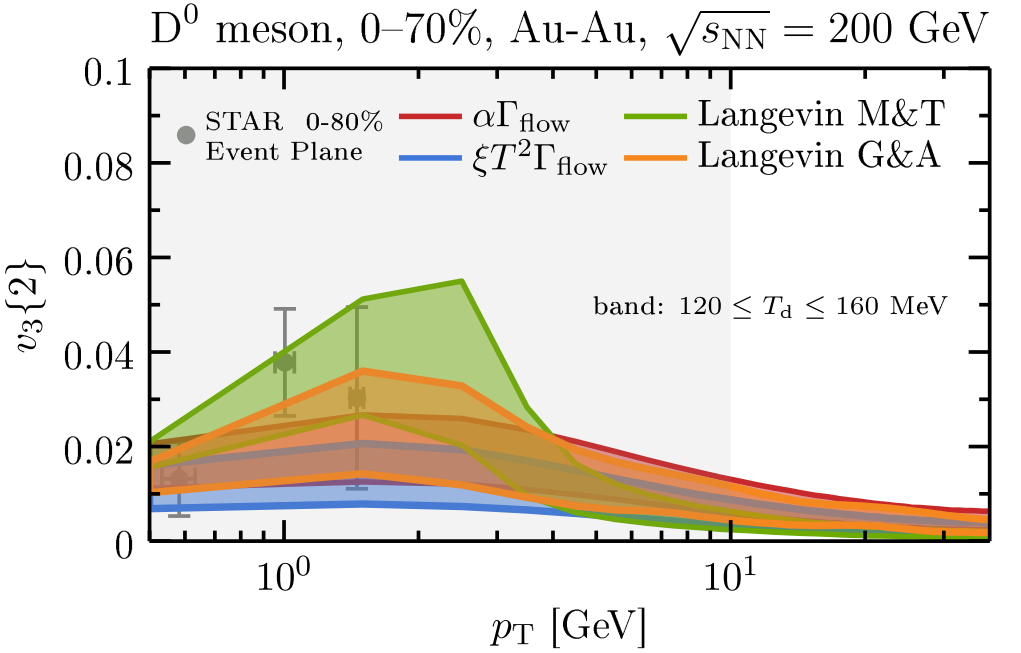}
    \includegraphics[width=0.47\textwidth]{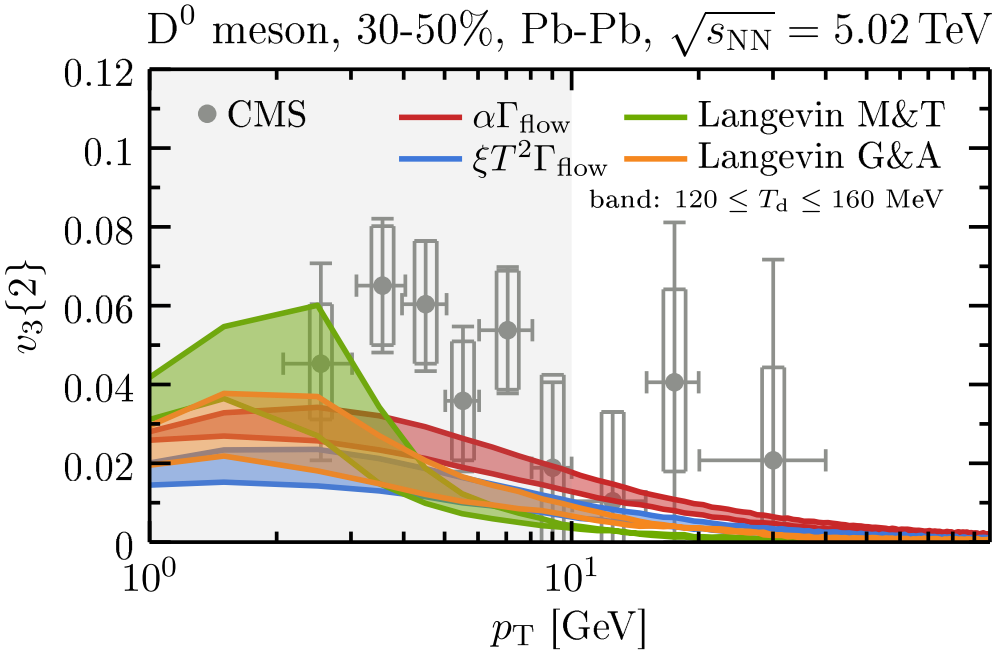}
    \caption{$\Dzero$ meson triangular flow coefficient $\vn3$ in the  0--70\% centrality range of  $\snnGeV[200]$ $\AuAu$ (left) and in the 30--50\% centrality range of $\snn[5.02]$ $\PbPb$ collisions (right). Experimental data from the \STAR\ ($|y|<1$)~\cite{refId0} and \cms\ ($|y|<1$)~\cite{Sirunyan:2017plt} collaborations, respectively.}
    \label{fig:v3_models}
\end{figure}

As done previously for the elliptic flow coefficient, we first test how the different parametrizations of Langevin and the selected energy loss models affect the triangular flow coefficient $\vn3$ using the correlation between two particles. From the plots of Fig.\ \ref{fig:v3_models} one can observe much larger bands when studying the same $\Td$ range in comparison with the results for $\vn2$. This wider band effect is consistent with a time hierarchy where $\vn2$ is built up first in the evolution of the system and $\vn3$ would be generated later. In this sense, a larger value of $\Td$ would hadronize the heavy quarks before they had enough time to build up a significant $\vn3$ flow.  Despite this difference, the general behavior is maintained among the different models with the constant energy loss model leading to higher values of $\vn3$ in the high $\pt$ regime, while the low $\pt$ region is dominated by the M\&T Langevin parametrization, leading to the best agreement with experimental data at $\snnGeV[200]$ despite the low number of data points. As with the $\vn2$ results, the models tend to underestimate the data at $\snn[5.02]$.

\begin{figure}[t]
    \centering
    \includegraphics[width=0.47\textwidth]{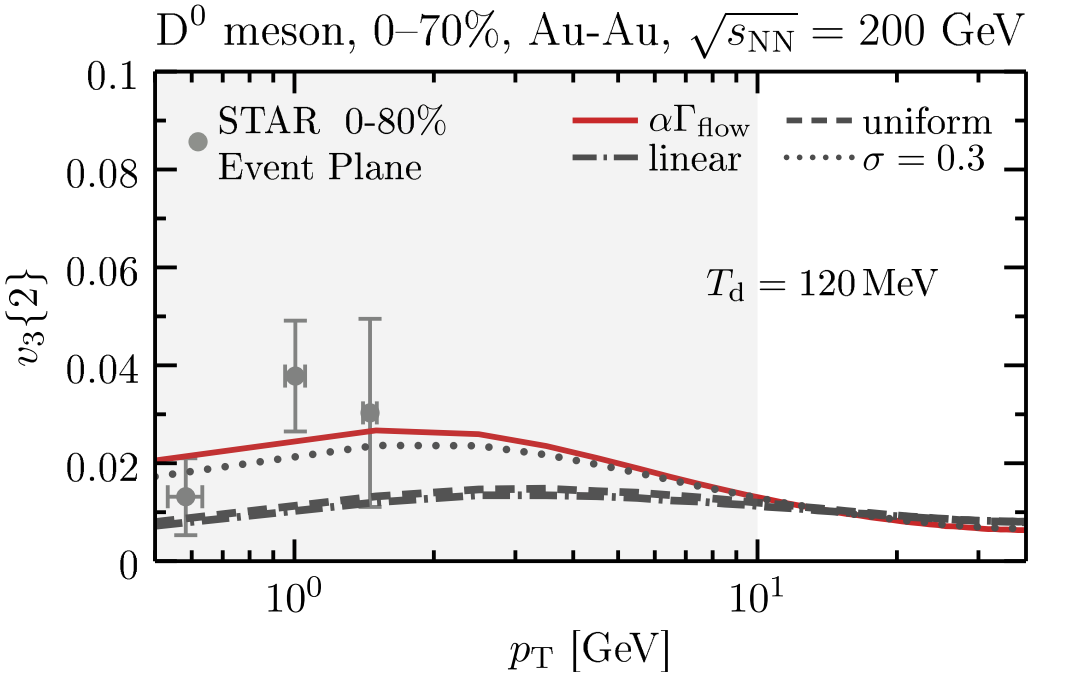}
    \includegraphics[width=0.46\textwidth]{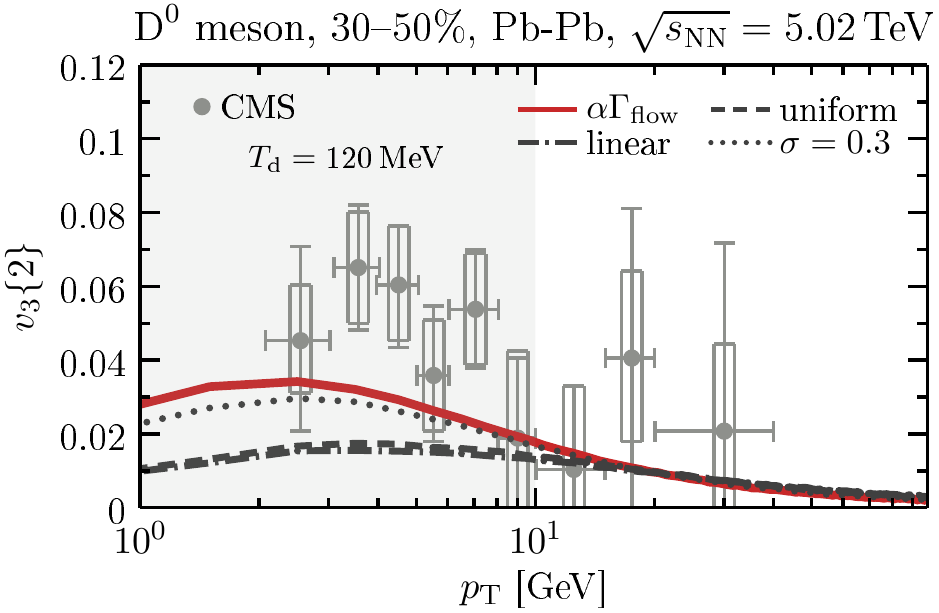}
    \caption{$\Dzero$ meson triangular flow coefficient $\vn3$ in the 0--70\% centrality range of $\snnGeV[200]$ $\AuAu$ collisions  (left) and in the 30--50\% centrality range of $\snn[5.02]$ $\PbPb$ collisions (right).  Results for the constant energy loss model with and without different types of energy loss fluctuations.  Experimental data from the \STAR\ ($|y|<1$)~\cite{refId0} and \cms\ ($|y|<1$)~\cite{Sirunyan:2017plt} collaborations, respectively.}
    \label{fig:v3_egylossfluc}
\end{figure}

Energy loss fluctuations are also studied and the corresponding results are shown in Fig.\  \ref{fig:v3_egylossfluc} in which we observe a similar trend to that found for $\vn2$.  While the Gaussian fluctuations slightly decrease the flow coefficient values in the lower $\pt$ regime, stronger fluctuations such as the linear and the constant ones seem to have a much bigger effect. In addition, even though their functional form differ greatly, the overall effect of energy loss fluctuations on $\vn3$ is not very different for the different cases considered. In the large $\pt$ regime the effect of the fluctuations seems to be negligible.

\begin{figure}[h]
    \centering
    \includegraphics[width=0.49\textwidth]{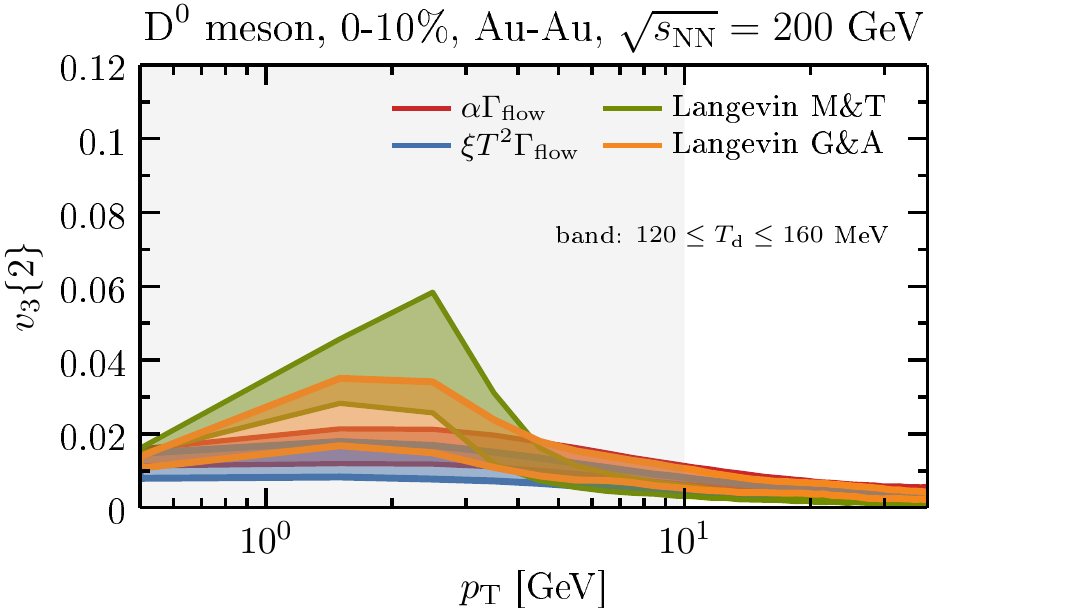}
    \includegraphics[width=0.49\textwidth]{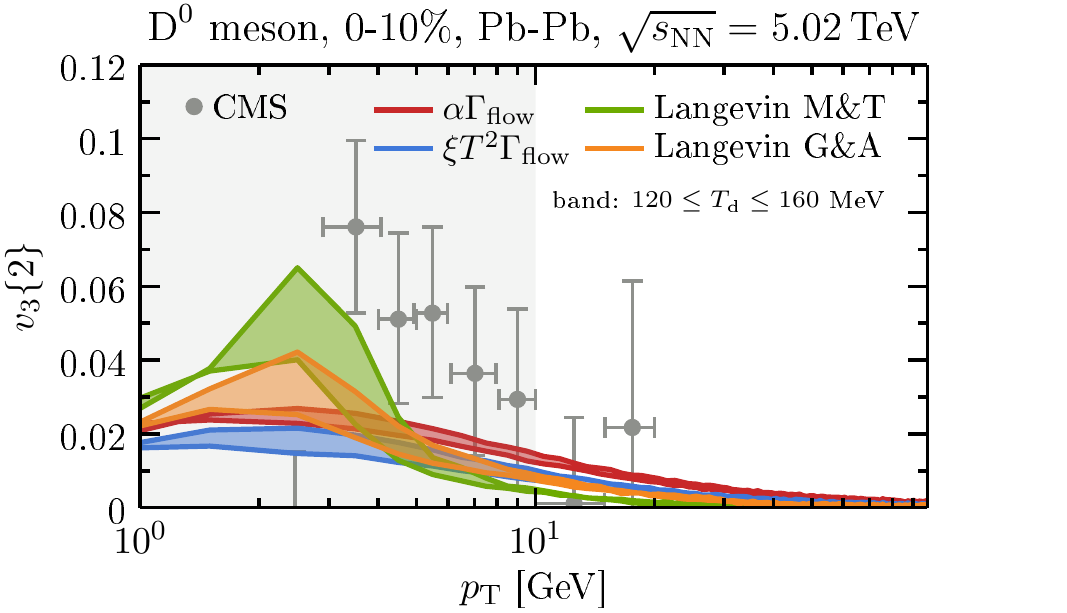}
    \caption{$\Dzero$ meson triangular flow coefficient $\vn3$ in the 0--10\% centrality class of $\snnGeV[200]$ $\AuAu$ collisions (left) and in $\snn[5.02]$ $\PbPb$ collisions (right). Experimental data from \cms\ ($|y|<1$) collaboration~\cite{Sirunyan:2017plt}.}
    \label{fig:v3_central}
\end{figure}

The corresponding results for $\vn3$ in central collisions are shown in Fig.\ \ref{fig:v3_central}. We notice the scale between the different models remain unchanged, in agreement with all the calculations for $\vn2$ and $\vn3$ shown before.  The simulation results still underestimate the experimental data for the largest beam energy collisions in agreement with the results observed in Fig.\ \ref{fig:v3_models}. One feature that is specific for these $\vn3$ calculations is that there seems to be very little centrality dependence for this observable, differing from the $\vn2$ observations. In that respect, all models seem to agree with the observed conclusion from data.

\begin{figure}[h]
    \centering
    \includegraphics[width=0.50\textwidth]{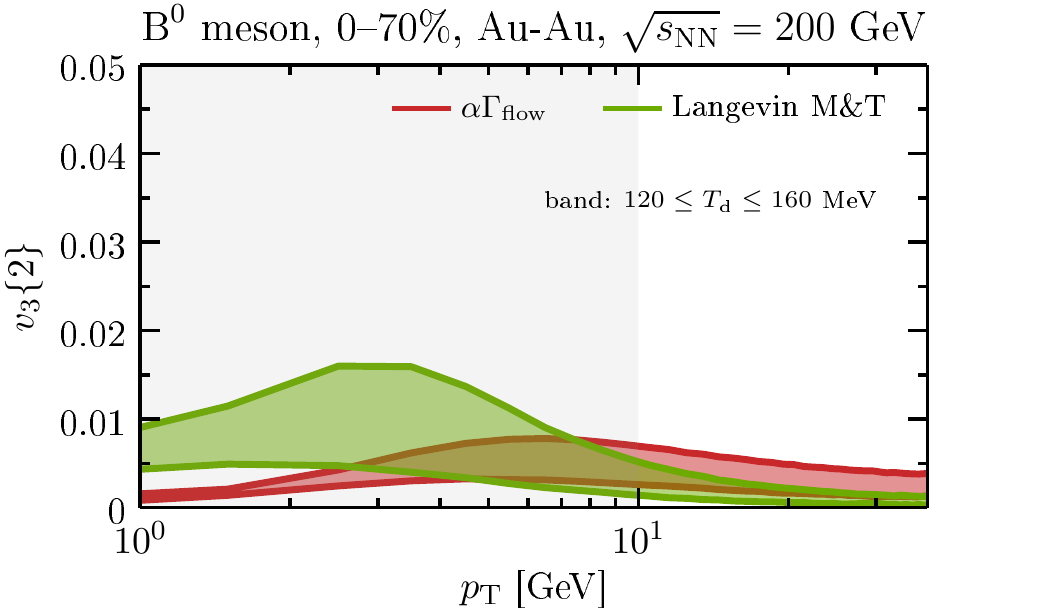}
    \caption{$\Bzero$ meson triangular flow coefficient $\vn3$ in the 0--70\% centrality range of $\snnGeV[200]$ $\AuAu$ collisions.}
    \label{fig:v3_Bmeson}
\end{figure}

In Fig.\ \ref{fig:v3_Bmeson} we show a prediction for the triangular flow of $\Bzero$ mesons using the constant energy loss model and the M\&T Langevin parametrization. We observe a much lower $\vn3$ in comparison with the results for the $\Dmeson$ meson, which also occurs for $\vn2$, suggesting that the mass hierarchy may propagate throughout the higher flow harmonics.

\subsubsection{4th order flow coefficient from two-particle cumulants}

\begin{figure}[!htb]
  \centering
  \includegraphics[width=0.49\textwidth]{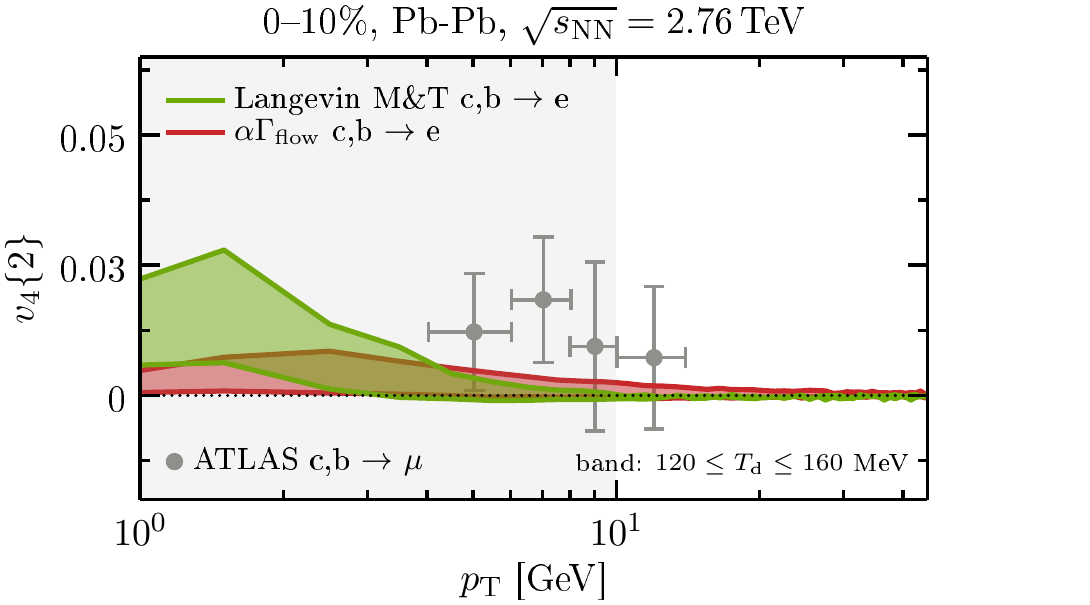}
  \includegraphics[width=0.49\textwidth]{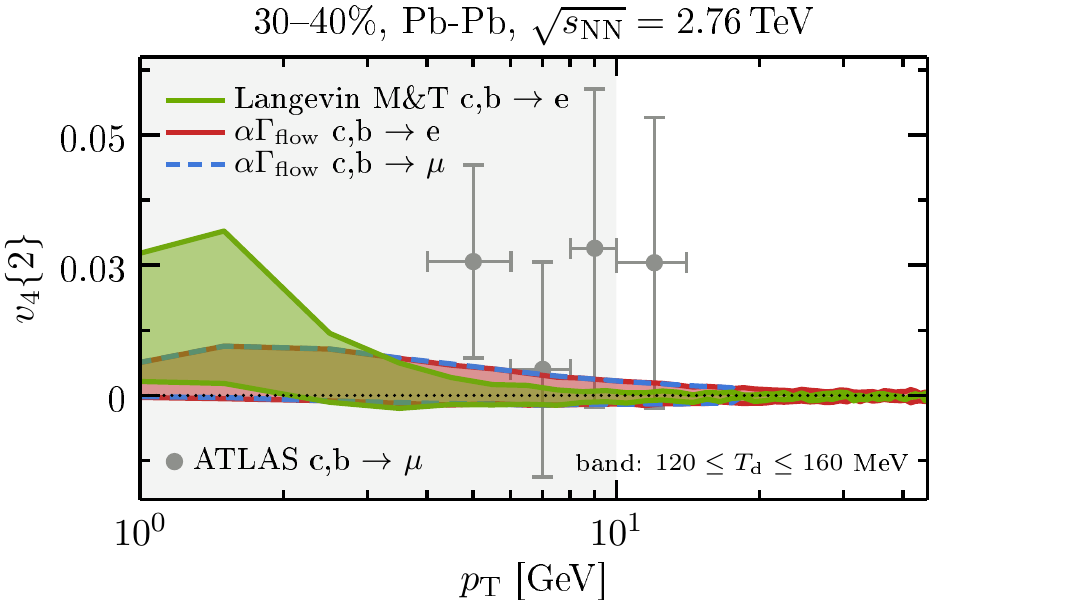}
  \caption{Heavy flavor electron (muon) flow coefficient $\vn4$ in the 0--10\% (left) and  30--40\% (right) centrality classes of $\snn[2.76]$ $\PbPb$ collisions. The gray area indicates the $\pt$ region where coalescence may be important. Experimental data from the ATLAS ($|y|<2$) collaboration~\cite{Aaboud:2018bdg}.}
  \label{fig:v4_e_LHC13040}
\end{figure}

In Fig.\ \ref{fig:v4_e_LHC13040} we compare the results for the 4th order flow coefficient from 2-particle correlations, $v_4\{2\}$, involving heavy flavor electrons and muons in the 0--10\% (left) and $30-40\%$ (right) centrality classes of PbPb 2.76 TeV collisions, obtained using the constant energy loss and the M\&T Langevin model. As already observed for the $\raa$ in Fig.\ \ref{fig:Raa_emuon3040}, one can first note that the electron and muon channels give equivalent results down to 1 GeV, so that comparing our electron results with muon data at mid-rapidities is fine. The $v_4$ results behave similarly to $v_2$ and $v_3$: they all decrease with increasing $\pt$ for $\pt\gtrsim 2$ GeV and, compared to the energy loss models, the Langevin framework leads to higher (smaller) values at low (high) $\pt$. Both models lie within the experimental data uncertainties although, as for the other flow coefficients, the $v_4$ results seem to slightly underestimate the data. We observe almost no variation of $v_4$ with centrality, as we did for $v_3$ in the previous section, which shows that $v_4$ and $v_3$ stem from geometrical fluctuations.

\begin{figure}[!htb]
  \centering
  \includegraphics[width=0.51\textwidth]{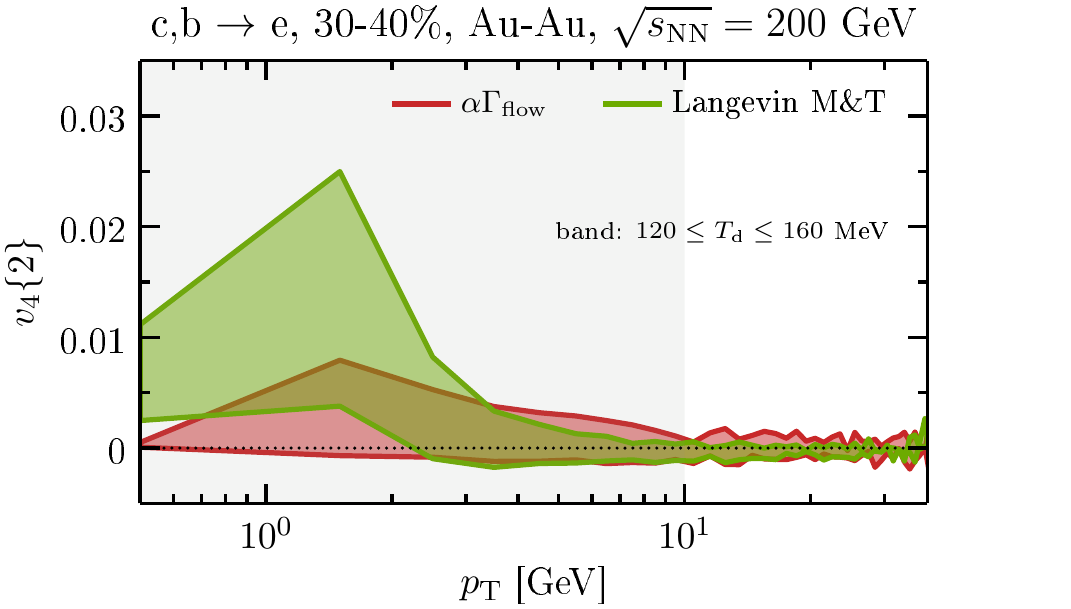}
 \includegraphics[width=0.48\textwidth]{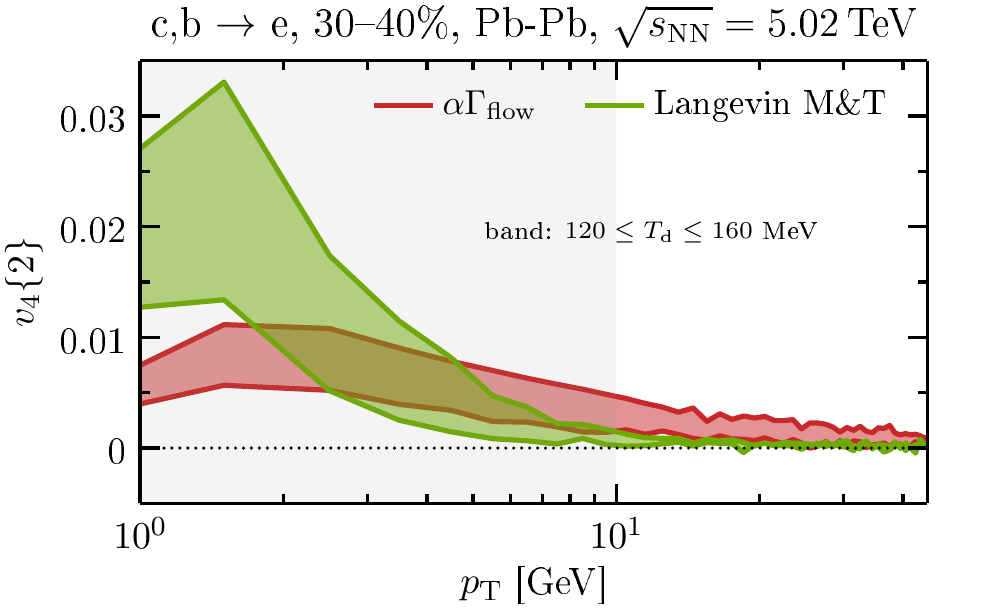}
  \caption{Heavy flavor electron flow coefficient $\vn4$ in the  30--40\% centrality class of $\snnGeV[200]$ $\AuAu$ collisions (left) and $\snn[5.02]$ $\PbPb$ collisions (right). }
  \label{fig:v4_e_RHICLHC23040}
\end{figure}

\begin{figure}[!htb]
  \centering
  \includegraphics[width=0.48\textwidth]{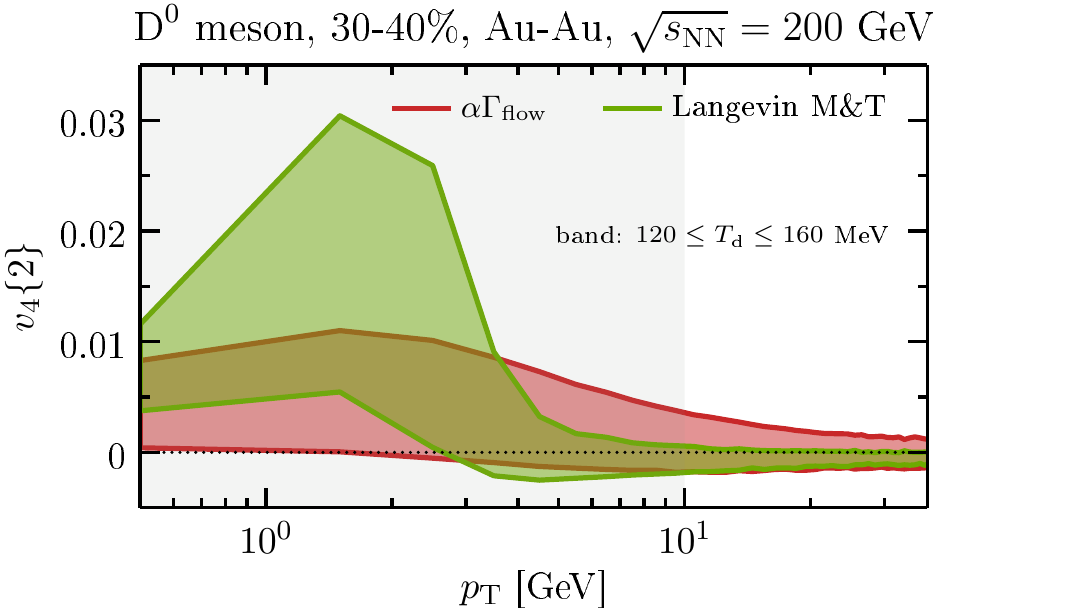}
 \includegraphics[width=0.48\textwidth]{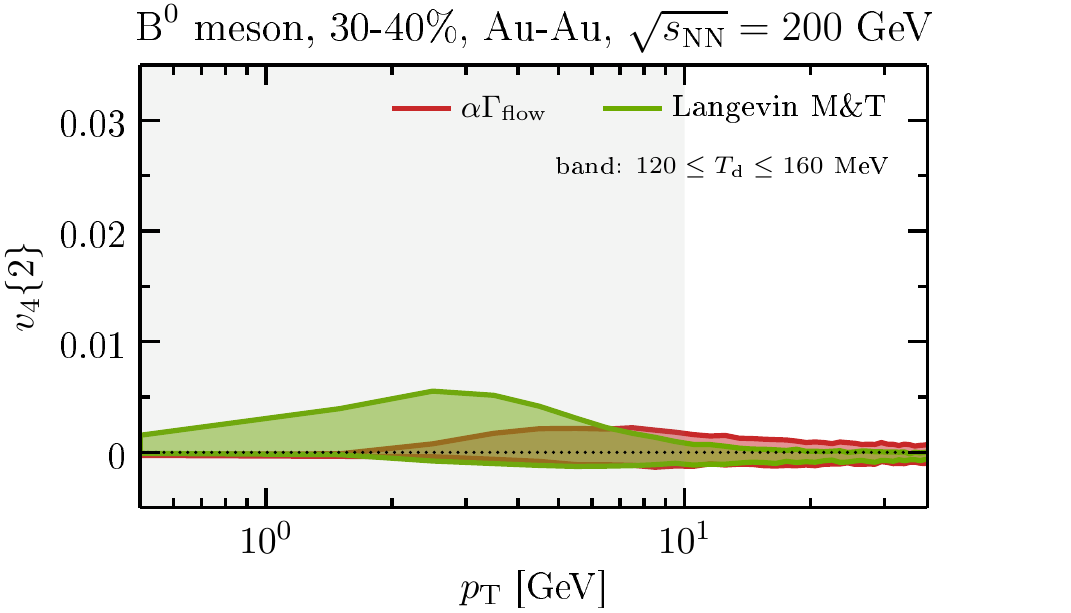}
  \caption{  $\Dzero$ (left) and $\Bzero$ (right) meson flow coefficient $\vn4$ in the 30--40\% centrality class of  $\snnGeV[200]$ $\AuAu$ collisions. }
  \label{fig:v4_DB_RHIC3040}
\end{figure}

\begin{figure}[!htb]
  \centering
  \includegraphics[width=0.48\textwidth]{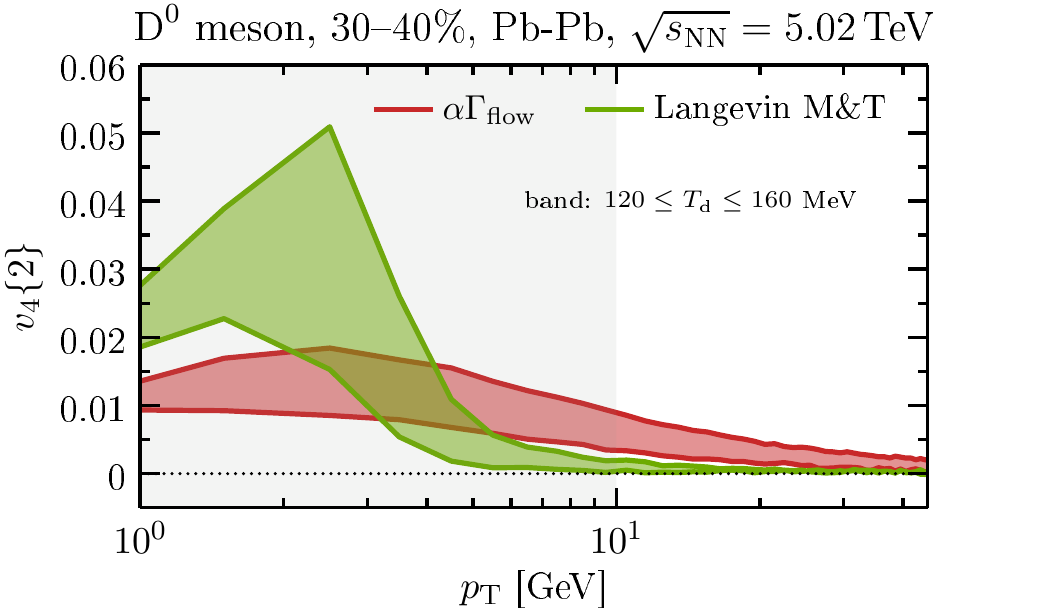}
 \includegraphics[width=0.48\textwidth]{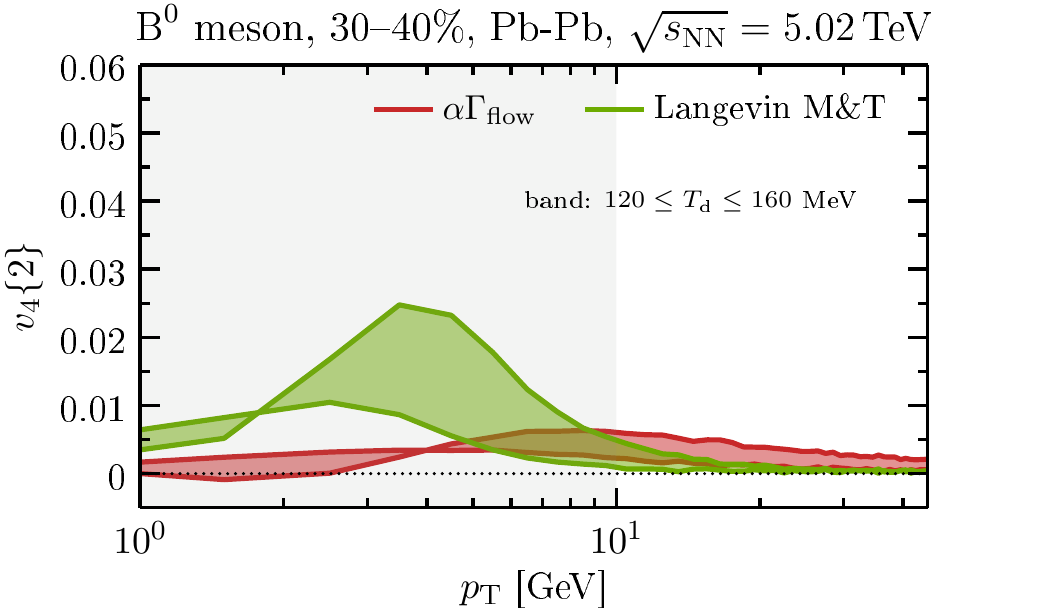}
  \caption{  $\Dzero$ (left) and $\Bzero$ (right) meson flow coefficient $\vn4$ in the 30--40\% centrality class of $\snn[5.02]$ $\PbPb$ collisions. }
  \label{fig:v4_DB_LHC23040}
\end{figure}

Despite the similarities with the other coefficients $v_4$ exhibits a new feature in $\snn[2.76]$ $\PbPb$ collisions: for the decoupling temperature $\Td=160$ MeV the heavy flavor electron/muon $v_4$ is negative on the whole $\pt$ range for both transport models. The corresponding experimental data is either positive or negative but it is mostly compatible with zero within the error bands~\cite{Aaboud:2018bdg}. As shown in Fig.\  \ref{fig:v4_e_RHICLHC23040}, this is also the case in 200 GeV $\AuAu$ collisions but not in $\snn[5.02]$ $\PbPb$ collisions. Additionally, as shown in Fig.\ \ref{fig:v4_DB_RHIC3040} for 200 GeV $\AuAu$ collisions (note the different situation in Fig.\ \ref{fig:v4_DB_LHC23040} for 5.02 TeV $\PbPb$ collisions), both the $\Dzero$ and $\Bzero$ meson $v_4$ are negative when $\Td=160$ MeV with the $\Dzero$ meson $v_4$ being larger in absolute values. 

\begin{figure}[!htb]
  \centering
  \includegraphics[width=0.45\textwidth]{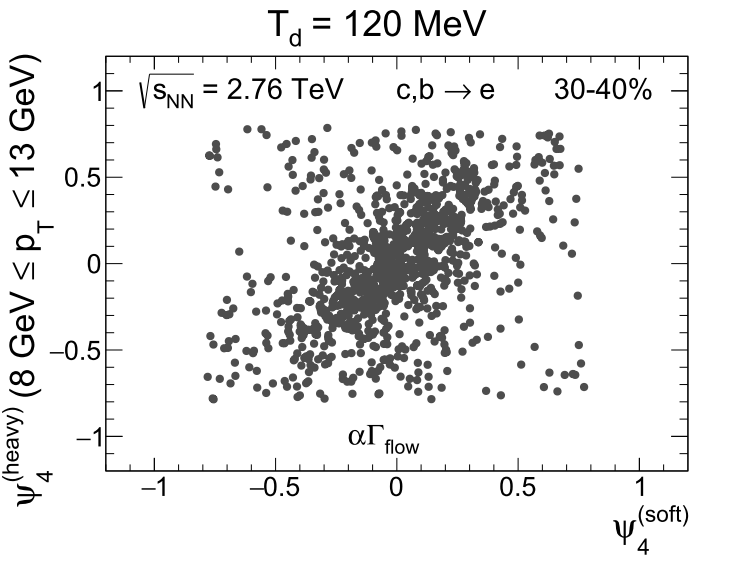}
  \includegraphics[width=0.45\textwidth]{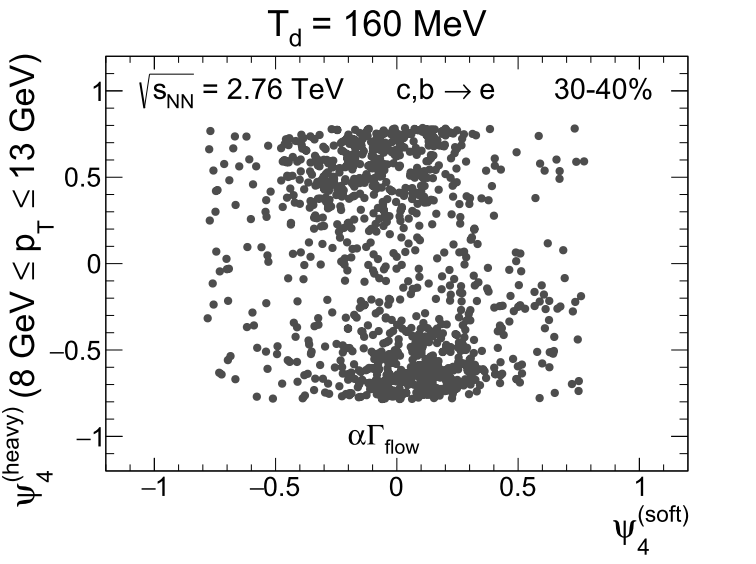}
  \caption{Event-by-event heavy flavour electron event plane angles $\psin4^\text{(heavy)}$ obtained with the constant energy loss model vs. all charged particles event plane angles $\psin4^\text{(soft)}$.}
  \label{fig:psi4_e_LHC13040}
\end{figure}

The reason for these negative values is an anti-correlation between the heavy flavor and the bulk azimuthal anisotropy angles when $\Td=160$ MeV, $\psi _4^{\rm (heavy)}$ and $\psi _4^{\rm (soft)}$, respectively. Indeed, as shown in Fig.\ \ref{fig:psi4_e_LHC13040} the event-by-event $\psi _4^{\rm (heavy)}(\psi _4^{\rm (soft)})$ distributions are anti-correlated when $\Td=160$ MeV (centered on $\psi _4^{\rm (heavy)}=\pm \pi/2$ when $\psi _4^{\rm (soft)}=0$) whereas they are correlated when $\Td=120$ MeV (centered on $\psi _4^{\rm (heavy)}=0$ with a linear correlation to $\psi _4^{\rm (soft)}$). For $\Td=160$ MeV in 200 GeV $\AuAu$ and 2.76 TeV $\PbPb$ collisions the heavy quark interaction with the bulk medium is therefore not long enough for the heavy quark flow vector $V_4^{\rm (heavy)}$ to be in phase with the bulk $V_4^{\rm (soft)}$. This phenomenon was also observed in light quark models for $n>3$~\cite{Jia:2012ez}. The $\psi _4$ observable will be further explored in the next section.
Finally, by comparing the different flow harmonics we see a clear hierarchy $v_2 > v_3 > v_4$ for any collision energy as was also observed by the experiments~\cite{Aaboud:2018bdg}.

\subsubsection{Correlation between the flow of heavy mesons and all charged particles}

In addition to looking at the correlations between the heavy and the soft sectors from the point of view of the cumulants, it is also interesting to study the direct correlation between these sectors on an event-by-event basis using event-shape engineering techniques.  Since the soft sector flow coefficients are directly related to the event eccentricities, this study can provide information on the role of the initial anisotropy on final observables for the heavy flavor sector.  This work uses the same approach that has been previously used in~\cite{Noronha-Hostler:2016eow} when investigating soft and hard sector correlations.  We define the correlations by binning the distribution of integrated flow harmonics in the soft sector, $\vnn^\text{(soft)}$, and then evaluating the corresponding flow coefficient of $\Bzero$ and $\Dzero$ mesons for each bin, $\vnn^\text{(heavy)}$.  The result can be understood as the probability that an event within a given soft sector flow class will correspond to a particular value of $\vnn^\text{(heavy)}$.  One can study the slope of these correlations with respect to different collision conditions leading to $\vnn$ fluctuations.  In that respect, if no $\vnn$ fluctuations were to be observed, the plot would show a flat horizontal line.

\begin{figure}[h]
  \centering
  \includegraphics[width=0.36\textwidth]{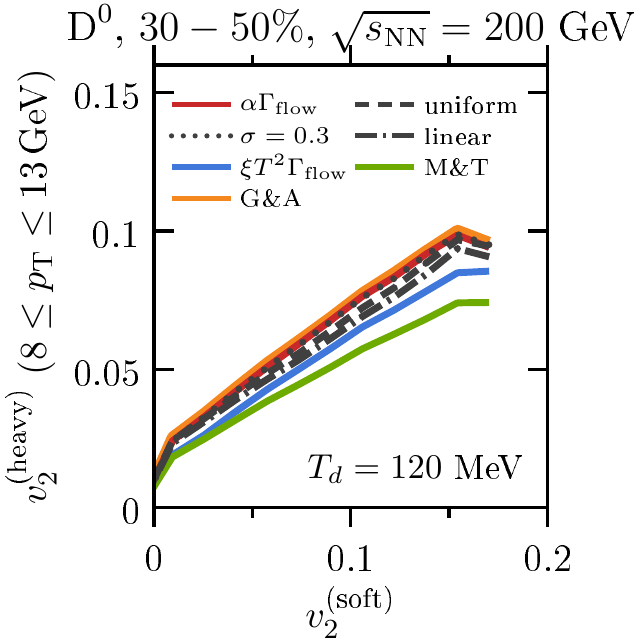}
  \includegraphics[width=0.35\textwidth]{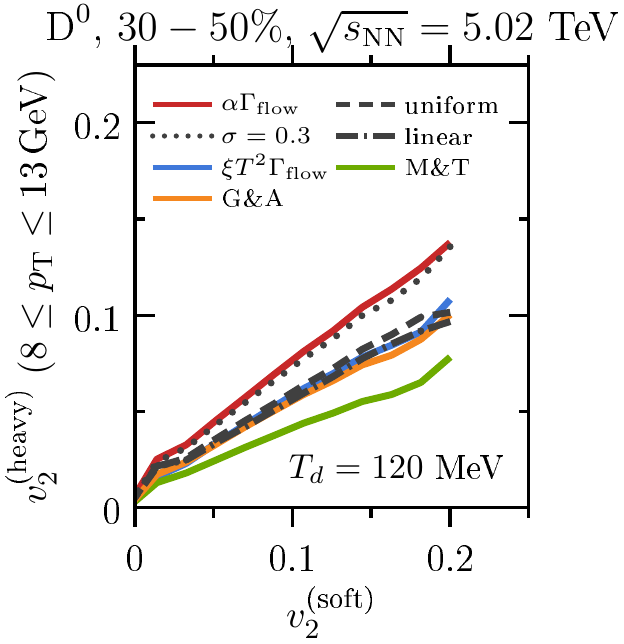}
  \caption{Correlations between the elliptic flow of $\Dzero$ mesons and the elliptic flow of all charged particles in $\snnGeV[200]$ $\AuAu$ collisions (left) and $\snn[5.02]$ $\PbPb$ collisions (right), computed using the different transport models.}
  \label{fig:corr_v2_RHICxLHC}
\end{figure}

Fig.\ \ref{fig:corr_v2_RHICxLHC} shows the correlation between the $\Dzero$ meson $\vn2$ and the corresponding quantity for all charged particles computed using different transport models and different beam energies.  We first notice that, regardless of the transport model used for the calculation, all results indicate that the correlation between the heavy and soft sectors is linear, although the slope of the lines might be different, reflecting the different $\vnn$ already observed in previous sections for a given $\pt$ range.  It is worth noticing that the hierarchy between the different transport models is not maintained for different energies, which may be due to the specific implementation details of each model.

\begin{figure}[h]
  \centering
  \includegraphics[width=0.35\textwidth]{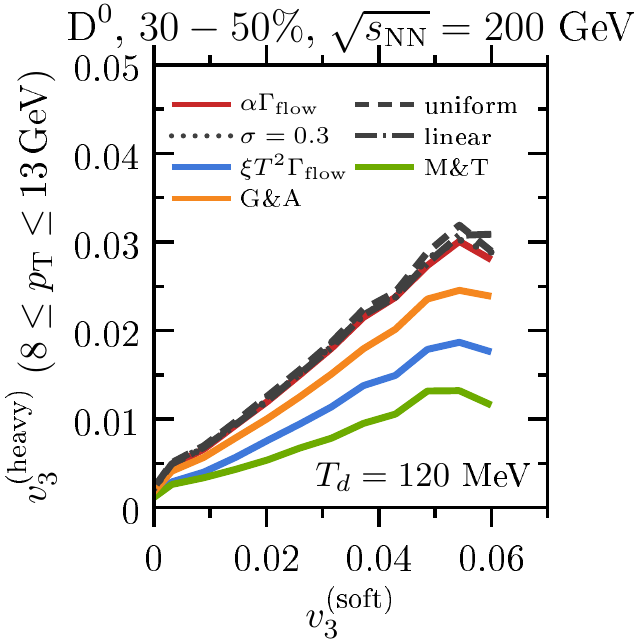}
  \includegraphics[width=0.39\textwidth]{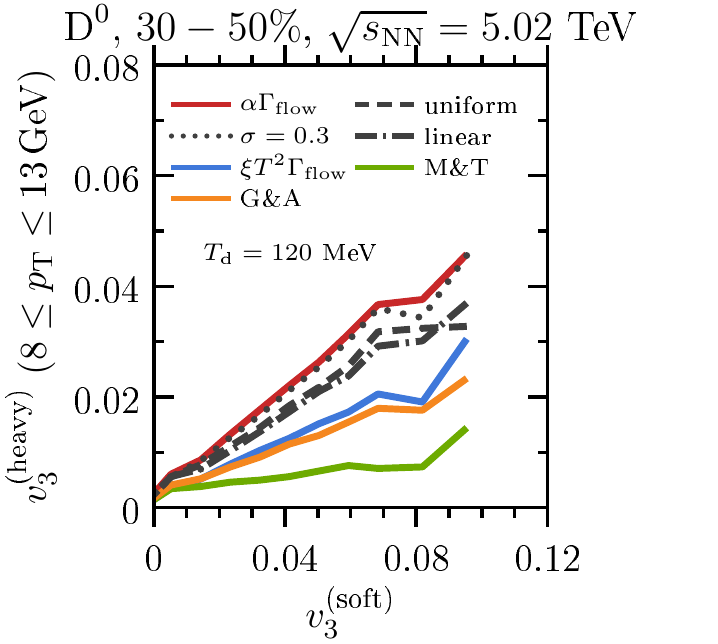}
  \caption{Correlations between triangular flow of $\Dzero$ mesons and that of all charged particles  in $\snnGeV[200]$ $\AuAu$ collisions (left) and in $\snn[5.02]$ $\PbPb$ collisions (right), computed using different transport models.}
  \label{fig:corr_v3_RHICxLHC}
\end{figure}

\begin{figure}[h]
  \centering
  \includegraphics[width=0.38\textwidth]{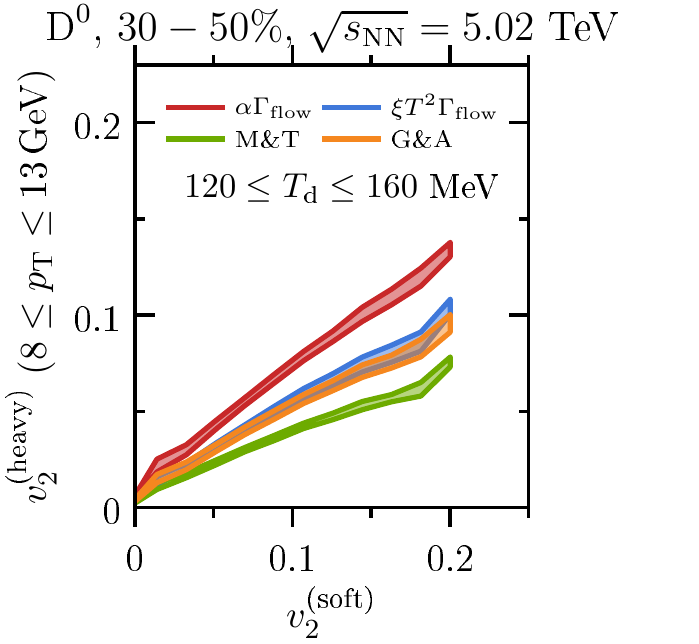}
  \includegraphics[width=0.39\textwidth]{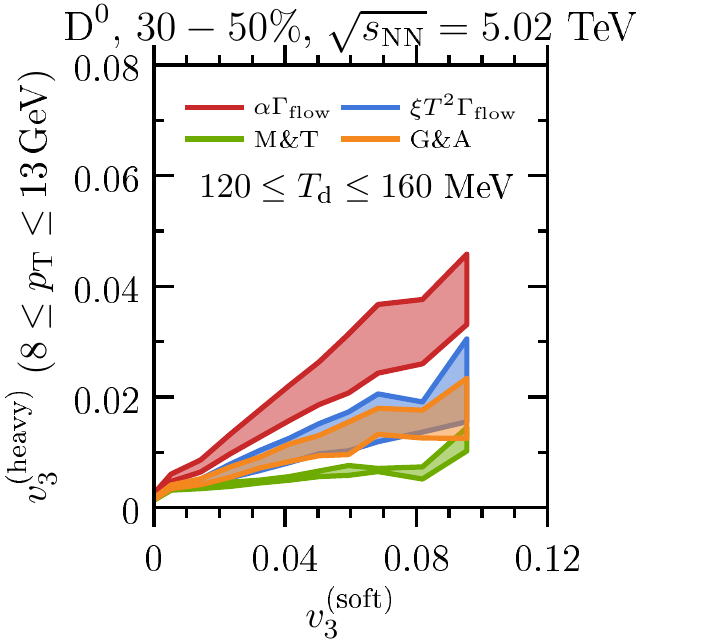}
  \caption{Decoupling temperature dependence of the correlations between $\Dzero$ mesons and all charged particles in the 30--40\% centrality class of $\snn[5.02]$ $\PbPb$ collisions. Elliptic flow is shown on the left and  triangular flow on the right.}
  \label{fig:corr_v2xv3_LHC2}
\end{figure}

The same approach can be used to investigate the same type of correlations involving higher order flow coefficients to check  if the linear correlation observed for $\vn2$ is still maintained at higher orders. In Fig.\ \ref{fig:corr_v3_RHICxLHC} it is possible to observe the same effect, although a deviation occurs when the soft $\vn3$ becomes large.  These deviations are related to the limited statistics of the event leading to a wider distribution of the heavy flavor sector $\vn3^\text{(heavy)}$.  Furthermore, we note that the hierarchy among the values obtained for the different transport models is similar to that found in the case of elliptic flow, except for the constant energy loss model at $200$~GeV.

\begin{figure}[h]
  \centering
  \includegraphics[width=0.39\textwidth]{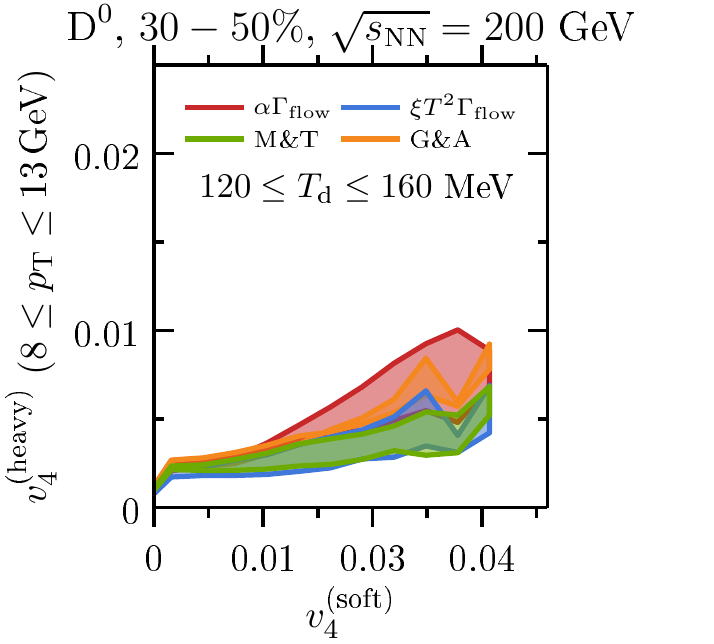}
  \includegraphics[width=0.39\textwidth]{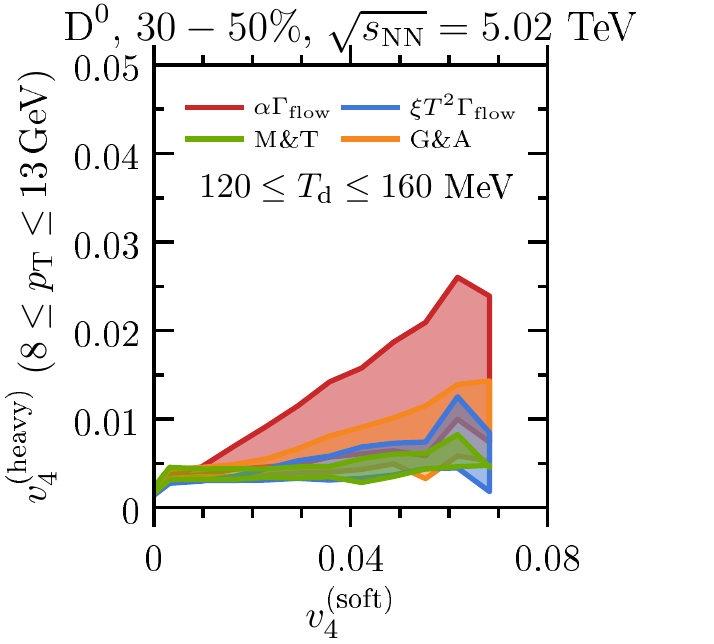}
  \caption{Correlations between the $\vn4$ of $\Dzero$ mesons and the $\vn4$ of all charged particles  in $\snnGeV[200]$ $\AuAu$ collisions (left) and $\snn[5.02]$ $\PbPb$ collisions (right), computed using different transport models.}
  \label{fig:corr_v4_RHICxLHC}
\end{figure}

One of the main questions involving heavy flavor quarks in the \qgp\ concerns their coupling with the expanding medium. In our framework this is also encoded in the decoupling parameter $\Td$, which defines a temperature scale below which heavy quarks are considered to not be coupled with the medium anymore.  This parameter affects the path length experienced by the heavy quark and can therefore greatly affect the results obtained for the flow coefficients.  The lower the decoupling temperature, the longer the heavy quarks are under the influence of the medium in the transport models considered in this paper.  In Fig.\ \ref{fig:corr_v2xv3_LHC2}, a range of decoupling temperatures is studied for every transport model considering the flow of $\Dzero$ mesons.  We notice that a variation of the decoupling temperature affects $\vn3$ more than it affects $\vn2$. Since $\vn2$ is built up quickly, the difference in the slopes due to the decoupling temperature is very small, especially at very large collision energies which require more time to cool down below $\Td$.  On the other hand, higher order harmonics and lower collision energies are more strongly affected by this parameter. In Fig.\ \ref{fig:corr_v4_RHICxLHC}, the correlation between the $\vn4$ of $\Dzero$ mesons and that of all charged particles is shown for the two collision energies considered.  We obtain even wider bands compared with the $\vn3$ correlation results, further indicating the hierarchy described before.  These results also agree with previous observations regarding the hierarchy of flow harmonics determined using the different models.

\begin{figure}[h]
  \centering
  \includegraphics[width=0.38\textwidth]{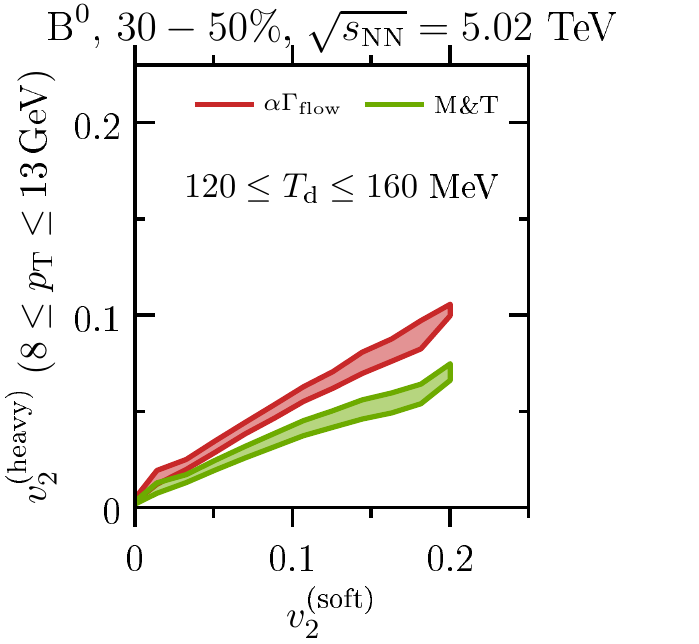}
  \includegraphics[width=0.39\textwidth]{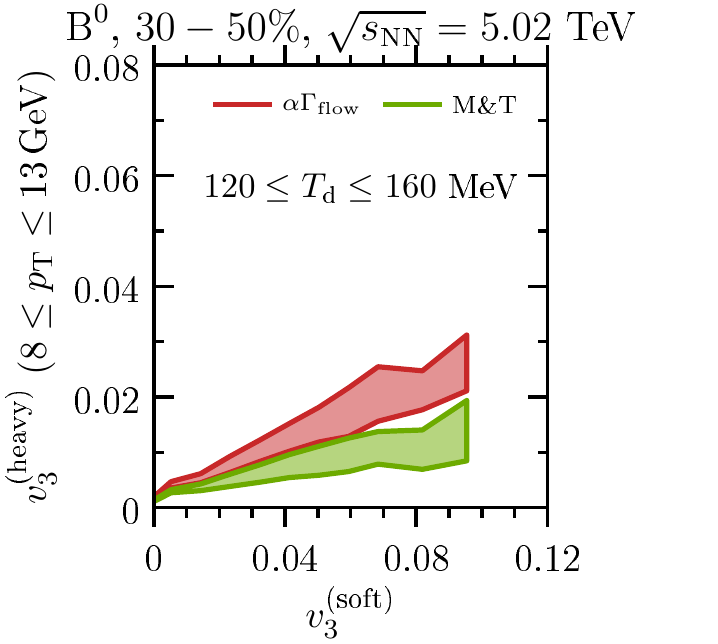}
  \caption{Decoupling temperature dependence for correlations between $\Bzero$ mesons and all charged particles in $\snn[5.02]$ $\PbPb$ collisions. Elliptic flow is shown on the left and triangular flow on the right.}
  \label{fig:corr_v2_v3_Bmeson}
\end{figure}

All the analysis performed on $\Dzero$ mesons can also be performed on $\Bzero$ mesons and the results are very similar.  The plots in Fig.\ \ref{fig:corr_v2_v3_Bmeson} show results for $\Bzero$ mesons at different decoupling temperatures and for different models.  In comparison with previous results for $\Dzero$ mesons we notice that the difference between transport models for $\Bzero$ mesons is less pronounced, though the decoupling temperature seems to play a bigger role in this case.  Since this observable is integrated over $\pt$, both of these effects may be due to the mass difference between the mesons and a direct comparison in the same $\pt$ range may be misleading.\\

Not only the flow harmonics correlations are worth looking into.  One can also study the correlations between their respective event plane angles $\psinn$.  This quantity gives information about the alignment between  the event plane angles of the soft and the heavy sector.  It is convenient to represent this correlation as a cosine term such as the ones that appear on the equations for the cumulants  so they can be easily related to the $\pt$-differential results introduced earlier.

\begin{figure}[h]
  \centering
  \includegraphics[width=0.45\textwidth]{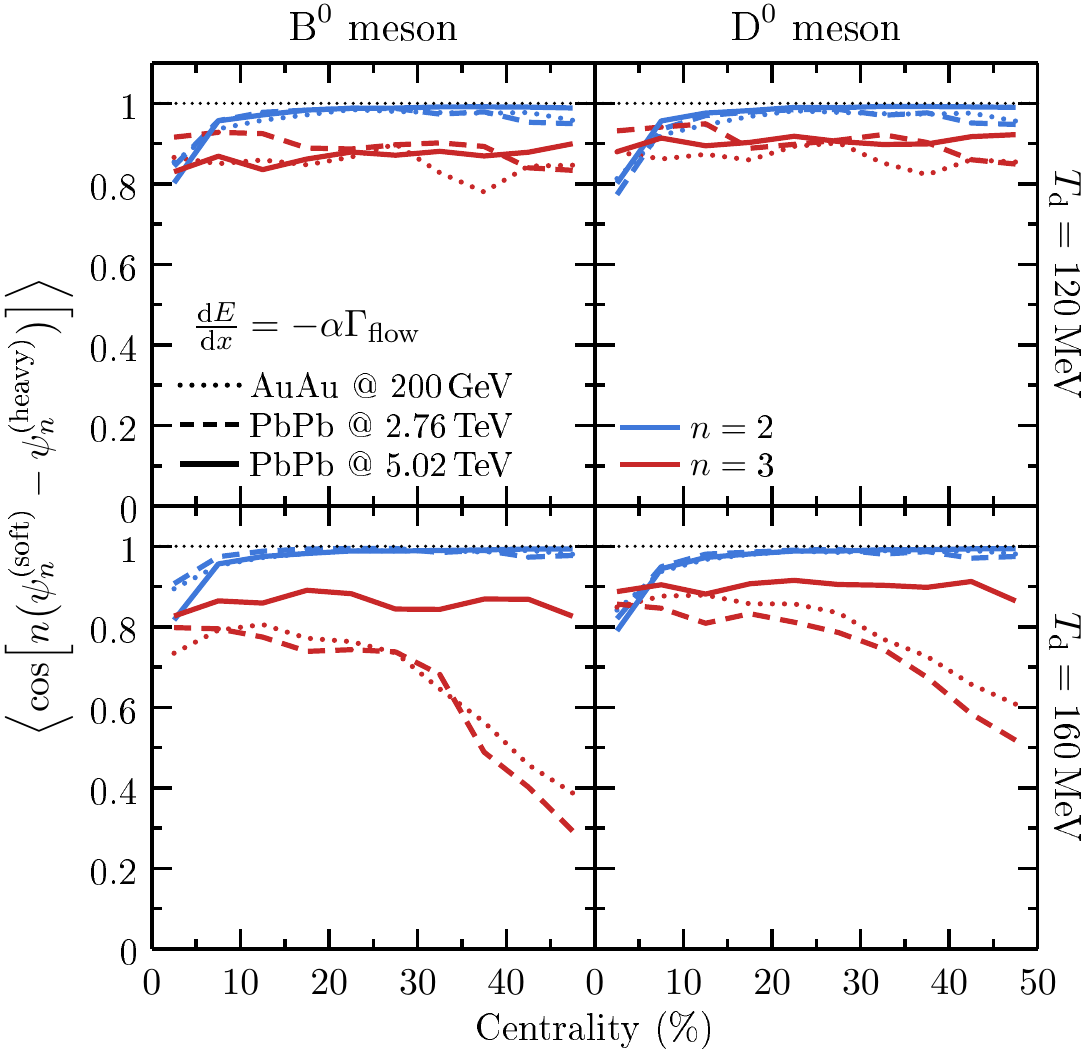}
  \caption{Correlation of the event plane angles $\psinn$ between the heavy and soft sectors obtained with the constant energy loss model for $\Bzero$ (left) and $\Dzero$ mesons (right) as a function of  centrality. Results for different collision energies are compared for decoupling temperatures $\Td = \SI{120}{MeV}$ (top) and $\Td = \SI{160}{MeV}$ (bottom) and for $n=2$ and $n=3$.}
  \label{fig:cosPsiConst}
\end{figure}

\begin{figure}[h]
  \centering
  \includegraphics[width=0.45\textwidth]{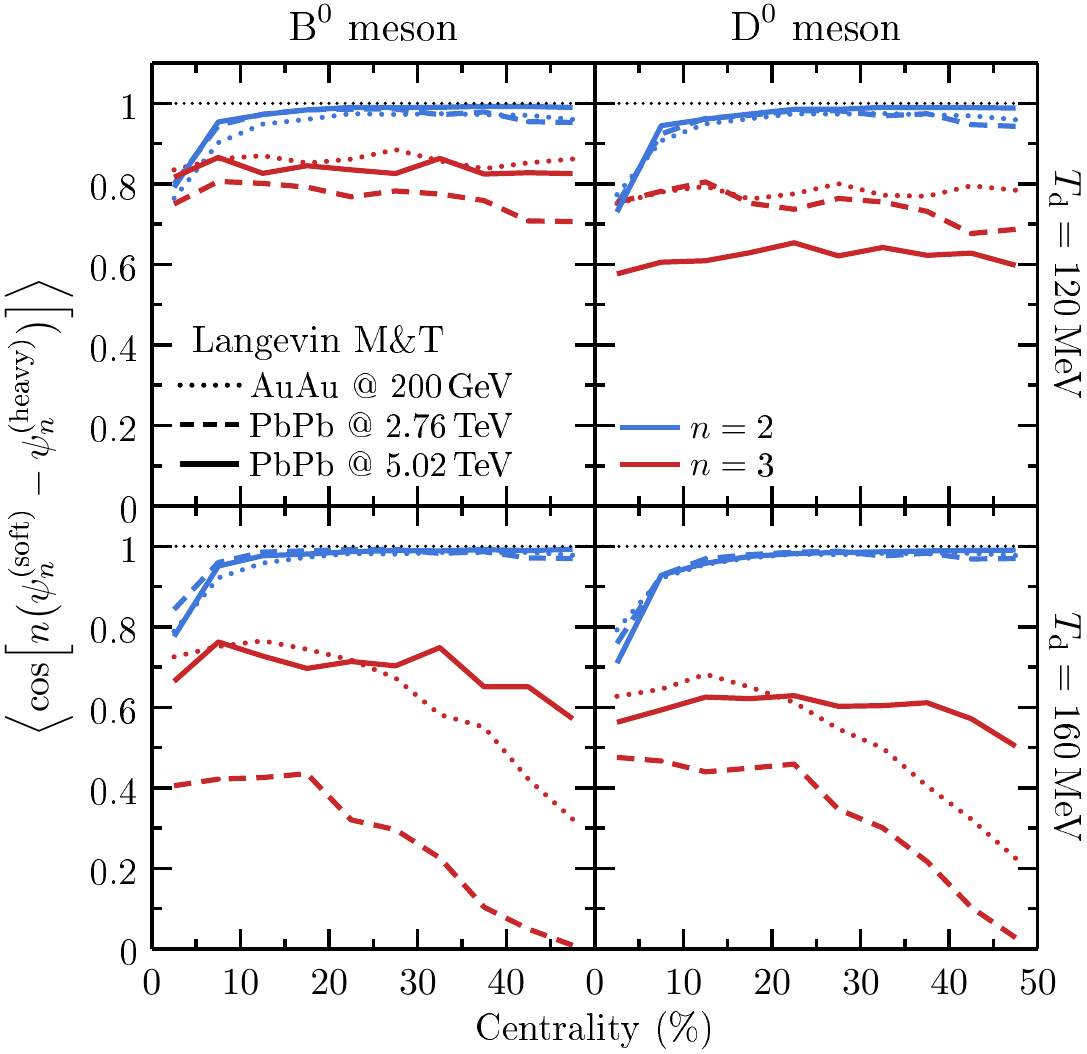}
  \caption{Correlation of the event plane angles $\psinn$ between the heavy and soft sectors obtained with the Langevin dynamics (Moore and Teaney parametrization) for $\Bzero$ (left) and $\Dzero$ mesons (right) as a function of centrality. Results for different collision energies are compared for decoupling temperatures $\Td = \SI{120}{MeV}$ (top) and $\Td = \SI{160}{MeV}$ (bottom) and for $n=2$ and $n=3$.}
  \label{fig:cosPsiMT}
\end{figure}

We first introduce a comparison for different collision energies of the event plane angles correlations in Fig.\ \ref{fig:cosPsiConst} in which both $\Bzero$ and $\Dzero$ mesons are studied using different values of the parameter $\Td$.  The second order event plane angle is consistently aligned with $\big\langle\cos\big[n\big(\psin2^\text{(soft)} - \psin2^\text{(heavy)}\big)\big]\big\rangle > \num{0.95}$ across most of the centrality range.  This is observed for all the different settings using the constant energy loss model.  The triangular event plane correlations show a different picture, though.  While the alignment is maintained for the low decoupling temperature $\Td = \SI{120}{MeV}$, deviations for non-central collisions are found if this temperature is increased to $\Td = \SI{160}{MeV}$.  This effect is not observed for the highest collision energy, but it does not seem to depend on the collision energy otherwise.  Furthermore, this result is consistent with the large suppression of $\vn3$ observed previously for $\Td = \SI{160}{MeV}$.

Turning to the M\&T Langevin parametrization, the same analysis for the event plane angles is shown in Fig.\ \ref{fig:cosPsiMT}.  The elliptic flow event plane angles are shown to be heavily aligned with that of charged particles, in the same manner as the previous results for the constant energy loss model.  In this case, however, the result for the triangular flow event plane angles is lower even for $\Td = \SI{120}{MeV}$.  For the larger decoupling temperature the correlation behaves similarly to the results for the constant energy loss in which the angles are less correlated for non-central collisions.  The collision energy dependence is different in this case  and the curves do not overlap in the same way we observe in Fig.\ \ref{fig:cosPsiConst}.  Looking back at Figs.\ \ref{fig:corr_v3_RHICxLHC}, \ref{fig:corr_v2xv3_LHC2}, and~\ref{fig:corr_v2_v3_Bmeson} we notice that this parametrization leads to the lowest values of $\vn3$ and the low event plane angle correlation is therefore consistent with that observation.

\begin{figure}[h]
  \centering
  \includegraphics[width=0.45\textwidth]{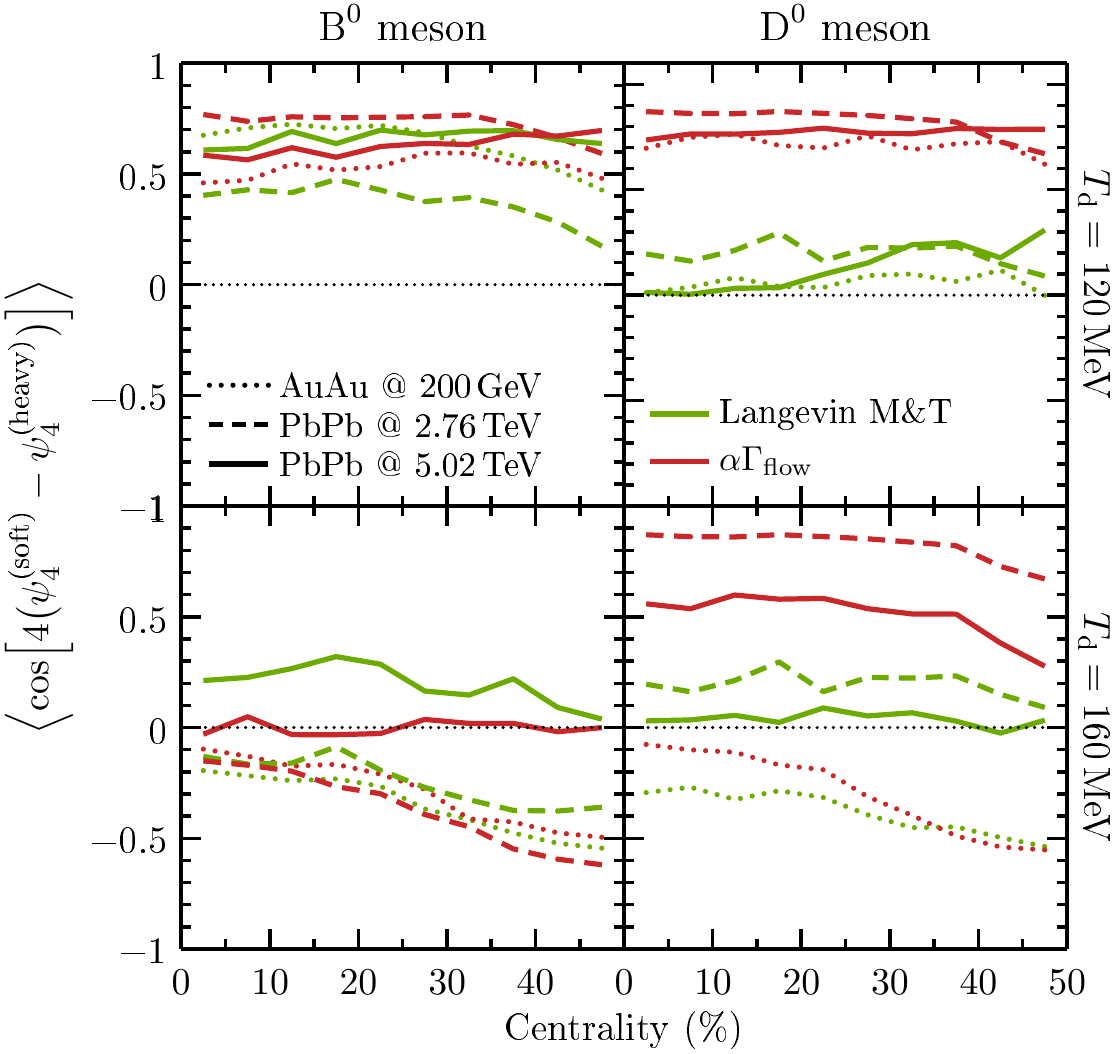}
  \caption{Correlation of the event plane angles $\psin4$ between the heavy and soft sectors obtained with the constant energy loss model and Langevin dynamics (using the Moore and Teaney parametrization) for $\Bzero$ (left) and $\Dzero$ mesons (right) as a function of centrality. Results for different collision energies are compared for decoupling temperatures $\Td = \SI{120}{MeV}$ (top) and $\Td = \SI{160}{MeV}$ (bottom).}
  \label{fig:psi4}
\end{figure}

It is interesting to check if the decorrelation keeps increasing when considering event plane angles of higher order flow harmonics.  The plots in Fig.\ \ref{fig:psi4} summarize the results for the fourth order flow harmonics for both transport models studied.  It is noticeable how these event plane angles are shown to be much less correlated with that of charged particles in comparison to the results found using lower order event plane angles shown before. In addition, the same general trend is maintained for $\Dzero$ mesons: the results are less correlated for the highest decoupling temperature and for the Langevin parametrization in contrast to the constant energy loss model.  Results for $\Bzero$ mesons however do not discriminate the transport models in the same manner and there seems to be no collision energy dependence as well.

\begin{figure}[h]
    \centering
    \includegraphics[width=0.4\textwidth]{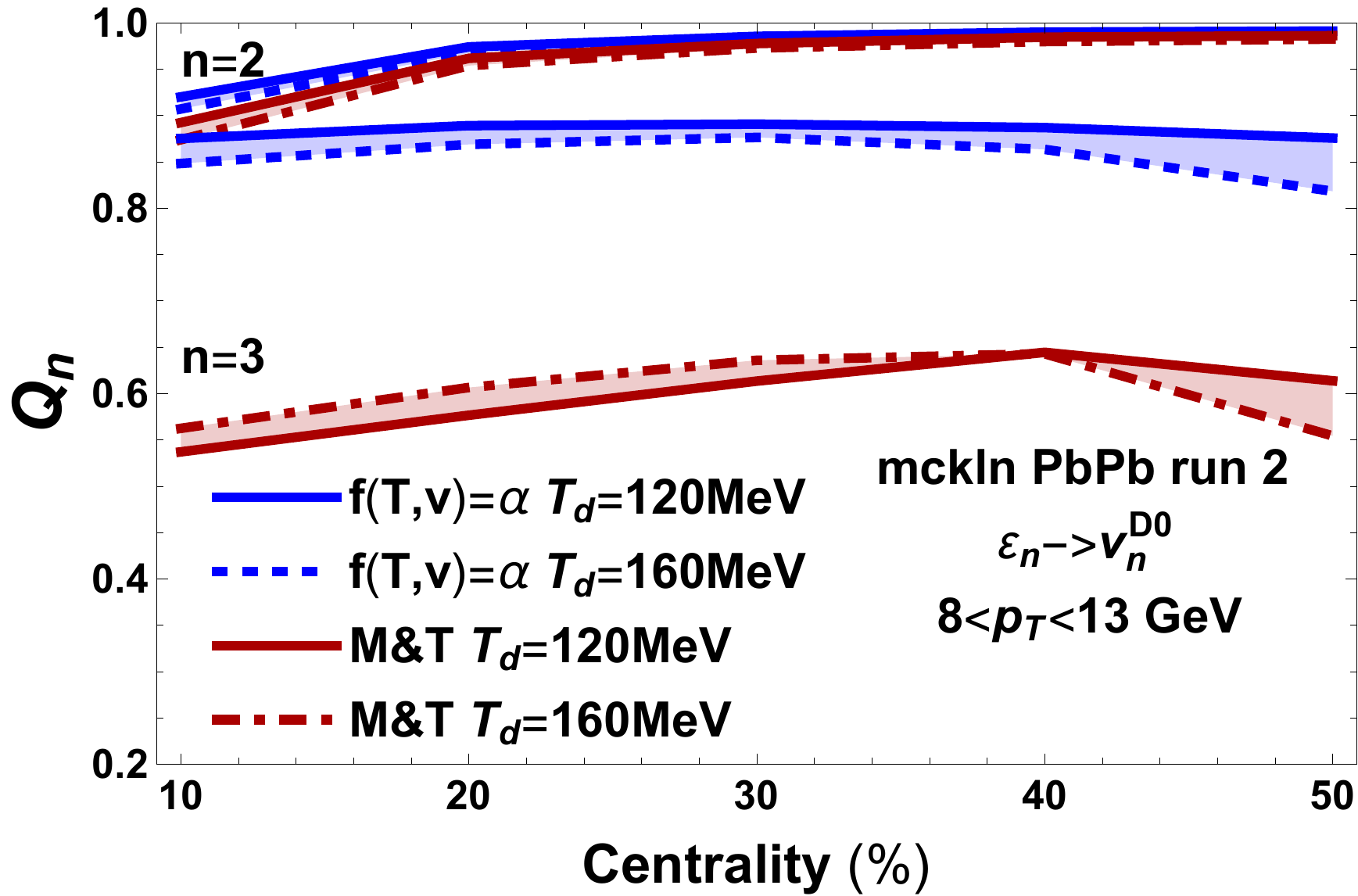}
    \caption{Centrality dependence of the Pearson coefficients $Q_2$ and $Q_3$ for the constant energy loss model and the Langevin model (Moore and Teaney parametrization) in $\PbPb$ collisions at $\snn[5.02]$ for $\Td = 120$--\SI{160}{MeV}. The integrated $\vnn^{heavy}$ is defined in the range 8 GeV$ < \pt  <$ 13 GeV.}
    \label{fig:pearson}
\end{figure}

Our previous results show the presence of linear scaling between $\vnn^\text{(soft)}$ and $\vnn^\text{(heavy)}$.  Considering that the former is also related to the eccentricities in the initial conditions, it is expected that this would also hold for the heavy sector.  In Fig.\ \ref{fig:pearson} we show the results for the centrality dependence of the Pearson coefficients $Q_2$ and $Q_3$~\cite{Betz:2016ayq} associated with $\Dzero$ mesons computed using the constant energy loss model and the M\&T parametrization of the Langevin model in  $\PbPb$ collisions at $\snn[5.02]$ for $\Td = 120$--\SI{160}{MeV}. We find that the centrality dependence of these quantities is generally weak. Concerning elliptic flow, we see that both models display large values of $Q_2$, with the constant energy loss model giving larger values than the Langevin description. In this case, the results do not vary appreciably when the decoupling temperature is varied in the range 120--\SI{160}{MeV}. However, while $Q_3$ displays large values $>0.8$ for the constant energy loss model, the same cannot be said about the Langevin result where $Q_3$ drops to $\sim 0.6$. This shows that  in general  $\vnn^\text{(heavy)}$ is more strongly correlated  to $\vnn^\text{(soft)}$ in the constant energy loss model than in the Langevin description. This may be expected due to the noise term in the Langevin description. Moreover, we see that the results do not change significantly with the decoupling temperature, unless in the case of more peripheral collisions.

\section{The effect of initial conditions: Trento vs.\ \mckln}  \label{SecIV}

As discussed in Sec.\ \ref{sec:back}, while \mckln\ and Trento can both reproduce  $v_2\left\{2\right\}$ of the soft sector well, Trento generally does best when one considers multiple beam energies when it comes to the ratio  $v_2\{4\}/v_2\{2\}$ (specifically at \lhc\ energies, while for \rhic\ \mckln\ does quite well). This implies that \mckln\ initial conditions can capture the mean of the $v_2$ distribution well but they do not have a wide enough $v_2$ distribution compared to experimental data. On the other hand, Trento can capture both the mean and the width of the distribution well.

The question still remains how these differences translate into the heavy flavor sector. Previous studies have looked at the influence of the choice of initial conditions  on $\langle v_2\rangle$ for smoothed, averaged initial conditions  in~\cite{Cao:2013ita} and on $v_2\{2\}$~\cite{Noronha-Hostler:2016eow,Xu:2018gux} defined using event-by-event initial conditions but the influence on the actual $v_2$ fluctuations has not yet been considered. However, as shown in Sec.\ \ref{sec:back}, this is precisely the sector where we expect the largest differences in the initial conditions.

In \ref{SubSection:FreeParameters} it was explained how that the parameters are tuned using the calculation of $\raa$ in central collisions. Thus, there are slight differences in the parameters for Trento vs. \mckln\ initial conditions. Finally, we note that in the figures below coalescence has not been included, which will be done in Sec.\  \ref{Section: coalescence}, so we do not anticipate a perfect match to experimental data. Rather, we focus here on the qualitative aspects of the results.

\subsection{Nuclear modification factor}

\begin{figure}[!htb]
  \centering
  \includegraphics[width=0.47\textwidth]{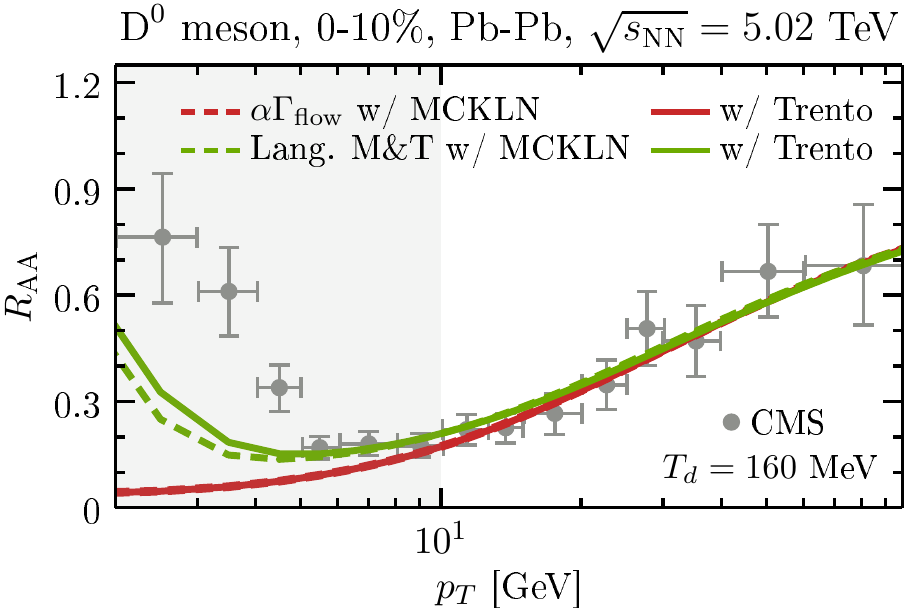}
  \caption{$\Dzero$ meson nuclear modification factor $\raa$ in the 0--10\% centrality class of $\snn[5.02]$ $\PbPb$ collisions obtained with \mckln\ (dashed lines) or Trento (solid lines) initial conditions. The gray area indicates the $\pt$ region where coalescence and initial/final state effects may be important. Experimental data from the CMS ($|y|<1$)~\cite{Sirunyan:2017xss} collaboration.}
  \label{RAAmcklnVsTrento}
\end{figure}

Based on the previous study in Ref.~\cite{Noronha-Hostler:2016eow}, we did not expect large differences in $\raa$ when comparing different initial conditions and, indeed, we find that to be true. We note, however, in ~\cite{Noronha-Hostler:2016eow} only the energy loss scenario was considered, which we find to have essentially no dependence on the initial conditions. However, the Langevin model does see a slight enhancement in $\raa$ at low $\pt$ when Trento initial conditions are considered. Since there is a tuning parameter to the $\raa$, and Trento initial conditions do exhibit a slightly different behavior in the Langevin model, we anticipate that the Langevin model may see differences in the $v_n$'s as well.

\subsection{Two-particle $v_n$ cumulants}

\begin{figure}[!htb]
  \centering
 \includegraphics[width=0.5\textwidth]{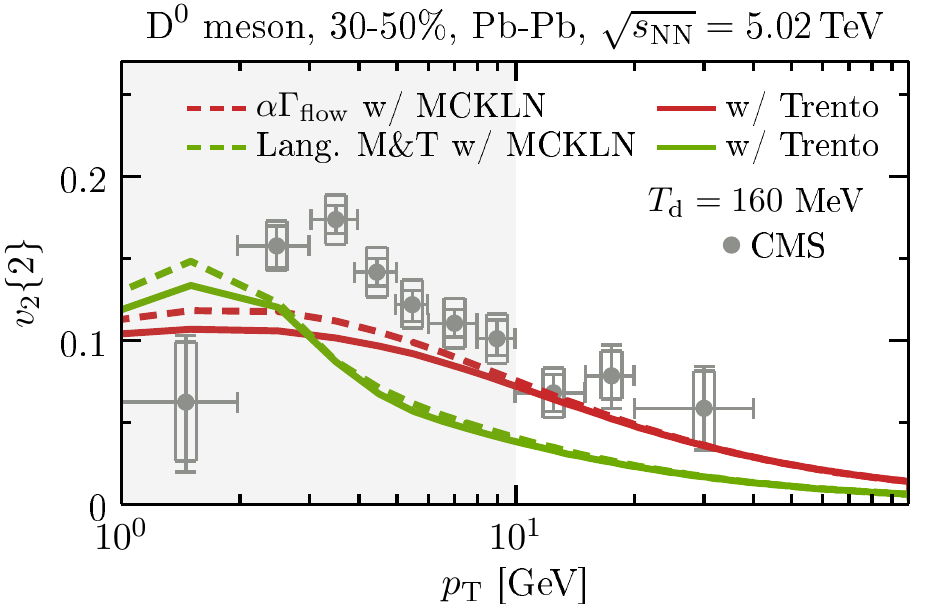}
  \caption{ $\Dzero$ meson elliptic flow coefficient $\vn2$ in the 30--50\% centrality class of $\snn[5.02]$ $\PbPb$ collisions obtained with \mckln\ (dashed lines) or Trento (solid lines) initial conditions. Experimental data from the CMS ($|y|<1$)~\cite{Sirunyan:2017plt} collaboration.}
  \label{v2mcklnVsTrento}
\end{figure}

Fig.\ \ref{v2mcklnVsTrento} shows $v_2\{2\}(\pt)$ at a fixed decoupling temperature. In general, we find that \mckln\ initial conditions produce a larger $v_2\{2\}$ than Trento initial conditions and that this effect is clearest in the low $\pt$ sector.
\begin{figure}[!htb]
  \centering
 \includegraphics[width=0.5\textwidth]{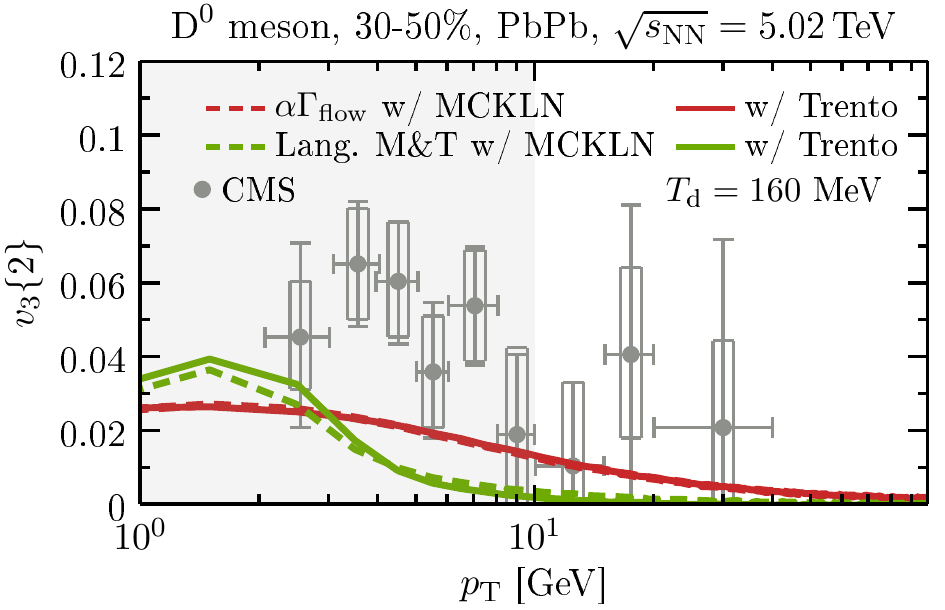}
  \caption{$\Dzero$ meson triangular flow coefficient $\vn3$ in the 30--50\% centrality class of $\snn[5.02]$ $\PbPb$ collisions obtained with \mckln\ (dashed lines) or Trento (solid lines) initial conditions. Experimental data from the CMS ($|y|<1$)~\cite{Sirunyan:2017plt} collaboration.}
  \label{v3mcklnVsTrento}
\end{figure}
Given that \mckln\ initial conditions have a slightly larger $\varepsilon_2$ than that found in Trento, one might  conclude that this is why we see an enhancement in $v_2\{2\}(\pt)$ for \mckln\ initial conditions.
We would like to point out, however, that the origin of this result is more complicated than that. In Fig.\ \ref{fig:v22} we demonstrated that by varying the hydrodynamic parameters we were able to find similar $v_n\{2\}$ for \mckln\ and Trento initial conditions in the soft sector. This affect is achieved by including a larger $\eta/s$ in the hydrodynamic backgrounds for the mckln initial conditions compared to those from TRENTO. Thus, initial conditions with {\it similar} results in the soft sector can lead to {\it different} results in the heavy flavor sector.

In Fig.\ \ref{v3mcklnVsTrento} $v_3\{2\}$ results for Trento and \mckln\ initial conditions are shown. While the energy loss scenario for heavy quarks shows no sensitivity to the initial conditions, we find that Trento enhances $v_3\{2\}$ for Langevin.

\subsection{Elliptic flow from multi-particle cumulants}

\begin{figure}[!htb]
  \centering
\includegraphics[width=0.48\textwidth]{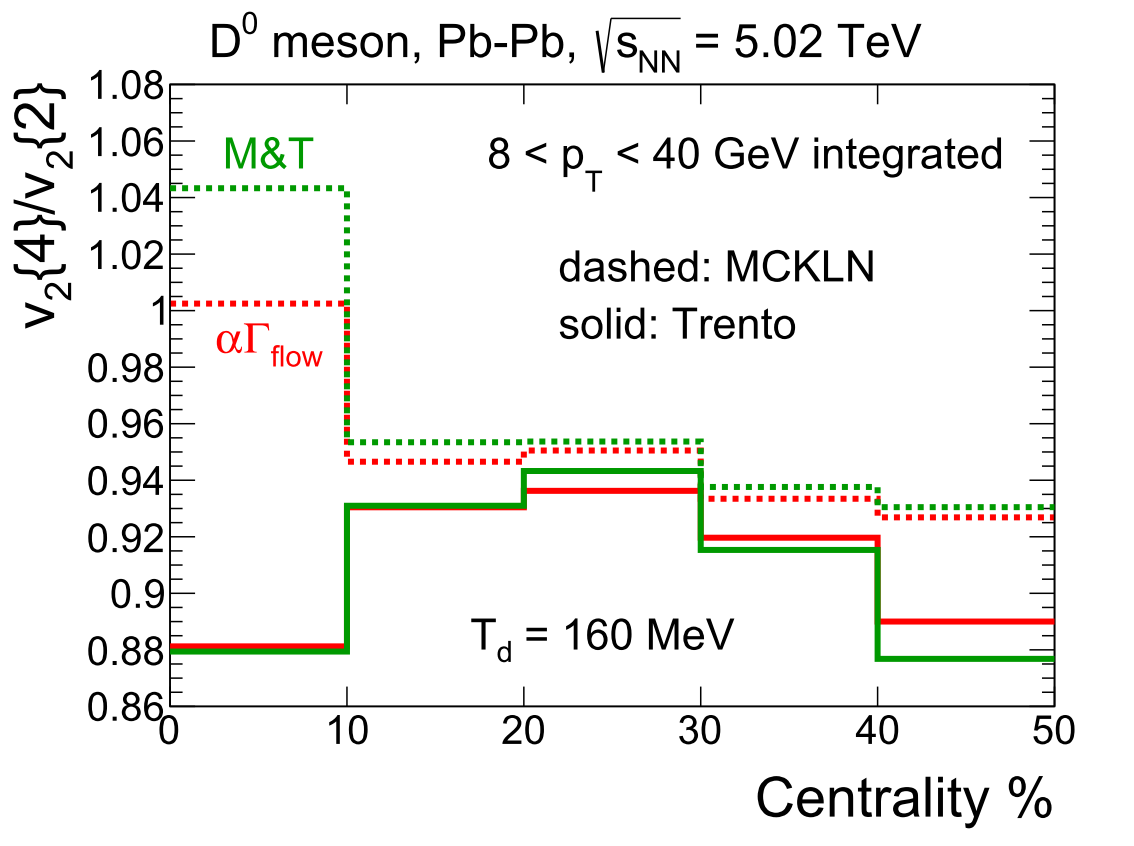}
  \caption{ $\Dzero$ $v_2\{4\}/v_2\{2\}$ ratio integrated over $8 < \pt < 40$ GeV for $\snn[5.02]$ $\PbPb$ collisions obtained with \mckln\ (dashed lines) or Trento (solid lines) initial conditions. }
  \label{fig:trento_mckln_fluc}
\end{figure}

While there are certain caveats to the 2-particle $v_2$ correlation when it comes to differences in the initial conditions, we do not expect those to strongly affect the ratio between the 4-particle and the 2-particle correlation. Thus, differences that arise in $v_2\{4\}/v_2\{2\}$ from the initial conditions may be very useful to constrain initial conditions (as was previously done in the soft sector in~\cite{Giacalone:2017uqx}). In Fig.\ \ref{fig:trento_mckln_fluc} a comparison between the integrated $v_2\{4\}/v_2\{2\}$ for \mckln\ vs. Trento initial conditions is shown for our two best fitting dynamical models. As previously discussed in Sec.\ \ref{sec:multi}, central collisions and peripheral collisions appear to be the best testing beds for different model parameters. This holds true for the initial conditions as well. In fact, central collisions are especially interesting because   $v_2\{4\}/v_2\{2\}>1$ for \mckln\ initial conditions (regardless of all dynamical parameters) and $v_2\{4\}/v_2\{2\}<1$ for Trento initial conditions. Thus, we find that one must incorporate the correct initial conditions {\it first} before being able to determine systematic differences from the dynamics of heavy quarks. Because of such a dramatic effect, we strongly encourage experimentalists to investigate these observables in upcoming runs.

\section{The effect of coalescence}  \label{Section: coalescence}

In this section we explore the effect of heavy-light quark coalescence on $\Dmeson$ meson production. We use a hybrid coalescence plus fragmentation model~\cite{Cao:2015hia} based on the widely used ``instantaneous approach'' of coalescence~\cite{Dover:1991,Lin:2003jy, Fries:2003kq,Greco:2003vf,Oh:2009zj}. Within this approach, the probability distribution ${\rm P}_{\rm coal}[q,Q\rightarrow M]$ that a heavy meson of momentum ${\bf p}_M$ is formed by coalescence of a heavy quark with momentum ${\bf p}_Q$ and a light quark of momentum ${\bf p}_q$ is given by
\begin{eqnarray}
{\rm P}_{\rm coal}[q,Q\rightarrow M]({\bf p}_Q, {\bf u})=N\int \dd^3 {\bf p}_q \,f_M({\bf p}_q,{\bf p}_Q)\, \, n_q({\bf p}_q, {\bf u}, \Td)\, \delta({\bf p}_M-{\bf p}_q-{\bf p}_Q),
\end{eqnarray}
where $N$ is a global normalization factor discussed below, $n_q$ is the momentum distribution of the light quarks at the time of hadronization, and $f_M$ is the probability density obtained from the projection of the two quark states onto the meson state. The different meson states are obtained from a simple harmonic oscillator model. The quarks being relativistic, the projection is performed in their center-of-mass frame following a Lorentz boost of their phase space coordinates from the global frame to the center-of-mass frame. Assuming that the light quarks have a uniform spatial distribution in the medium cell where the heavy quark is located, one can average the probability density over the spatial coordinates, which then gives
\begin{eqnarray}\label{WignerMeson}
f_M({\bf p}_q,{\bf p}_Q) = g_M\,h_M\,\frac{(2\sqrt{\pi}\sigma)^3}{(2\pi)^3}\,e^{-{\bf p}_{\rm rel}^{\, 2}\sigma^2},
\end{eqnarray}
where $g_M$ is the usual color-spin-isospin statistical factor for the two spin 1/2 quarks to form a color neutral meson $M$, and $h_M$ are new ``thermal'' factors discussed below. The width $\sigma=1/\sqrt{\mu\omega}$ is given by the angular frequency of the harmonic oscillator $\omega$ and the reduced mass of the two quark system $\mu = m_q m_Q /(m_q+m_Q)$. The mass of the light quarks are assumed to be their constituent masses\footnote{If the effective or quasi-particle masses~\cite{Levai:1997yx, Plumari:2011mk} for the light quarks is taken to be in the range $m_{u,d,s}\sim 300$-600~MeV, no significant differences in the final values of ${\rm P}_{\rm coal}$ are found.} $m_{u,d}=300$ MeV and $m_s = 460$ MeV~\cite{BorkaJovanovic:2010yc} to take into account the non-perturbative effects of QCD near the cross-over transition temperature $T_c$, while the heavy quark masses are taken to $m_c=1.27$ GeV and $m_b=4.19$ GeV. The value of the angular frequency is evaluated to be $\sim 0.3$ GeV for $\Dzero$, $\Dplus$, $\romanup{D}_s$ and $\Bmeson^-$ mesons\footnote{Note that for $\Bzero$ and $\Bmeson_s^-$ we obtain $\omega\sim 0.5$ GeV and $\sim 0.6$ GeV, respectively, which is quite different from the other meson ground states.} from their vacuum charge radii obtained within the light cone model~\cite{Hwang:2001th}. Nevertheless, it is known that at $T\sim T_c$ the charge radii must be (much) larger than in the vacuum. Motivated by a (limited) comparison to the case of quarkonia in which the radii of the $J/\psi$ and $\Upsilon(1S)$ are evaluated from lattice QCD potentials to be $\sim 3$ times larger at $T_c$ than in the vacuum, we set $\omega = 0.1$ GeV ($\omega$ being inversely proportional to the radius) and assume this value to be valid for all the hadrons considered within this basic model. The relative momentum of the two quarks including relativistic corrections
 \begin{eqnarray}
{\bf p}_{\rm rel}=\frac{m_q\,{\bf p}_Q'-m_Q\,{\bf p}_q'}{m_q+m_Q},
\end{eqnarray}
is defined via the momenta in the center-of-mass frame (denoted with primed coordinates). The momentum density distribution of the light quarks $n_q$ is assumed to be thermal in the local rest frame of the medium cell considered being given by the Fermi-Dirac distribution
\begin{eqnarray}\label{FermiDiracDistr}
n_q({\bf p}_q, {\bf u}, \Td)= \frac{g_q}{e^{\sqrt{{{\bf p}_{q}^{\rm cell}}^2+m_q^2}/\Td}+1} =  \frac{g_q}{e^{p_{q}.u/\Td}+1},
\end{eqnarray}
where ${\bf p}_{q}^{\rm cell}$ is the light quark momentum in the local rest frame of the medium cell, $g_q=6$ is the statistical factor that takes into account the spin and color degeneracy of the light quarks, $p_{q}.u$ is the 4-product between the light quark 4-momentum in the global frame and the fluid 4-velocity $u=(\gamma_u,\gamma_u {\bf u})$ with $\gamma_u=\frac{1}{\sqrt{1-{\bf u}^2}}$. Thanks to this fluid velocity dependence of $n_q$, the derived coalescence probabilities will depend not only on the heavy quark momentum but also on the local flow and on the angle between them.

By extension, the probability distribution ${\rm P}_{\rm coal}[q_1,q_2,Q\rightarrow B]$ that a heavy baryon of momentum ${\bf p}_B$ is formed by coalescence of a heavy quark with momentum ${\bf p}_Q$ and two light quarks of momenta ${\bf p}_{q_1}$ and ${\bf p}_{q_2}$ is given by
\begin{eqnarray}
{\rm P}_{\rm coal}[q_1,q_2,Q\rightarrow B]({\bf p}_Q, {\bf u})=N\int \dd^3 {\bf p}_{q_1}\dd^3 {\bf p}_{q_2} \,f_B({\bf p}_{q_1}, {\bf p}_{q_2}, {\bf p}_Q)\, \, n_{q_1}({\bf p}_{q_1}, {\bf u}, \Td)\,\times \nonumber\\ \times \, n_{q_2}({\bf p}_{q_2}, {\bf u}, \Td)\, \delta({\bf p}_B-{\bf p}_{q_1}-{\bf p}_{q_2}-{\bf p}_Q),
\end{eqnarray}
where $N$ is the global normalization factor, $f_B$ is the probability density obtained by combining the two light quarks first and then using their center of mass to recombine with the heavy quark,
\begin{eqnarray}
f_B({\bf p}_{q_1},{\bf p}_{q_2},{\bf p}_Q) = g_B\,h_B\,\frac{(2\sqrt{\pi})^6 (\sigma_1\sigma_2)^3}{(2\pi)^6} \, e^{-\vec{p}_{\rm rel 1}^{\, 2}\sigma_1^2-\vec{p}_{\rm rel 2}^{\, 2}\sigma_2^2},
\end{eqnarray}
where $g_B$ is the usual color-spin-isospin degeneracy factor for the three spin 1/2 quarks to form a color neutral baryon $B$, and $h_B$ are the ``thermal'' factors discussed below. Note that the resulting ${\rm P}_{\rm coal}[q_1,q_2,Q\rightarrow B]$ is perfectly identical if we combine the two light quarks first or if we combine the heavy quark and one of the light quarks first. The widths $\sigma_1=1/\sqrt{\mu_1\omega}$ and $\sigma_2=1/\sqrt{\mu_2\omega}$ are given by the angular frequency of the harmonic oscillator $\omega$, the reduced mass of the two light quark system $\mu_1 = m_{q_1} m_{q_2} /(m_{q_1}+m_{q_2})$, and the reduced mass of the system formed by the two light quarks center-of-mass and the heavy quark $\mu_2 = (m_{q_1}+m_{q_2})m_Q /(m_{q_1}+m_{q_2}+m_Q)$. The relative momenta with relativistic corrections are
\begin{eqnarray}
{\bf p}_{\rm rel 1}=\frac{m_{q_2}{\bf p}_{q_1}'-m_{q_1}{\bf p}_{q_2}'}{m_{q_1}+m_{q_2}}
\end{eqnarray}
and
\begin{eqnarray}
{\bf p}_{\rm rel 2}=\frac{m_Q(\vec{p}_{q_1}'+{\bf p}_{q_2}')-(m_{q_1}+m_{q_2}){\bf p}_Q'}{m_{q_1}+m_{q_2}+m_Q},
\end{eqnarray}
defined\footnote{The relative momenta, as defined in~\cite{Cao:2015hia}, i.e.\ with $E_q'$ replacing $m_q$ in the definitions, have the disadvantage of breaking the ``combining order symmetry'', i.e.\ ${\rm P}_{\rm coal}[q_1,q_2,Q\rightarrow B]$ is different if we combine the two light quarks first or the heavy quark and a light quark. It is then difficult to justify why one type of ordering should be better than the other.} via the momenta in the baryon center-of-mass frame (denoted with primed coordinates). Similarly to the mesons, the momentum density distributions of the light quarks $n_{q_1}$ and $n_{q_2}$ are assumed to be thermal in the local rest frame of the considered medium cell and given by the Fermi-Dirac distribution \eqref{FermiDiracDistr}.

For the charmed mesons (baryons) we consider all the symmetric $J^P=0^-$ and $1^-$ states ($J^P=1/2^+$ and $3/2^+$ states), i.e.~the $\Dzero$, $\Dplus$, $\Dmeson_s^+$ ($g_M=1/36$ each), $\Dmeson^{*0}$, $\Dmeson^{*+}$, $\Dmeson_s^{*+}$ ($g_M=1/12$ each) mesons and the $\Lambda_c^+$, $\Sigma_c^0$, $\Sigma_c^+$, $\Sigma_c^{++}$ ($g_B=1/108$ each), $\Sigma_c^{*0}$, $\Sigma_c^{*+}$, $\Sigma_c^{*++}$ ($g_B=1/18$ each), $\Xi_c^0$, $\Xi_c^+$, $\Xi_c^{'0}$, $\Xi_c^{'+}$ ($g_B=1/54$ each), $\Xi_c^{*0}$, $\Xi_c^{*+}$ ($g_B=1/27$ each), $\Omega_c^{0}$ ($g_B=1/108$), $\Omega_c^{*0}$ ($g_B=1/54$) baryons. The anti-symmetric states cannot be included as the considered Wigner distribution (\ref{WignerMeson}) is symmetric, being thus usually neglected in coalescence models.

The global normalization factor $N$ is usually chosen such that ${\rm P}_{\rm coal}[c\rightarrow {\rm any\,hadron}]({\bf p}_Q=0)=1$ assuming that a quark with zero momentum does not hadronize via fragmentation but via coalescence only\footnote{See a criticism of this assumption in~\cite{Song:2015ykw}.}. In this work we extend this idea by assuming instead that ${\rm P}_{\rm coal}[c\rightarrow {\rm any\,hadron}]({\bf p}_Q,{\bf u})=1$ when the heavy quark velocity vector is equal to the one from the local medium flow, i.e.~when the heavy quark is moving together with the light quarks surrounding it. In this way we avoid the low ${\bf p}_Q$ ``saturation'' of ${\rm P}_{\rm coal}$ when $|{\bf u}|>0$ obtained in~\cite{Cao:2015hia}. This saturation seems unjustified to the present authors: a heavy quark with a velocity much lower than the typical velocities of the surrounding light quarks may not coalesce with unity probability.\\

Recently, the prompt heavy hadron ratios $\Dplus/\Dzero$, $\Dmeson^{*+}/\Dzero$, $\Dmeson_s^{+}/\Dzero$, $\Dmeson_s^{+}/\Dplus$, and $\Lambda_c^{+}/\Dzero$ have been measured in heavy ion collisions and compared to proton-proton collisions~\cite{Acharya:2018hre,STARprelHFE,ALICEprelLambda,Acharya:2018ckj}. They provide important information about the hadronization process in heavy ion collisions and can serve as a benchmark to test and calibrate coalescence models. First, the low and intermediate $\pt$ enhancement of the $\Lambda_c^{+}/\Dzero$ ratio tends to confirm the presence of heavy-light quarks coalescence in heavy ion collisions when $\pt<7$ GeV. The $\pt$ and collision system independence of the $\Dplus/\Dzero$ and $\Dmeson^{*+}/\Dzero$ ratios show a certain universality in the way the different hadronization processes distribute the quarks between the meson states. Finally, the enhancement of $\Dmeson_s^{+}/\Dplus$ at low and intermediate $\pt$ confirms the thermal enhancement of strange quark production. Within the ``basic'' coalescence model (i.e.~without the $h_M$ and $h_B$ factors) the prompt $\Dplus/\Dzero$ and $\Dmeson^{*+}/\Dzero$ ratios are only based on the color-spin-isospin degeneracy factors, the rest of the probabilities canceling out, e.g.
\begin{eqnarray}\nonumber
\frac{\Dplus_{\rm prompt}}{\Dzero_{\rm prompt}} = \frac{\Dplus_{\rm dir.}+\Dmeson^{*+}_{\rm dir.}\, Br(\Dmeson^{*+}\rightarrow \Dplus)}{\Dzero_{\rm dir.}+\Dmeson^{*0}_{\rm dir.}\, Br(\Dmeson^{*0}\rightarrow \Dzero)+\Dmeson^{*+}_{\rm dir.}\, Br(\Dmeson^{*+}\rightarrow \Dzero)} = \frac{g_\Dmeson + 0.307 g_{\Dmeson^{*}}}{g_\Dmeson + 1.677 g_{\Dmeson^{*}}} \approx 0.32
\end{eqnarray}
where ``prompt'' means including feed-downs from excited states and ``dir.'' means the direct production (without feed-downs) that we obtain from the coalescence model. The different ratio values obtained within the ``basic'' coalescence model are summarized in table\ \ref{Tab:TableRatios} and compared to low $\pt$ experimental data in AA collisions. The $\Dplus/\Dzero$ and $\Dmeson_s^{+}/\Dplus$ ratios particularly miss the experimental data. Additionally the $\Lambda_c^{+}/\Dzero$ ratio fits LHC preliminary data~\cite{ALICEprelLambda} but underestimate RHIC preliminaries~\cite{STARprelHFE}. To improve the fit to the $\Dplus/\Dzero$ ratio, one could naturally think of decreasing the angular frequency $\omega$ for the excited states --- as their radii should be larger than the ground states --- but the $\Dplus/\Dzero$ ratio is then observed to decrease, which is the opposite of what was expected.

\begin{table}[h!]
\begin{center}
    \begin{tabular}{|C{3.2cm}||C{4.5cm}|C{2.7cm}|C{2.7cm}|}
    \hline
Approximate values at low $\pt$ & Experimental \cite{Acharya:2018hre,STARprelHFE,ALICEprelLambda,Acharya:2018ckj,Gladilin:2014tba,Cacciari:2003zu} & ``basic'' coalescence & with extra thermal factors \\
    \hline
    \hline
direct $\qcharm\rightarrow\Dzero$ & in pp: 0.17  & 0.06  & 0.10 \\
    \hline
prompt $\qcharm\rightarrow\Dzero$ & in pp: 0.55  & 0.36  & 0.32  \\
    \hline
$\Dplus/\Dzero$ & 0.47  & 0.32  & 0.45  \\
    \hline
$\Dmeson^{*+}/\Dzero$ & 0.45  & 0.5  & 0.4  \\
    \hline
$\Dmeson_s^{+}/\Dzero$ & 0.35  & 0.31  & 0.34 \\
    \hline
$\Dmeson_s^{+}/\Dplus$ & 0.75  & 0.99 & 0.76 \\
    \hline
$\Lambda_c^{+}/\Dzero$ & @ \rhic: 1.5 - @ \lhc: 0.7  & 0.74  & 0.94 \\
    \hline
    \end{tabular}
\caption {\label{Tab:TableRatios}
\small The approximate low $\pt$ branching ratios or prompt hadronic ratios obtained with different configurations of the coalescence model (using a typical $|{\bf u}|\approx 0.6$) compared to the low $\pt$ experimental data in AA collisions (or pp collisions if specified).}
\end{center}
\vspace{-0.4cm}
\end {table}

To improve the fit to the ratios and, subsequently, the predictive power of coalescence in our study one can instead take a step aside from the basic model and include \emph{by hand} a new element. One of the weak points of the basic model is the complete absence of hadron masses\footnote{A consequence of this is the non conservation of the $4^{\text{th}}$ component of the system's 4-momentum during the coalescence process. We do not intend to address this problem with these new elements.} in the formalism whereas it is clear that the formation of excited states with larger masses requires more energy from the combining quark than to form the ground state. Inspired by the thermal model of hadronization~\cite{Oh:2009zj}, we include the ``thermal'' factors $h_{H=M,B}=\exp[-(m_{H}-m_{{H}_0})/\Td]$, where $H$ is the hadron state for which we want to compute ${\rm P}_{\rm coal}$, $m_{H}$ is its mass, and $m_{{H}_0}$ is the mass of the corresponding ground state ${H}_0$ built with the same quark content as $H$, e.g.~$\Dzero$ for $\Dmeson^{*0}$ or $\Lambda_c^+$ for $\Sigma_c^+$ and $\Sigma_c^{+*}$. Therefore, for the same quark content, an excited state $H$ has $g_{H}\exp[-m_{H}/\Td]/(g_{{H}_0}\exp[-m_{{H}_0}/\Td])$ less probability to form than its corresponding ground state ${{H}_0}$. With these new factors the ratios are, thus, not anymore based on spin and color considerations alone but they also depend on the hadron masses, leading to a more relevant statistical hierarchy between the different energy states of a combination of quark flavors. Note that thanks to these new factors a natural justification arises for the non consideration of more excited states (such as the $J=2$ mesons and $J=5/2$ baryons) and antisymmetric states (such as the $J^P=0^+$ mesons and $J^P=1/2^-$ baryons), as they are now suppressed due to their larger mass. The resulting ratios are summarized in Tab.\ \ref{Tab:TableRatios} and fit well the experimental data. Additionally, a larger $\Lambda_c^{+}/\Dzero$ ratio is obtained with the new factors, giving more confidence in the possibility of fitting both \rhic\ and \lhc\ final data.

\begin{figure}[!htb]
  \centering
\includegraphics[width=0.45\textwidth]{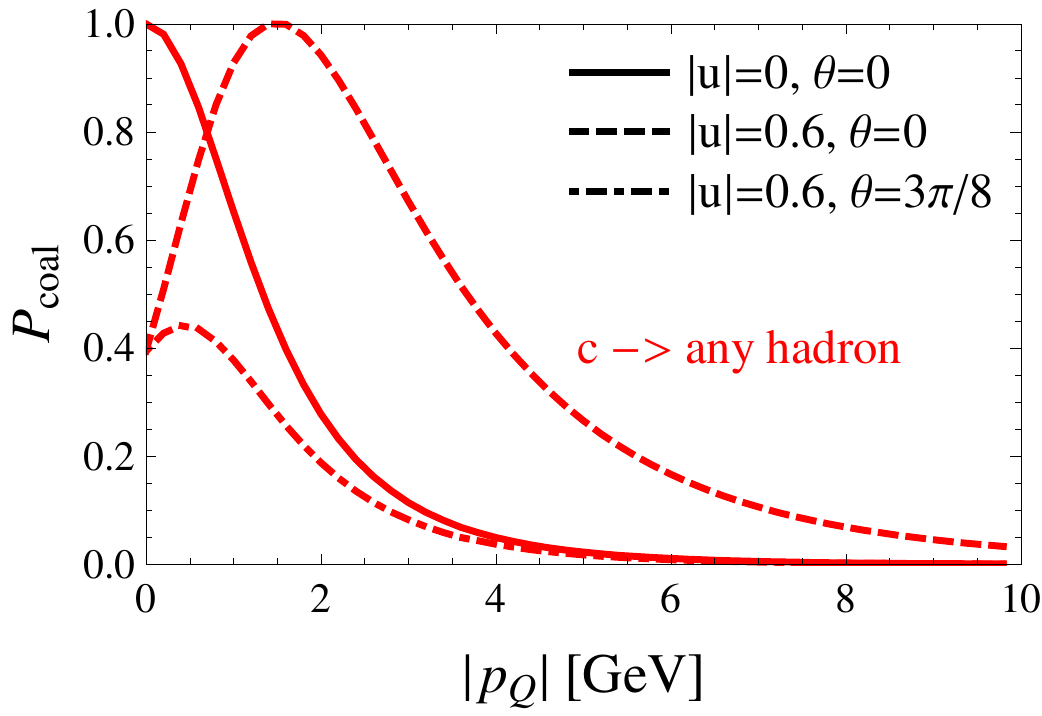}
\hspace{5mm}
\includegraphics[width=0.46\textwidth]{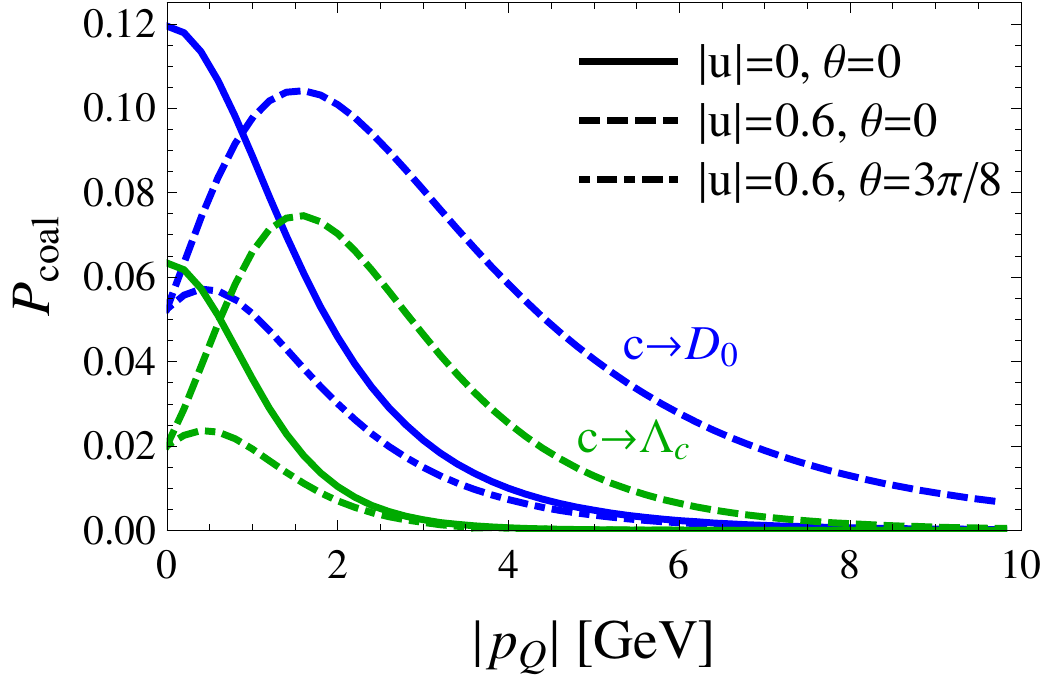}
\label{fig:fewPcoal1}
  \caption{Probabilities for a charm quark of momentum $|{\bf p}_Q|$ to coalesce into any hadron (left) and into a $\Dzero$ meson or a $\Lambda_c^+$ baryon (right). These probabilities are shown for 3 different cases: in a static medium ($|{\bf u}|=0$ --- plain lines), when the charm quark momentum is collinear with the nonzero velocity of the medium cell ($|{\bf u}|=0.6$ and $\theta = 0$ --- dashed lines) and when the charm quark momentum is not collinear with the nonzero velocity of the medium cell ($|{\bf u}|=0.6$ and $\theta = 3\pi/8$ - dot-dashed lines). $\theta$ is defined as the angle between the heavy quark momentum and the medium cell velocity.}
\end{figure}

We emphasize again that the motivation for these new thermal factors is purely phenomenological, i.e., to improve the ratio fits and therefore the quality of the coalescence probabilities, and we do not assume that they are theoretically justified from the beginning. The probabilities of coalescence into any hadron and into $\Dzero$ mesons and $\Lambda_c^{+}$ baryons are shown in Fig.\ \ref{fig:fewPcoal1} for different medium cell velocities $|{\bf u}|$ and $\theta$ angles between ${\bf u}$ and ${\bf p}_Q$. Larger $|{\bf u}|$ and, thus, larger average $|{\bf p}_q|$ tend to shift the maximum of probability towards larger $|{\bf p}_Q|$. An increasing $\theta$ angle tends to lower the probability amplitudes and restrict the coalescence to heavy quarks with lower momenta. One can finally note that with increasing $|{\bf u}|$ the meson production from coalescence tends to decrease whereas the baryon production increases.


The choice between fragmentation and coalescence in heavy ion collisions is performed for each heavy quark through a simple Monte Carlo procedure using the probabilities of coalescence ${\rm P}_{\rm coal}$ described above and of fragmentation found in~\cite{Cacciari:2003zu} for the rest of the probability $1-{\rm P}_{\rm coal}$. If the coalescence into a $\Dzero$ meson is drawn for a heavy quark of momentum ${\bf p}_Q$, a light quark of momentum ${\bf p}_q$ is generated according to the momentum space density distribution $n_q$ in the frame of the medium cell and then boosted back into the global frame where it can possibly combine with the given heavy quark according to the probability density $f_M({\bf p}_q,{\bf p}_Q)$ using a Monte Carlo procedure. If they do not recombine, another light quark is generated until the meson is formed. The $\Dzero$ meson momentum ${\bf p}_{\Dzero}$ is finally given by the momentum of the heavy-light quark system ${\bf p}_Q+{\bf p}_q$.

\subsection{Nuclear modification factor}

\begin{figure}[!htb]
  \centering
  \includegraphics[width=0.55\textwidth]{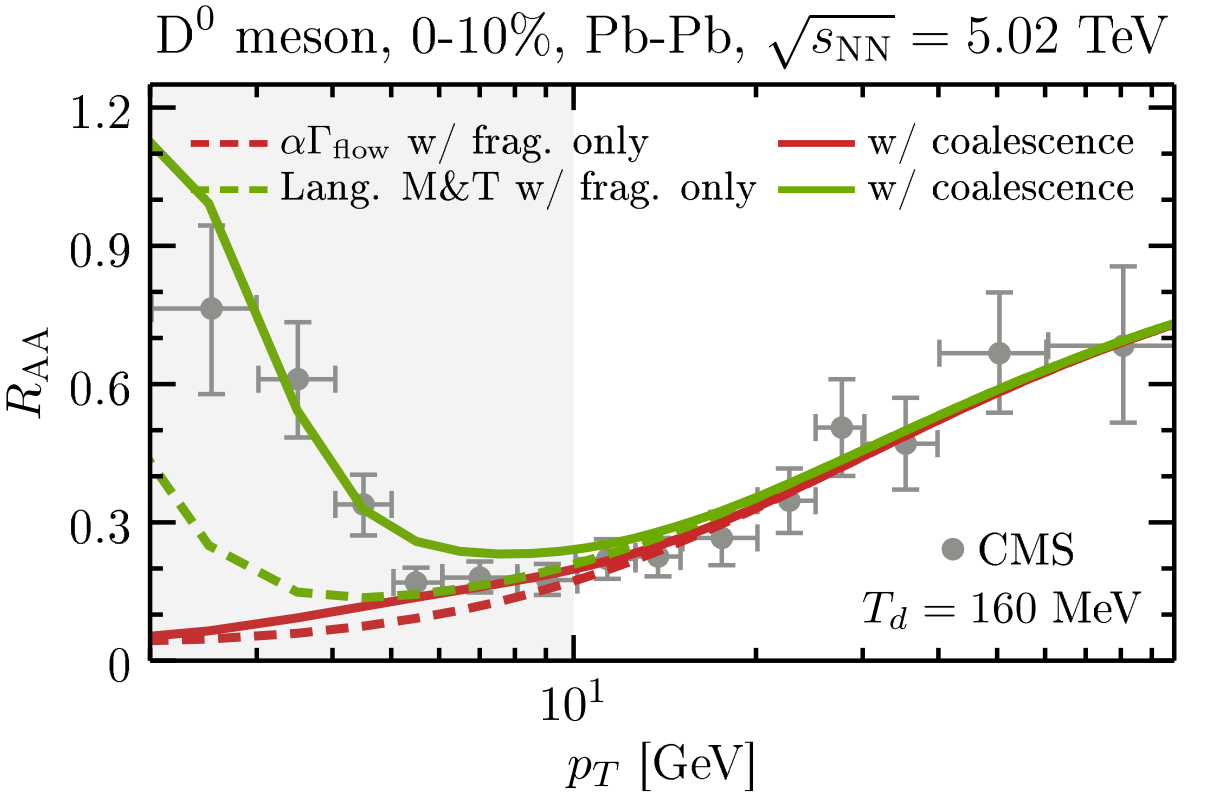}

  \caption{Direct $\Dzero$ meson nuclear modification factor $\raa$ in the 0--10\% centrality range of $\snn[5.02]$ $\PbPb$ collisions obtained with a hadronization based on fragmentation only (dashed lines) or including coalescence (solid lines). The gray area indicates the $\pt$ region where initial/final state effects may also be important. Prompt experimental data from the CMS ($|y|<1$) collaboration~\cite{Sirunyan:2017xss}.}
  \label{fig:RAAcoal}
\end{figure}

In Fig.\ \ref{fig:RAAcoal} we show the effect of coalescence on the nuclear modification factor. The inclusion of coalescence in the M\&T Langevin model remarkably improves the fit to the $\pt<5$ GeV data leading to a good description of the entire $\pt$ range. For the constant energy loss model the inclusion of coalescence only amounts to a very small increase in $\raa$ for $\pt<10$ GeV, which does not fix the large discrepancy at low $\pt$ already observed using only fragmentation. Note however that these results correspond to direct $c\rightarrow\Dzero$ production whereas the experimental data in this section correspond to prompt $\Dzero$ production. The probabilities of $c\rightarrow\Dstar$ being different within the fragmentation and coalescence frameworks (see Tab.\ \ref{Tab:TableRatios}), the prompt results might be a bit different at low $\pt$. The effect of coalescence on $\raa$ can be decomposed into a small $\raa$ shift towards larger values - most probably due to a smaller direct $c\rightarrow\Dzero$ production with coalescence than with fragmentation (see Tab.\ \ref{Tab:TableRatios}) - and more importantly into a $\pt$ shift of the low $\pt$ $\raa$ values of around $\sim 2$ GeV towards higher $\pt$. The latter occurs because fragmentation produces a hadron with less momentum than the heavy quark while in coalescence there is a momentum gain from the light quark ``thermal'' momentum and mass. With only fragmentation the low $\pt$ $\raa$ for the energy loss model becomes almost $\pt$ independent and, thus, the $\pt$ shift cannot be observed. In contrast, with only fragmentation the low $\pt$ $\raa$ of the Langevin model exhibits a large decrease in $\raa$ such that the $\pt$ shift is visible and has a strong impact.

\begin{figure}[!htb]
  \centering
    \includegraphics[width=0.5\textwidth]{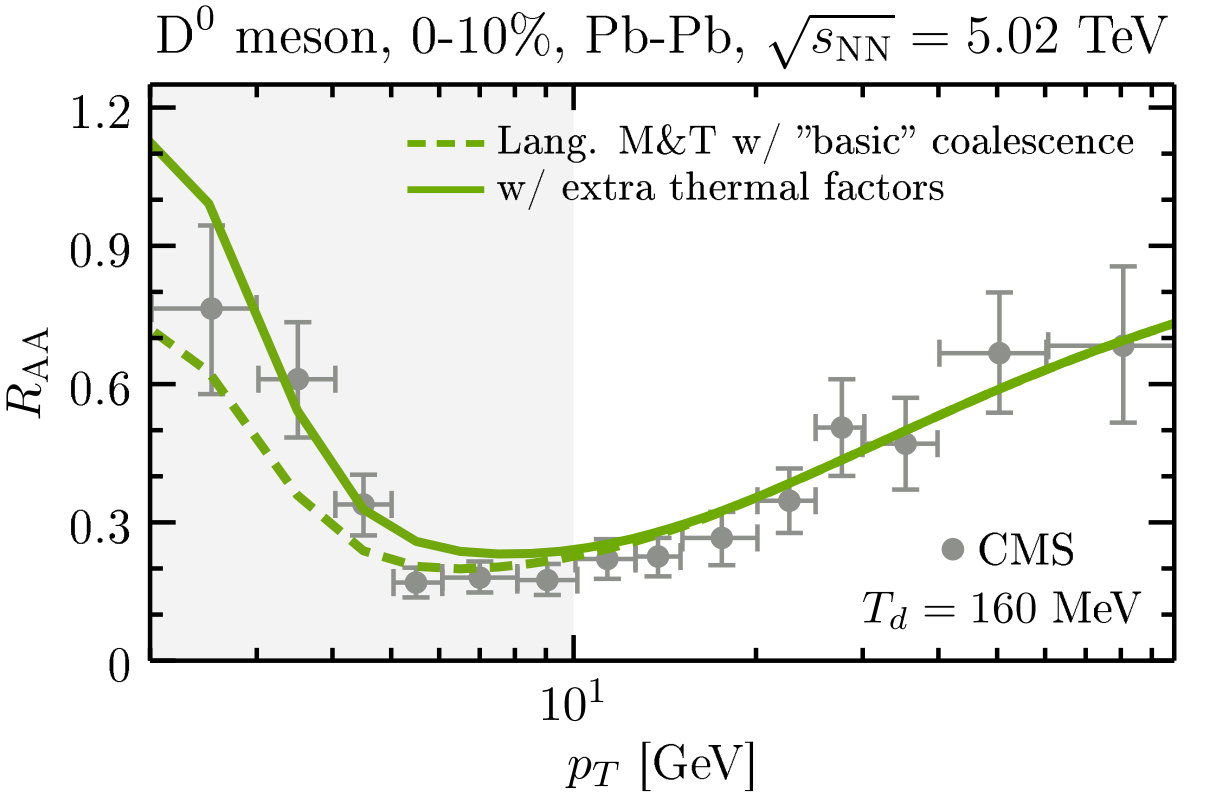}
  \caption{Direct $\Dzero$ meson nuclear modification factor $\raa$ in the 0--10\% centrality range of $\snn[5.02]$ $\PbPb$ collisions obtained with different configurations of the coalescence model: the ``basic'' model i.e. without the thermal factors (dashed lines) and the one including the thermal factors (solid lines). Prompt experimental data from the CMS ($|y|<1$) collaboration~\cite{Sirunyan:2017xss}.}
  \label{fig:RAAcoalBasic}
\end{figure}

The difference between the $\raa$ obtained with and without the thermal factors in the coalescence model is shown in Fig.\  \ref{fig:RAAcoalBasic}. The extra thermal factors logically increase the low $\pt$ $\raa$ due to a larger $\Dzero$ formation probability (see Tab.\ \ref{Tab:TableRatios}).   

%
\subsection{Elliptic and triangular flow coefficients with two-particle cumulants}

\begin{figure}[!htb]
  \centering
 \includegraphics[width=0.47\textwidth]{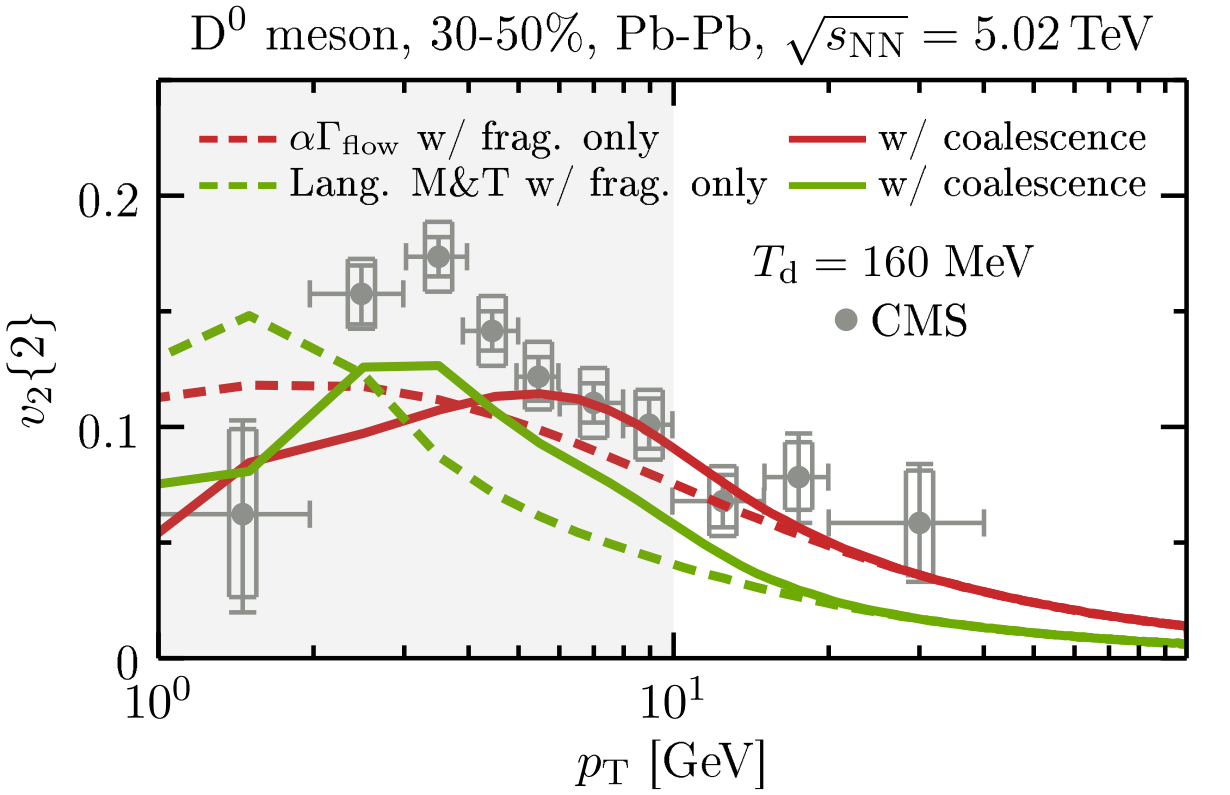}
\hspace{3mm}
 \includegraphics[width=0.48\textwidth]{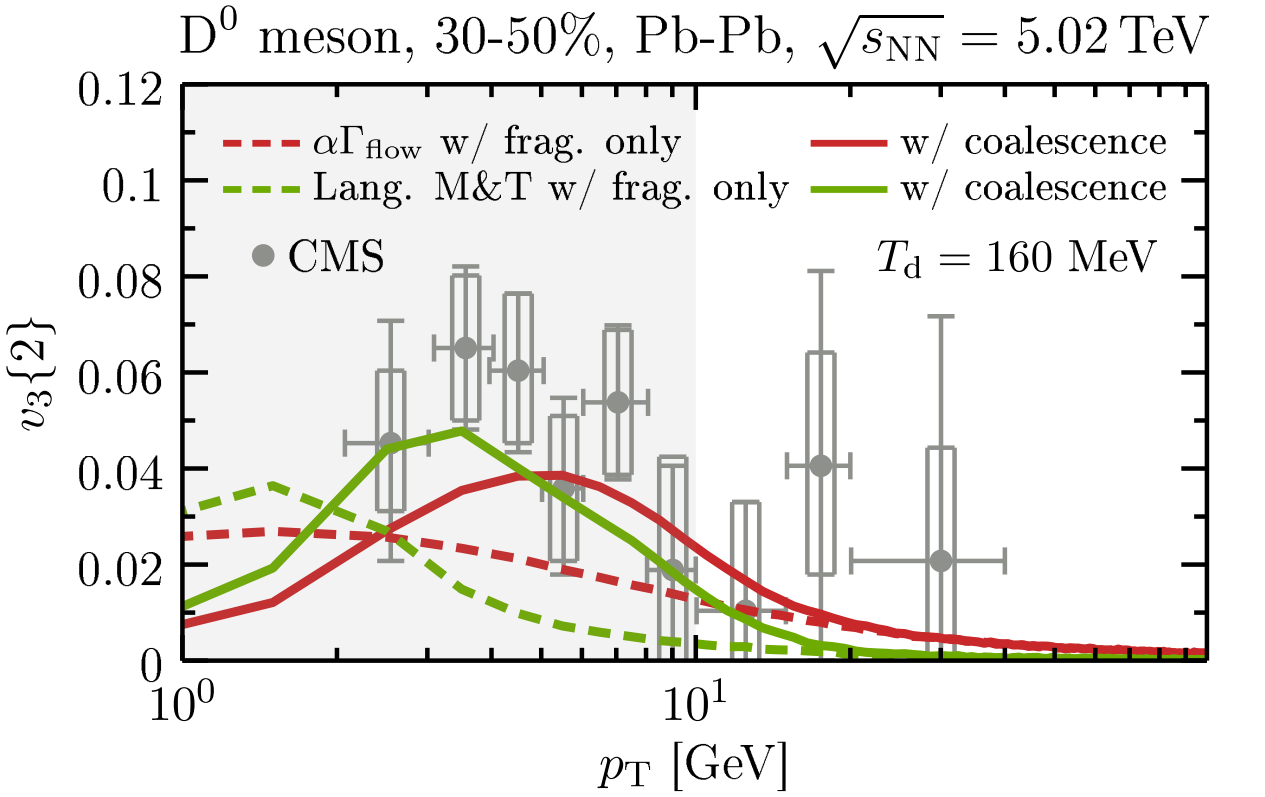}
  \caption{Direct $\Dzero$ meson elliptic (left) and triangular (right) flow coefficients $\vn2$ and $\vn3$ in the 30--50\% centrality class of $\snn[5.02]$ $\PbPb$ collisions obtained with a hadronization based on fragmentation only (dashed lines) or including coalescence (solid lines). Prompt experimental data from the CMS ($|y|<1$) collaboration~\cite{Sirunyan:2017plt}.}
  \label{fig:v2v3coal}
\end{figure}

Regarding $v_2$ and $v_3$, as shown in Fig.\ \ref{fig:v2v3coal} we do not find a single model that can quantitatively capture all the experimental data though coalescence generally improves the description. The M\&T Langevin model with coalescence has the best fit for $\pt\lesssim4$ GeV but underpredicts the data at higher $\pt$, whereas the constant energy loss model with coalescence performs best from $\pt\gtrsim5$ GeV but underpredicts the data at lower $\pt$. Similarly to what we observed for $\raa$, coalescence tends to shift the low $\pt$ peaks in $v_n(\pt)$ to higher $\pt$. Additionally, coalescence also tends to non-negligibly increase the low $\pt$ $v_3$, i.e., when combining the light quarks one communicates part of the medium's triangular flow to the heavy quarks.

\begin{figure}[h]
  \centering
 \includegraphics[width=0.5\textwidth]{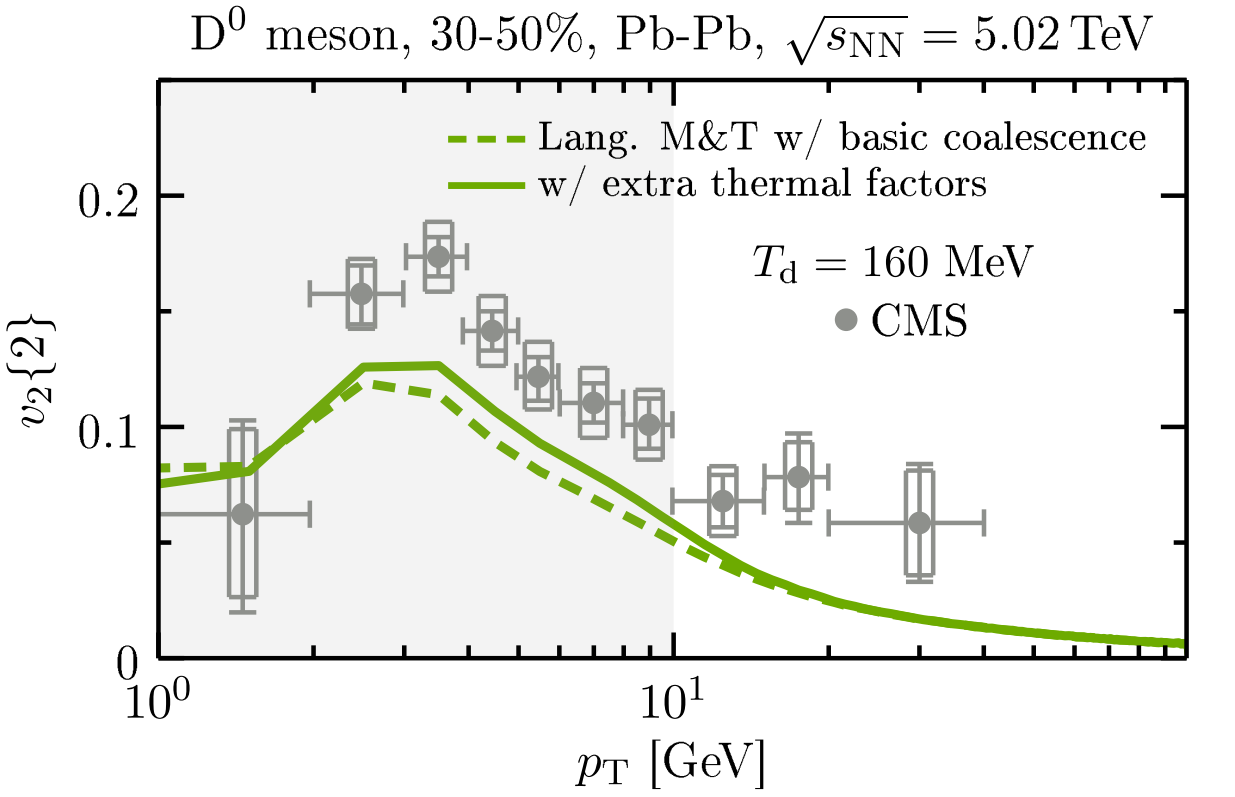}
  \caption{Direct $\Dzero$ meson elliptic flow coefficients $\vn2$ in the 30--50\% centrality class of $\snn[5.02]$ $\PbPb$ collisions obtained with different configurations of the coalescence model: the ``basic'' model, i.e. without the thermal factors (dashed lines) and the one including the thermal factors (solid lines). Prompt experimental data from the CMS ($|y|<1$) collaboration~\cite{Sirunyan:2017plt}.}
  \label{fig:v2coalbasic}
\end{figure}

The difference between the $v_2$ obtained with and without the thermal factors in the coalescence model is shown in Fig.\  \ref{fig:v2coalbasic}. The extra thermal factors are observed to change the $v_2$ except at $\pt\approx 1.5$ GeV where the coalescence probability is maximal (see Fig.\ref{fig:fewPcoal1}) and the fragmentation probability close to 0. The $v_2$ obtained through the coalescence and fragmentation processes being different, this effect can be explained by the different relative proportions of $\Dzero$ coming from coalescence and fragmentation in the two configurations of the coalescence model. More generally, these non-negligible variations of the observables with the coalescence model configuration underline the key role played by the $c\rightarrow\Dmeson$ coalescence probability which value depends on various features of the coalescence models, e.g. the angular frequency $\omega$ and the considered set of hadron states\footnote{Set of hadron states chosen differently for instance in~\cite{Oh:2009zj} and~\cite{Cao:2015hia}.}.

\subsection{Heavy mesons and all charged particles flow correlation}

\begin{figure}[!htb]
  \centering
 \includegraphics[width=0.36\textwidth]{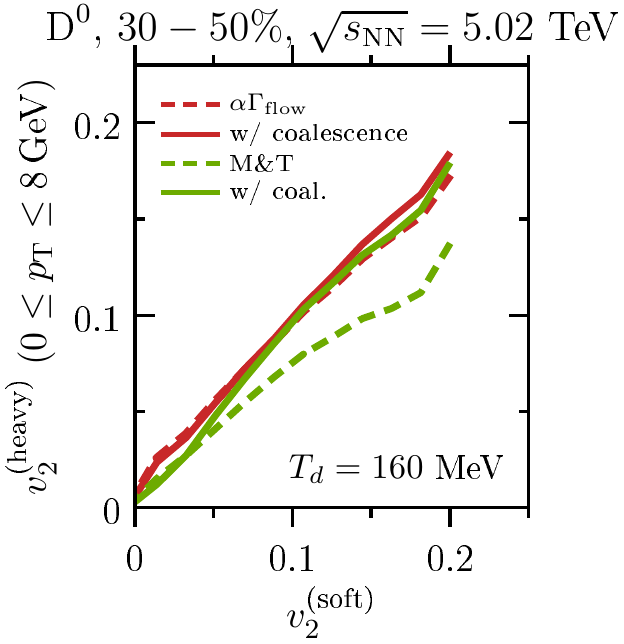}
\hspace{5mm}
 \includegraphics[width=0.36\textwidth]{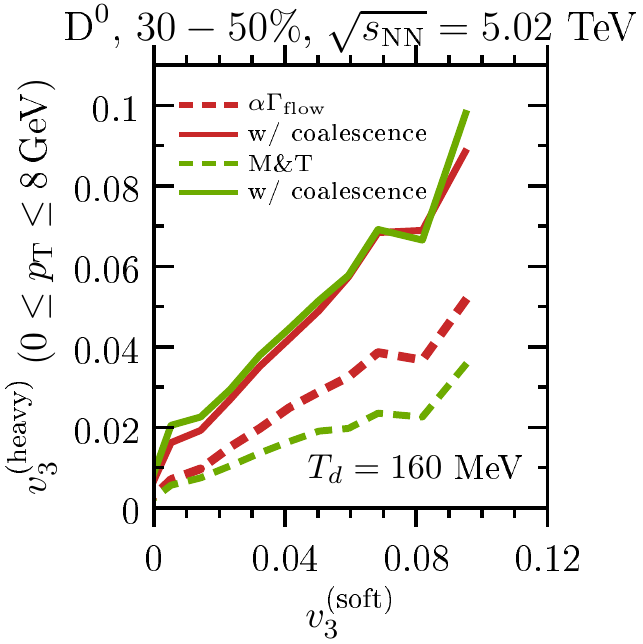}
  \caption{Correlations between the elliptic (left) (respectively triangular (right)) anisotropies of $\Dzero$ mesons and that of all charged particles in $\snn[5.02]$ $\PbPb$ collisions for the transverse momentum range 0-8 GeV comparing two transport models with and without coalescence.}
  \label{fig:Corrcoal}
\end{figure} 

With the addition of coalescence one can explore the low $\pt$ soft-hard flow correlations with more confidence. As seen in Fig.\ \ref{fig:Corrcoal}, coalescence increases the linear correlation between the heavy meson anisotropies $v_n^{\rm (heavy)}$ and the all charged particles elliptic flow $v_n^{\rm (soft)}$ proportionally to the gain in $\pt$ integrated $v_n^{\rm (heavy)}$ due to coalescence. The increase is then more important for the M\&T Langevin model in the $\pt$ range 0--8 GeV.

\begin{figure}[!htb]
  \centering
 \includegraphics[width=0.40\textwidth]{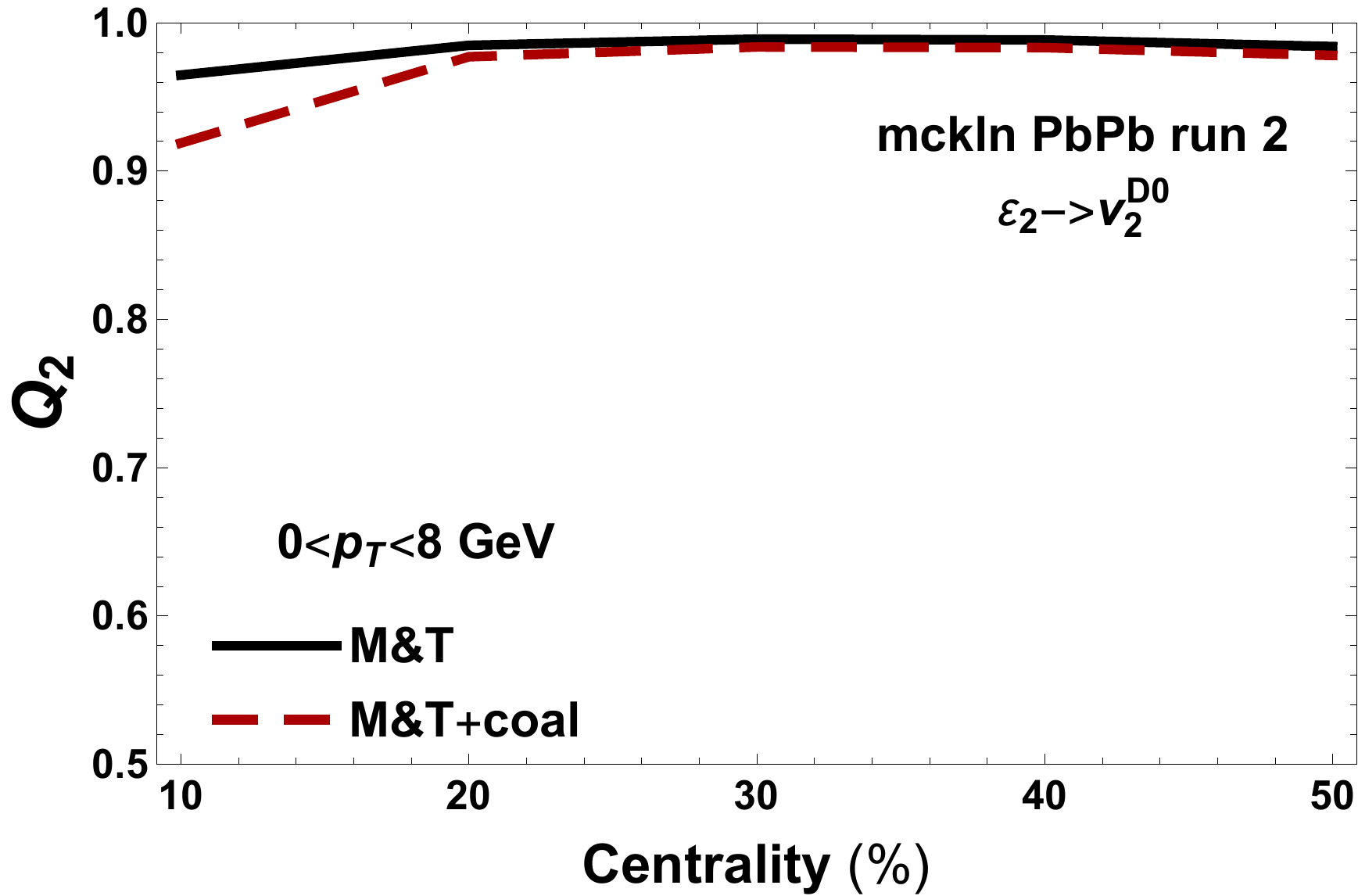}
\hspace{6mm}
 \includegraphics[width=0.40\textwidth]{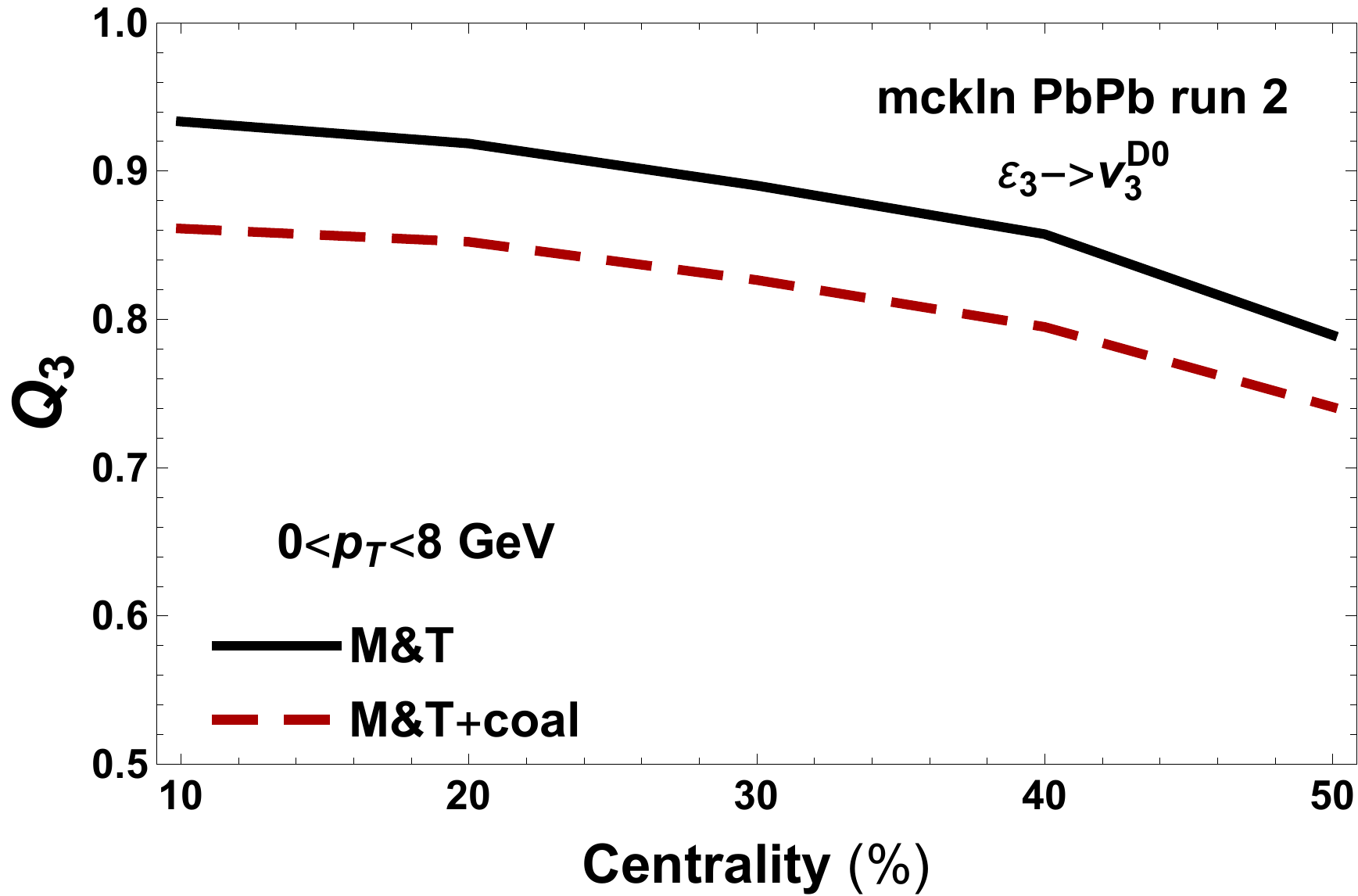}
  \caption{Effect of coalescence on the $\Dzero$ meson Pearson coefficients $Q_2$ (left) and $Q_3$ (right) defined in the transverse momentum range 0-8 GeV computed using the M\&T Langevin model for $\snn[5.02]$ $\PbPb$ collisions.}
  \label{fig:CorrQ2Q3}
\end{figure}

Fig.\ \ref{fig:CorrQ2Q3} shows how coalescence affects the Pearson coefficients associated with elliptic and triangular anisotropies, $Q_2$ and $Q_3$, computed using the M\&T Langevin model. One can see that the correlation weakens with the addition of coalescence, in contrast to what was observed in Fig.~\ref{fig:Corrcoal}. Coalescence tends to produce a somewhat more dispersed correlation on an event-by-event basis between the magnitudes $v_n^{\rm (heavy)}$ and $v_n^{\rm (soft)}$, and between the $\psi_n^{\rm (heavy)}$ and $\psi_n^{\rm (soft)}$ distributions leading to a slight event plane decorrelation.

\subsection{Elliptic flow from multi-particle cumulants}

\begin{figure}[!htb]
  \centering
\includegraphics[width=0.48\textwidth]{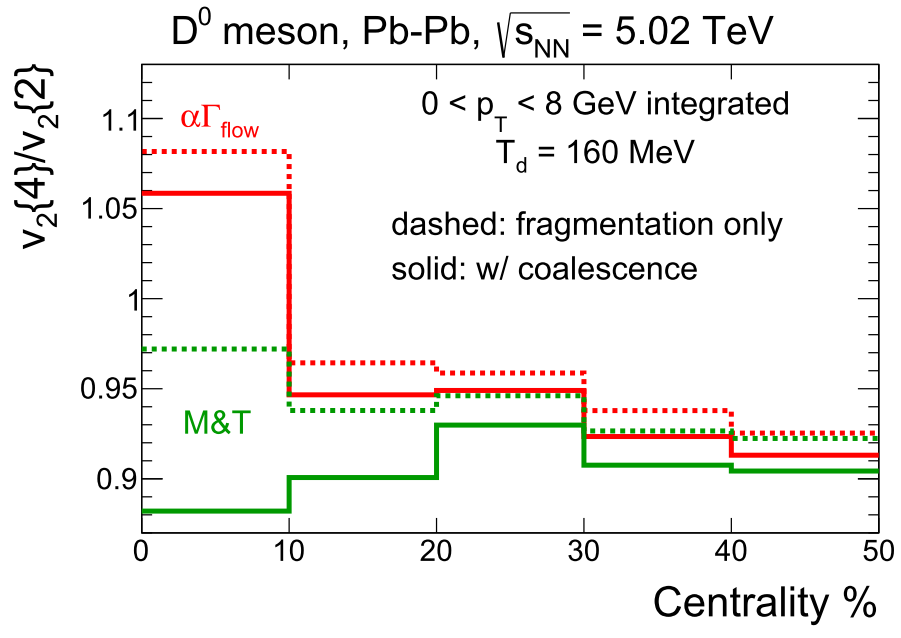}
 \includegraphics[width=0.48\textwidth]{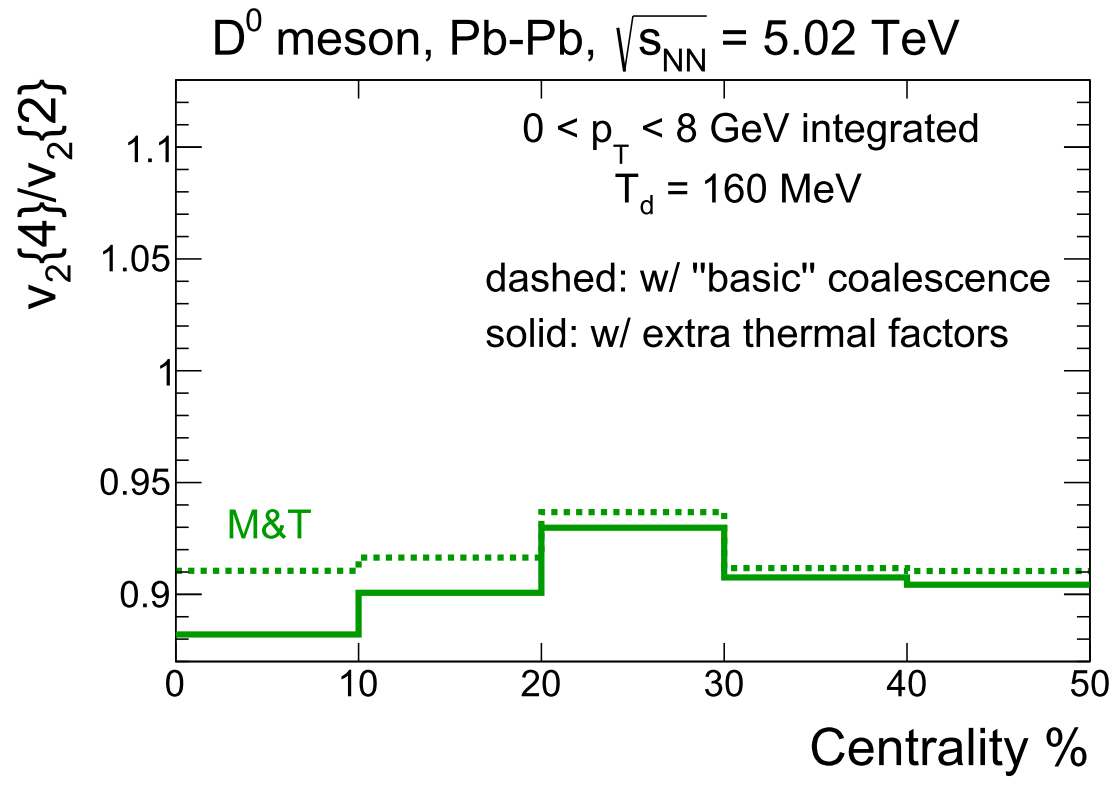}
  \caption{ Effect of coalescence on the $\Dzero$ $v_2\{4\}/v_2\{2\}$ cumulant ratio integrated over $0 < \pt < 8$ GeV for $\snn[5.02]$ $\PbPb$ collisions. \mckln\ initial conditions are used.}
  \label{fig:CumRatio_coal}
\end{figure}

We show in Fig.\ \ref{fig:CumRatio_coal} the influence of the hadronization process on the low $\pt$ integrated $v_2\{4\}/v_2\{2\}$ cumulant ratio as a function of centrality. The addition of coalescence is observed to generally decrease the low $\pt$ integrated cumulant ratio. Remembering Eq.\ \eqref{eqn:hardv24}, it is indeed consistent for the coalescence mechanism to reduce the cumulant ratio as it enhances the heavy quark anisotropy fluctuations with respect to the soft fluctuations obtained via the heavy-light quark combination. As shown on the left panel, the extra thermal factors lead to a larger decrease of the cumulant ratio that can be explained by a larger proportion of $\Dzero$ coming from coalescence than in the ``basic'' configuration.

\section{Conclusion} \label{conclusions}

In summary, in this paper we have presented the details behind \dabmod, a new heavy flavor model code that allows for a modular description of the dynamics of heavy quarks in heavy-ion collisions. The code can be run on top of any event-by-event relativistic hydrodynamic background and it samples the initial heavy quark distribution according to pQCD FONLL calculations. Heavy quark evolution is done either by solving relativistic Langevin equations or employing parametrized energy loss models (which include energy loss fluctuations). Finally, the heavy quarks can either be fragmented or coalesce to produce heavy flavor mesons (heavy meson decay is implemented via Pythia 8 to produce semileptonic channels). To improve the fits to the recently observed heavy hadron ratios at low $p_T$ and therefore the predictiveness of our coalescence model, we proposed an ``empirical'' inclusion to the usual probabilities of some new thermal-like factors (based on hadron masses) between hadrons of identical quark content. They have the additional asset of giving a natural explanation for the non-consideration of anti-symmetric and higher spin states (because of their larger mass which then suppresses their probabilities).

In this paper we have thoroughly checked the influence of a wide range of assumptions/parameters involved in the modeling of the heavy flavor sector in heavy ion collisions such as the inclusion or not of energy loss fluctuations, energy loss models vs. Langevin descriptions, heavy quark decoupling temperatures (which define when the heavy quarks decouple from the medium), heavy flavor meson vs. electrons/muon observables, choice of initial conditions for the background hydrodynamic evolution, as well as coalescence vs. fragmentation hadronization mechanisms. Similarly to~\cite{Betz:2014cza}, we find that energy loss fluctuations play a significant role in $v_n(\pt)$ calculations at $\pt<10$~GeV. In general, either a no energy loss fluctuation scenario (which we admit to be unrealistic) or a Gaussian distribution fare best compared to experimental data.
Our best fitting results at low $\pt$ stem from the Moore and Teaney inspired spatial diffusion coefficient within the Langevin formalism with the inclusion of coalescence. At high $\pt$ (roughly $\pt>5$ GeV), the energy loss model with a constant energy loss works the best when also coupled to coalescence. Thanks to the momentum brought by the light quark to the final meson, the addition of coalescence is observed to shift the low $p_T$ variations or lumps of the $\raa$ and $v_n$'s of around $\sim 2$ GeV towards higher $\pt$.

Unlike in~\cite{Andres:2019eus} where the effects of the initialization time were explicitly studied, in this work we kept a fixed heavy quark initialization time, $\tau_0=0.6$ fm, but we note that that is compatible with the best fit in~\cite{Andres:2019eus}. However, we remark that our two different setups for the initial conditions (\mckln\ vs. Trento) used different equations of state (the results computed using \mckln\ employed a now outdated equation of state from~\cite{Huovinen:2009yb} while the runs using Trento employed a newer one which contains the most up-to-date Particle Data booklet resonances from ~\cite{Alba:2017hhe}). In general, these correspond to different initialization temperatures $T_0$ and, thus, the \mckln\ events had a lower $T_0$ (even though $\tau_0$ was fixed) than the Trento events. Due to the significant run time that these different backgrounds take, we did not check this difference explicitly in this paper. However, we do plan in a later paper to investigate possible equation of state effects, especially its underlying assumptions such as whether charm quarks are thermalized or not~\cite{Borsanyi:2016ksw,Alba:2017hhe,Noronha-Hostler:2018zxc}. Additionally, it would be interesting to check the effects of a pre-equilibrium stage as in~\cite{Das:2017dsh}. Keeping in mind these limitations, the type of initial conditions is observed to have a little impact on the heavy flavor $\raa$ and $v_n\{2\}$'s within our model, with Trento leading to a slightly lower $v_2\{2\}$.

In this paper we investigated higher order flow coefficients involving heavy flavor. Similarly to the experimental data, the $v_3\{2\}$ is observed to be mostly centrality independent. An intriguing result was found concerning $v_4\{2\}$ of heavy flavor leptons in $\snnGeV[200]$ $\AuAu$ and $\snn[5.02]$ $\PbPb$ collisions: if the decoupling temperature parameter, $\Td$, is too large ($\Td=160$ MeV) then $v_4\{2\}$ becomes negative. This occurs when the heavy quark does not have enough time to interact with the medium, leading to an anti-correlation of the heavy flavor and bulk event planes. In~\cite{Aaboud:2018bdg} $v_4\{2\}$ was measured by the ATLAS collaboration and, with the current error bars, it is not yet clear what the overall sign of $v_4\{2\}$ is. However, for some centrality classes it does appear that it could be positive and in others negative. With future upgrades to the experiments it may be eventually possible to use $v_4\{2\}$ as a way to constrain the decoupling temperature and obtain an estimate of how long heavy quarks remain coupled to the expanding medium. This would be very useful on the theoretical front since $\Td$ produces a large systematic uncertainty in our calculations.

We also make predictions for multi-particle cumulants that have not yet been measured experimentally in the heavy flavor sector. We find that the ratio $v_2\{4\}/v_2\{2\}$ is different from the soft sector, and especially interesting in the most central collisions (0--10\%) and in peripheral collisions ($50-60\%$) because not only it is predominately sensitive to the initial conditions (which affects its overall magnitude) but it also has a different behavior across $\pt$ depending on the choice made to model heavy flavor evolution throughout the medium. Finally, $v_2\{4\}/v_2\{2\}$ has a rather small but non-trivial beam energy dependence so it would be especially interesting to see model comparisons to data from sPHENIX in a few years and also to data acquired at the top \lhc\ energies.

Additionally, we perform event-shape engineering calculations in the heavy flavor sector, which is the theory analog of what was experimentally done in~\cite{Acharya:2018bxo}. One caveat is that the results depend on the number of bins chosen within a centrality class (as well as the method of centrality binning). Thus, this must be set by an individual experimental collaboration depending on their available statistics before direct theory vs.~experimental comparisons can be made. In fact, we suggest the experiments to always publish the corresponding integrated  $v_n\{2\}$  of all charged particles for their event-shape engineered bins so more precise comparisons can be made. However, even with these caveats in mind, we clearly see linear correlations between soft and heavy $v_n$'s for both elliptical and triangular flow. We have observed as well the non-trivial correlations and decorrelations of the soft and heavy event planes $\psi_n$'s across centralities, colliding energies, heavy quark masses and models (e.g.~the energy loss models leading to more correlations than Langevin dynamics).

The development of \dabmod\ allows for a systematic comparison of a variety of approaches to heavy quark dynamics using state-of-the-art hydrodynamic backgrounds  that can describe the behavior of multi-particle cumulants in the soft sector of heavy-ion collisions. The predictions-data comparisons made throughout this paper tend to show that it is necessary to include more phenomena to the simulation, such as the initial shadowing or the final hadronic re-scattering, to improve its predictions (especially for the $v_n\{2\}$'s). With the upcoming era of sPHENIX the creation of \dabmod\ will allow for many future heavy flavor studies and constraints on our knowledge of heavy flavor dynamics. This will allow for many future comparisons involving different collisions systems, energies, and sizes (which will hopefully further constrain heavy flavor theoretical modeling) and become an invaluable tool to study heavy flavor in the upcoming \rhic\ and \lhc\ runs.

\section*{Acknowledgments}

We wish to thank Prof. Pol-Bernard Gossiaux and Shanshan Cao for their fruitful discussions and help. The authors thank Funda\c{c}\~ao de Amparo \`a Pesquisa do Estado de S\~ao Paulo (FAPESP) and Conselho Nacional de Desenvolvimento Cient\'ifico e Tecnol\'ogico (CNPq) for support. R.K. is supported by the Region Pays de la Loire (France) under contract No. 2015-08473. C.A.G.P. is supported by the NSFC under grant No. 11521064, MOST of China under Project No. 2014CB845404.  J.N.H. acknowledges the support of the Alfred P. Sloan Foundation, support from the US-DOE Nuclear Science Grant No. DE-SC0019175, and the Office of Advanced Research Computing (OARC) at Rutgers, The State University of New Jersey for providing access to the Amarel cluster and associated research computing resources that have contributed to the results reported here. J.N. is partially supported by CNPq grant 306795/2017-5 and FAPESP grant 2017/05685-2.

\bibliography{library}
\end{document}